\documentclass[prd,preprintnumbers,superscriptaddress,nofootinbib,floatfix,twocolumn,notitlepage,longbibliography]{revtex4-1}

\usepackage{textcase}
\usepackage{graphicx,epsfig,psfrag,amssymb}
\usepackage{multirow}
\usepackage{color,epsfig,amsmath,empheq}
\usepackage{bm}
\usepackage{mathrsfs,amsfonts}
\usepackage[tight,footnotesize]{subfigure}
\usepackage{hepunits}
\usepackage[utf8]{inputenc}
\usepackage[colorlinks=true
,urlcolor=blue
,anchorcolor=blue
,citecolor=blue
,filecolor=blue
,linkcolor=blue
,menucolor=blue
,linktocpage=true
,pdfproducer=medialab
,pdfa=true
]{hyperref}

\newcommand{\Ref}[1]{Ref.\:\cite{#1}}
\newcommand{\Refs}[1]{Refs.\:\cite{#1}}
\newcommand{\Fig}[1]{Fig.\:\ref{#1}}
\newcommand{\Figs}[2]{Figs.\:\ref{#1} and \ref{#2}}

\newcommand{\Tab}[1]{Table~\ref{#1}}

\newcommand{\Sec}[1]{Sec.\:\ref{#1}}

\newcommand{\App}[1]{App.\:\ref{#1}}

\newcommand{\Eq}[1]{Eq.\:(\ref{#1})}

\newcommand{\be}{\begin{eqnarray}}
\newcommand{\ee}{\end{eqnarray}}
\def\lsim{\mathrel{\rlap{\lower4pt\hbox{\hskip 0.5 pt$\sim$}}
\raise1pt\hbox{$<$}}}

\def\pt         {\mbox{$p_{\rm T}$}\xspace}
\newcommand{\tev}{\ensuremath{\mathrm{\: Te\kern -0.1em V}}\xspace}
\newcommand{\gev}{\ensuremath{\mathrm{\: Ge\kern -0.1em V}}\xspace}
\newcommand{\mev}{\ensuremath{\mathrm{\: Me\kern -0.1em V}}\xspace}

\def\pythia     {\mbox{\textsc{Pythia}}\xspace}
\def\herwig     {\mbox{\textsc{Herwig++}}\xspace}

\def\fjcontrib     {\mbox{\textsc{fjcontrib}}\xspace}
\def\dire     {\mbox{\textsc{Dire}}\xspace}
\def\vincia     {\mbox{\textsc{Vincia}}\xspace}

\def\powhegbox     {\mbox{\textsc{PowhegBox}}\xspace}

\def\Dbar    {{\kern 0.2em\overline{\kern -0.2em \mathrm{D}}{}}\xspace}

\def\jpsi     {{\ensuremath{{J\mskip -3mu/\mskip -2mu\psi\mskip 2mu}}}\xspace}
\def\kstarbar    {{\kern 0.2em\overline{\kern -0.2em K}{}^{*0}}\xspace}
\def\akt{anti-$k_{t}$\xspace}
\def\Akt{Anti-$k_{t}$\xspace}
\def\fcone{FlavorCone\xspace} 
\def\sdrop{SoftDrop\xspace}

\begin{document}

\title{Disentangling Heavy Flavor at Colliders}

\author{Philip Ilten}
\email{philten@cern.ch}
\affiliation{Laboratory for Nuclear Science, Massachusetts Institute of Technology, Cambridge, MA 02139, U.S.A.}

\author{Nicholas L. Rodd}
\email{nrodd@mit.edu}
\affiliation{Center for Theoretical Physics, Massachusetts Institute of Technology, Cambridge, MA 02139, U.S.A.}

\author{Jesse Thaler}
\email{jthaler@mit.edu}
\affiliation{Center for Theoretical Physics, Massachusetts Institute of Technology, Cambridge, MA 02139, U.S.A.}

\author{Mike Williams}
\email{mwill@mit.edu}
\affiliation{Laboratory for Nuclear Science, Massachusetts Institute of Technology, Cambridge, MA 02139, U.S.A.}

\begin{abstract}
We propose two new analysis strategies for studying charm and beauty quarks at colliders.
The first strategy is aimed at testing the kinematics of heavy-flavor quarks within an identified jet.
Here, we use the \sdrop jet-declustering algorithm to identify two subjets within a large-radius jet, using subjet flavor tagging to test the heavy-quark splitting functions of QCD.
For subjets containing a $\jpsi$ or $\Upsilon$, this declustering technique can also help probe the mechanism for quarkonium production.
The second strategy is aimed at isolating heavy-flavor production from gluon splitting. 
Here, we introduce a new \fcone algorithm, which smoothly interpolates from well-separated heavy-quark jets to the gluon-splitting regime where jets overlap.
Because of its excellent ability to identify charm and beauty hadrons, the LHCb detector is ideally suited to pursue these strategies, though similar measurements should also be possible at ATLAS and CMS.
Together, these \sdrop and \fcone studies should clarify a number of aspects of heavy-flavor physics at colliders, and provide crucial information needed to improve heavy-flavor modeling in parton-shower generators.
\end{abstract}

\preprint{MIT-CTP/4880}
\maketitle

\section{Introduction}
\label{sec:Introduction}

The production of charm and beauty quarks  at the Large Hadron Collider (LHC) is studied both as a fundamental probe of Standard Model (SM) phenomenology, and as an important component of searches for physics beyond the SM.
For example, heavy-flavor tagging is used to test the properties of the SM Higgs boson, whose largest branching fraction is to a pair of beauty quarks\,\cite{deFlorian:2016spz}. 
Similarly, identifying large-radius jets with double-flavor-tagged substructure enables searching for new physics scenarios involving high-\pt Higgs bosons\,\cite{ATLAS-CONF-2012-100,CMS-PAS-BTV-13-001,CMS-PAS-BTV-15-002,ATLAS-CONF-2016-039,Khachatryan:2016cfa,Aaboud:2016xco}.
To address SM backgrounds in both cases, it is essential to understand the mechanisms for heavy-flavor production at the LHC within quantum chromodynamics (QCD).
Of particular importance is the process of gluon splitting to heavy-quark pairs $g\to Q\bar{Q}$, where $Q$ denotes a $b$ or $c$ quark, which is challenging to study both theoretically and experimentally.

\begin{figure}[t]
\centering
	\subfigure[~Gluon Splitting]{\label{fig:LeadingOrderProduction:a}\includegraphics[scale=1.0]{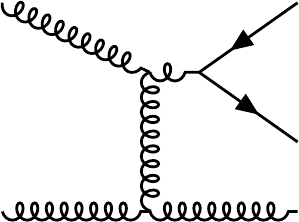}} \hspace{0.2in}
	\subfigure[~Flavor Excitation]{\label{fig:LeadingOrderProduction:b}\includegraphics[scale=1.0]{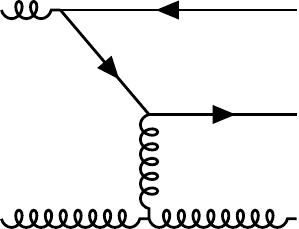}} \\
	\subfigure[~$s$-channel Flavor Creation]{\includegraphics[scale=1.0]{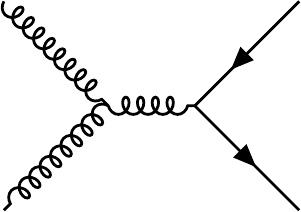}} \hspace{0.2in}
	\subfigure[~$t$-channel Flavor Creation]{\includegraphics[scale=1.0]{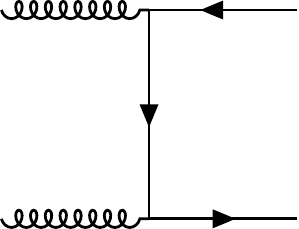}}
\caption{\footnotesize{Leading diagrams in QCD that contribute to heavy-flavor production at the LHC:
(a) {\em gluon splitting}, where a $Q\bar{Q}$ pair arises from a time-like off-shell gluon;
(b) {\em flavor excitation}, where $Q$ is excited from an incoming proton;
and
(c,d) {\em flavor creation}, where the $Q\bar{Q}$ pair originates directly from the hard scattering.
The precise $\alpha_s$ order at which these diagrams appear depends on whether one is working in a 3-, 4-, or 5-flavor scheme for parton distribution functions (PDFs).  Note that at higher orders, there is no gauge-invariant distinction between these categories.
}}
\label{fig:LeadingOrderProduction}
\end{figure}

In this article, we present two analysis strategies aimed at testing key features of heavy-flavor production  at the LHC.
First, we use a jet-declustering method to study heavy-flavor kinematics within identified jets, with the goal of testing the well-known but as-of-yet-unmeasured massive $1 \to 2$ splitting functions of QCD.
Second, we introduce a new jet algorithm designed to enable disentangling the various QCD-production processes for heavy flavor (see \Fig{fig:LeadingOrderProduction}), with an emphasis on studying the contribution from gluon splitting.
Both of these analyses can in principle be performed by any of the LHC experiments.
Here, we focus on the LHCb detector, which covers the pseudorapidity range $\eta \in [2,5]$, since its excellent heavy-flavor-identification capabilities offer the best short-term prospects.
In the appendices, we also present results for ATLAS and CMS, which cover $\eta \in [-2.5,2.5]$.
Qualitatively, the results of both proposed analyses are the same for LHCb and for ATLAS/CMS.

Our jet-declustering method is based on the \sdrop algorithm\,\cite{Larkoski:2014wba} and its precursor, the (modified) MassDrop tagger\,\cite{Butterworth:2008iy,Dasgupta:2013via,Dasgupta:2013ihk}.
Starting from a single large-radius jet, the declustering method strips away soft peripheral radiation and forces the groomed jet to have 2-prong substructure.
As shown in \Ref{Larkoski:2015lea}, the kinematics of the two resulting subjets match the famous Altarelli-Parisi $1 \to 2$ splitting functions  for massless QCD\,\cite{Altarelli:1977zs}.
\sdrop has been used by CMS\,\cite{CMS:2016jys} and STAR\,\cite{StarTalk} in the context of heavy-ion collisions, and a related strategy was proposed to test the dead cone effect for boosted top quarks\,\cite{Maltoni:2016ays}.
Here, we extend the analysis to QCD with heavy-flavor quarks,  exploiting the ability to flavor-tag subjets to test the splitting kinematics of $Q \to Q g$ and $g \to Q \bar{Q}$.
Because the \sdrop algorithm works equally well on tagged and untagged jets, we can compare our massive results directly to the massless case.
In addition, this method can be applied to quarkonium states like the $\jpsi$ and $\Upsilon$, potentially providing new insights into the puzzle of quarkonium polarization and fragmentation\,\cite{Schub:1995pu,McGaughey:1996wp,Sansoni:1996zs,Abe:1997jz,Brambilla:2010cs,Faccioli:2014cqa}.

Our gluon-splitting study is based on a new jet algorithm, referred to as \fcone, that identifies conical jets by centering the jet axes along the flight directions of well-identified flavor-tagged hadrons.
Unlike standard jet algorithms like \akt \cite{Cacciari:2008gp}, the \fcone method allows two jet axes to become arbitrarily close.
This feature, partially inspired by the XCone jet algorithm\,\cite{Stewart:2015waa,Thaler:2015xaa}, is ideal for studying gluon splitting to heavy flavor, where the two outgoing heavy quarks are often more collimated than the jet radius $R$.
In standard jet analyses, overlapping heavy-flavor jets are typically merged, with a precipitous drop in efficiency at angular scales smaller than $R$.
In the \fcone method, by contrast, heavy-flavor jet axes can be arbitrarily close, with the separate jet constituents determined by nearest-neighbor partitioning.
In this way, the \fcone algorithm enables a full exploration of the heavy-flavor production phase space, interpolating between the traditional regime of well-separated jets to the overlapping regime dominated by gluon splitting.

 Like standard approaches to studying high-\pt heavy-flavor production at the LHC, the \sdrop and \fcone strategies involve \emph{tagging} (sub)jets that contain heavy-flavor hadrons.
As we will see below, however, both the \sdrop and \fcone analyses require a definition of flavor tags that is more closely tied to heavy-flavor hadrons than typically required for tagging applications.
Specifically, it will be essential to reconstruct the \emph{flight directions} of heavy-flavor hadrons.
For \sdrop, these flight directions are used to define flavor-tagged subjet categories.
For \fcone, these flight directions directly determine the central jet axes.
In this way, the experimental requirements for---and challenges of---performing both analyses are largely shared.

As an alternative to (sub)jet tagging, one could perform exclusive \emph{reconstruction} of heavy-flavor hadrons.
From the experimental perspective, tagging is typically more efficient than reconstruction, since there are relatively few heavy-flavor decay modes that can be fully reconstructed.
From the theoretical perspective, analyses based on flavor-tagged jets are less sensitive to nonperturbative physics than those directly based on heavy-flavor hadrons.
To the extent that the typical jet scale $\pt R$ is larger than the heavy-flavor-hadron masses, the properties of heavy-flavor jets can be reliably calculated in (resummed) perturbative QCD, without the use of heavy-flavor fragmentation functions.
Of course, there are always nonperturbative corrections from hadronization and the underlying event, but jet-level measurements are generally expected to be closer to parton-level perturbative calculations.
In any case, jet-based and hadron-based analyses provide complementary information and both should be pursued when studying heavy flavor.

We validate the performance of these methods at the 13 TeV LHC using parton-shower generators.
Our primary focus is on \pythia 8.212\,\cite{Sjostrand:2006za,Sjostrand:2007gs,Sjostrand:2014zea}, which includes heavy-quark mass effects using matrix-element corrections\,\cite{Norrbin:2000uu}, and allows a leading-order classification of events into gluon-splitting and non-gluon-splitting topologies.
For the \sdrop study, we compare \pythia to \herwig~2.7.1\,\cite{Bahr:2008pv,Bellm:2015jjp} in order to test the robustness of the $1 \to 2$ subjet kinematics to different showering and hadronization models.\footnote{Because we are focusing on relatively low-\pt jets at LHCb, we generate {\em minimum bias} events, which precludes the use of NLO generators.}
For the \fcone study, we also consider alternative perturbative-shower results from \vincia~2.0.01\,\cite{Giele:2007di,Fischer:2016vfv} and \dire~0.900\,\cite{Hoche:2015sya}, as well as matched next-to-leading-order (NLO) results from \powhegbox~v2\,\cite{Nason:2004rx,Frixione:2007vw,Alioli:2010xd,Alioli:2010xa}, all using \pythia for hadronization.\footnote{These programs are not compatible with a common underlying event model; therefore, we turn off multiple parton interactions (MPI) in \pythia for the \fcone study to focus on perturbative physics.  In the case of \pythia, we tested that the addition of MPI does not impact our conclusions.}
Where needed, we use \textsc{FastJet}~3.1.2\,\cite{Cacciari:2011ma} for jet finding and the \textsc{RecursiveTools} \fjcontrib~1.024\,\cite{Cacciari:2011ma} for \sdrop.

The remainder of the article is organized as follows.
In \Sec{sec:Kinematics}, we show how \sdrop declustering can be used to study heavy-flavor kinematics within large-radius jets, including the kinematics of quarkonium production.
In \Sec{sec:Rate}, we define the \fcone jet algorithm and demonstrate how it can be used to disentangle heavy-flavor production processes in QCD.
We do not include detector-response effects on the distributions presented here, though we do discuss the prospects for applying these methods in the realistic LHCb environment in \Sec{sec:ImplementationAtLHCb}.
We conclude in \Sec{sec:conclusion}, leaving additional plots to the appendices.

\section{\sdrop Jet Declustering to Probe Heavy-Flavor Kinematics}
\label{sec:Kinematics}

The goal of our jet-declustering analysis is to study the collinear-splitting kernels of QCD appropriate for massive quarks.\footnote{For related work, see \Ref{Anderle:2017qwx}.}
These kernels form the basis for parton showers like \pythia, so we expect jet-declustering measurements will help improve theoretical predictions in the collinear regime.
We also present results for quarkonium production within an identified jet.
The current \pythia models for \jpsi and $\Upsilon$ production are known to be incomplete, so measurements of the quarkonium-splitting kinematics should provide valuable information.
In this section, we use the \sdrop algorithm along with heavy-flavor tagging to reveal the massive-quark splitting kernels.

\subsection{Review of \sdrop}

\sdrop is a jet-grooming technique that removes wide-angle soft radiation from a jet.
This algorithm is a generalization of the (modified) MassDrop tagger from \Refs{Butterworth:2008iy,Dasgupta:2013via,Dasgupta:2013ihk}, with an additional angular exponent $\beta$ that controls the degree of grooming.
In general, \sdrop reduces the dependence of the jet kinematics on other aspects of the full event, such as the underlying event, color correlations to the initial state, and pileup contamination.
Here, we will be primarily interested in using \sdrop to define $1 \to 2$ splitting kinematics.

\sdrop starts from a jet of radius $R$ that has been clustered with some jet algorithm, typically \akt.
Regardless of the clustering algorithm used to form the initial jet, one builds a Cambridge-Aachen (C/A) \cite{Dokshitzer:1997in,Wobisch:1998wt} clustering tree from the jet constituents.
Working backwards from the top of the tree, \sdrop recursively checks whether the two branches of the tree satisfy the following condition, set by the the grooming parameters $z_{\rm cut}$ and $\beta$:
\be
\frac{\min (p_{\rm T1}, p_{\rm T2})}{p_{\rm T1}+p_{\rm T2}} > z_{\rm cut} \left( \frac{R_{12}}{R} \right)^{\beta}\,, \label{eq:softdropcriterion}
\ee
where $p_{{\rm T}i}$ are the transverse momenta of the two branches and $R_{12}$ is their rapidity-azimuth separation.
If the condition in \Eq{eq:softdropcriterion} is not satisfied, then the softer of the two branches is dropped and the procedure is repeated on the next node down the C/A tree.
The procedure terminates once the \sdrop condition is satisfied, and the two final branches define the two \sdrop subjets.

The \sdrop algorithm has proven to be a valuable tool for the study of jets; see, for example, \Refs{Larkoski:2014bia,Dasgupta:2015yua,Kasieczka:2015jma,Adams:2015hiv,Luisoni:2015xha,Larkoski:2015npa,Aad:2015rpa,Dolen:2016kst,Frye:2016okc,Frye:2016aiz,Maltoni:2016ays,Moult:2016cvt,Dasgupta:2016ktv,Chien:2016led,Salam:2016yht}.
As already mentioned, \sdrop has been shown both theoretically \cite{Larkoski:2015lea} and experimentally \cite{CMS:2016jys,StarTalk} to expose the basic splitting functions of massless QCD.
Using a parton-shower analysis, we argue below that \sdrop can also be used to directly study the massive QCD splitting functions.

\subsection{Event Selection and Flavor Classification}

Because we want to compare the splitting kinematics for jets that contain different numbers of heavy-flavor-tagged hadrons, we define an event selection that is independent of the heavy-flavor content.
We start from large-radius merged jets without applying any flavor-tagged hadron requirements, and then use the following analysis workflow.
\begin{itemize}
\item We identify all flavor-tagged hadrons with $\pt > 2 \gev$  and treat their flight directions as {\em ghost} particles \cite{Cacciari:2008gn} for the purposes of jet clustering.\footnote{A ghost is a particle with infinitesimal energy, but well-defined direction, that is clustered for the purpose of (sub)jet heavy-flavor tagging.   See \Sec{sec:ImplementationAtLHCb} for a discussion of the experimental aspects of flavor-tagged hadrons.} For the case of charm tagging, we require that the $c$-hadron does not come from a $b$ decay.
\item We cluster the hadrons and the ghosts into \akt\ {\em fat} jets with $R=1.0$.
\item The hardest jet is required to have $\eta \in [3,4]$, so that the full nominal jet cone is within LHCb acceptance, and $\pt > 20 \gev$, which is a typical jet scale in LHCb.  In \App{app:CMSATLAS}, we show results relevant for ATLAS and CMS using a larger $p_{\rm T}$ threshold.
\item We apply the \sdrop jet-declustering algorithm to the hardest jet, taking the \sdrop parameters to be $\beta = 0$ and $z_{\rm cut} = 0.1$.  Note that with this choice of parameters \sdrop acts identically to the modified MassDrop tagger with $\mu=1$ \cite{Dasgupta:2013ihk}.
\item For each flavor-tagged hadron that is kept after \sdrop, we calculate
\be
\label{eq:ztagdef}
z_{\rm tag} = \frac{p_{\rm T}^{\rm tag}}{p_{\rm T1} + p_{\rm T2}},
\ee
where $p_{\rm T1}$ and $p_{\rm T2}$ are the transverse momenta of the two \sdrop subjets.  To count as a flavor tag in the classification scheme below, we require $z_{\rm tag} > 0.05$.
\end{itemize}
The resulting \sdrop subjets, and their flavor labels, form the basic objects of interest for subsequent analysis.
We choose $z_{\rm tag}$ to be half the value of $z_{\rm cut}$ in order to reduce kinematic dependence on the tagging condition, though one could further optimize the relationship between $z_{\rm tag}$ and $z_{\rm cut}$ to balance perturbative control against dependence on heavy-flavor fragmentation.
Experimentally, the method used to tag the hadron flavor must provide a measurement of the hadron flight direction, using, for example, the vector formed by connecting the $pp$-collision point to the hadron-decay vertex.

These selection requirements are loose, and require some care to implement properly in event generators.
Within the LHCb acceptance, we often find that the fat jet comes not from the hard-scattering process but from underlying event activity.
For this reason, we only test the \pythia and \herwig event generators, since they have full implementations of the underlying event including MPI.

\begin{figure*}[t]
\centering
\begin{tabular}{cccc}
\multirow{2}{*}{\subfigure[]{\includegraphics[width=0.24\textwidth, trim = 0 0 0 1.5in]{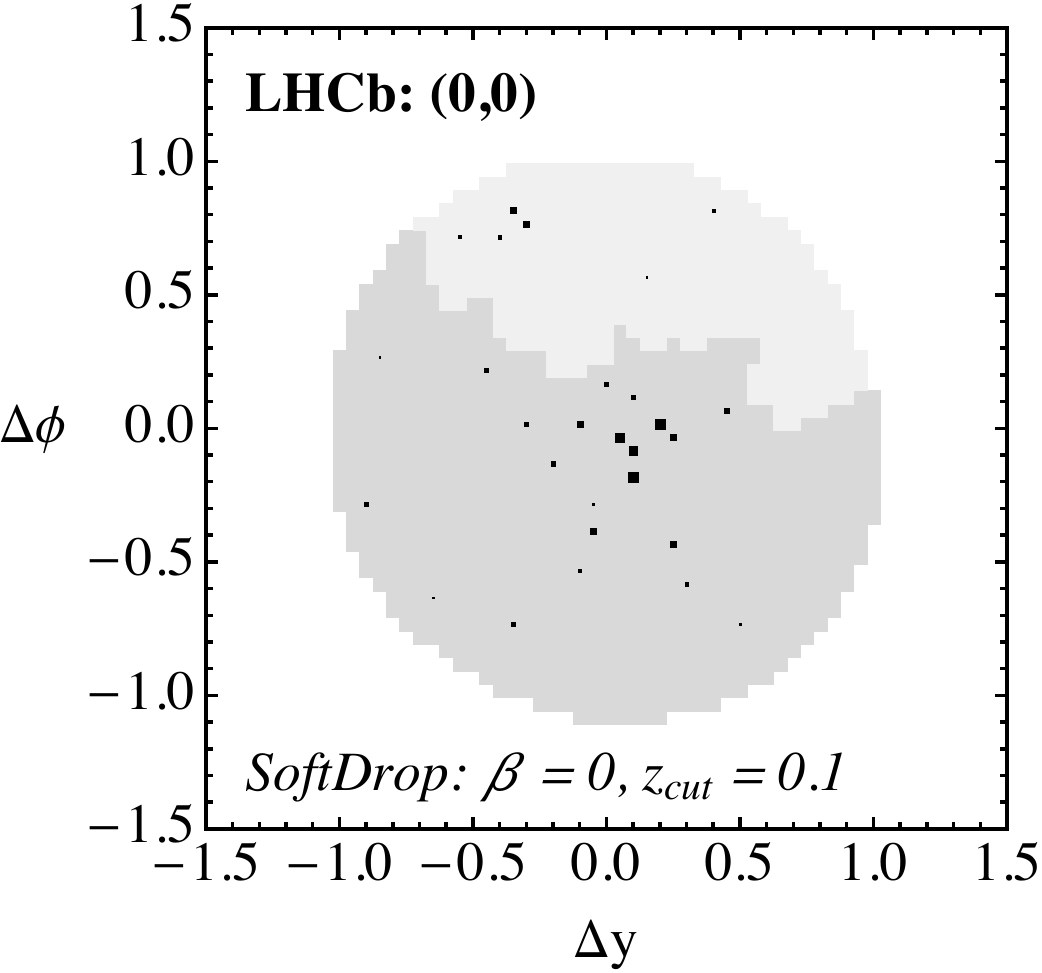}}}&
\subfigure[]{\includegraphics[width=0.24\textwidth]{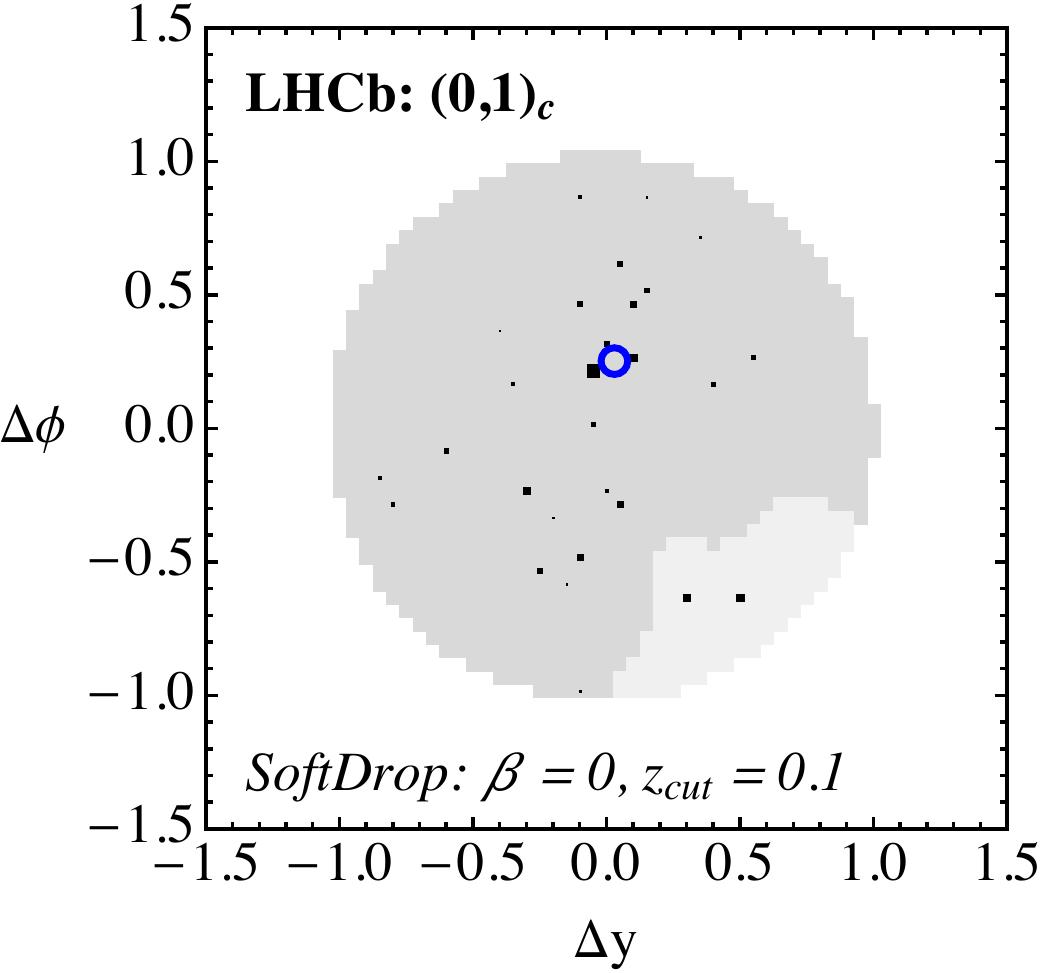}}&
\subfigure[]{\includegraphics[width=0.24\textwidth]{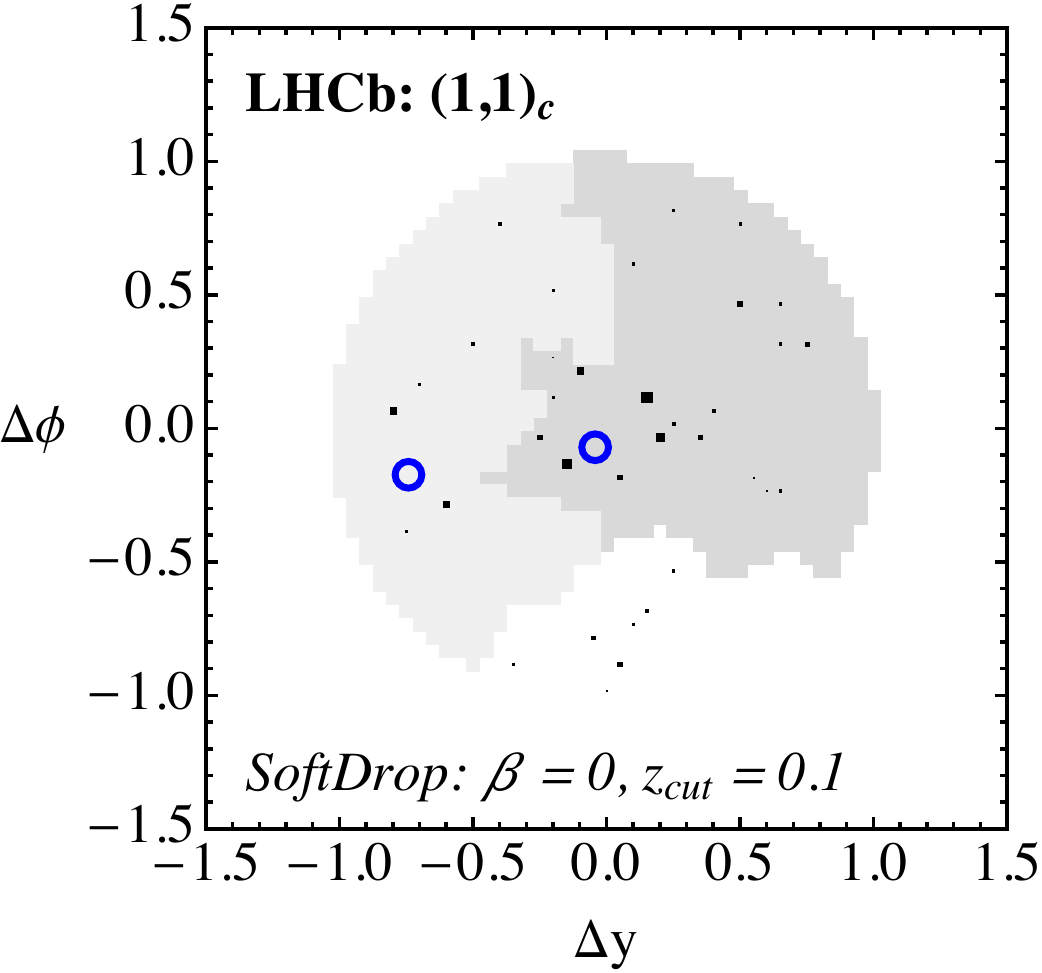}}&
\subfigure[]{\includegraphics[width=0.24\textwidth]{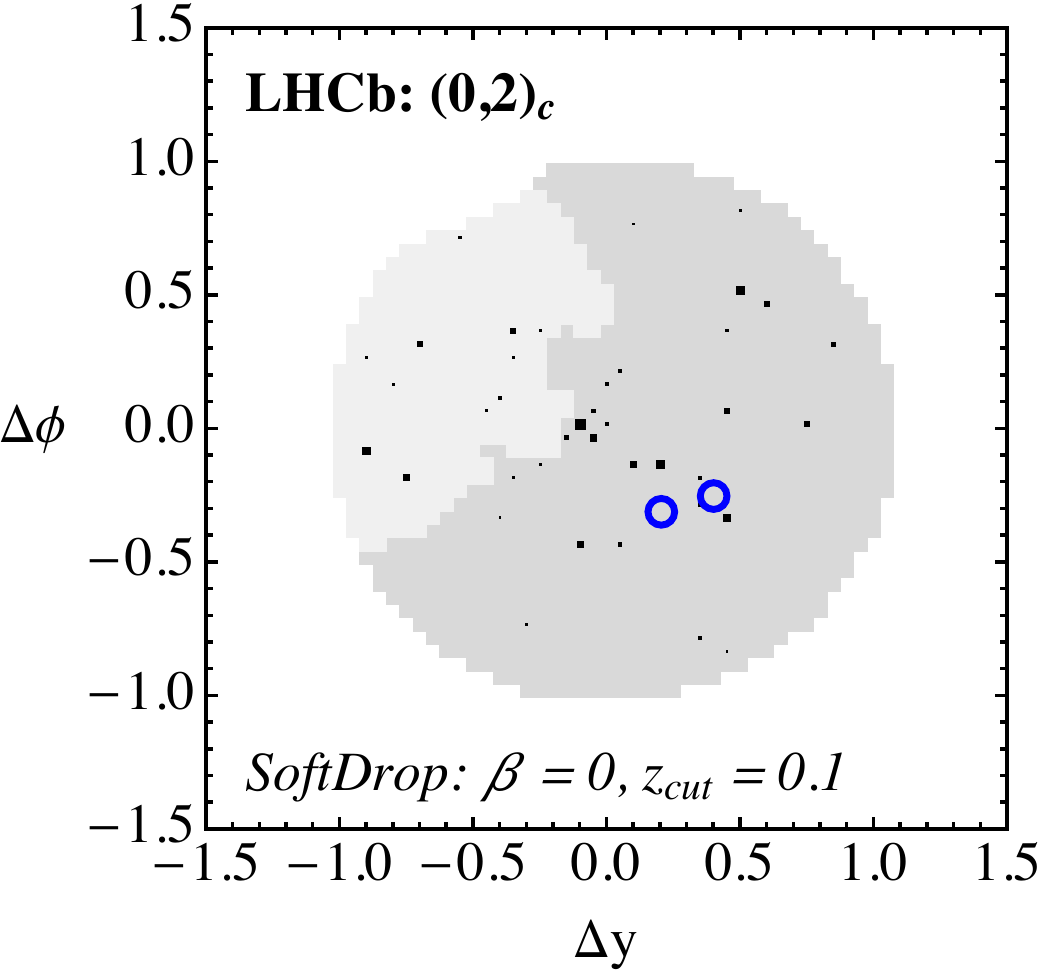}}\\
&
\subfigure[]{\includegraphics[width=0.24\textwidth]{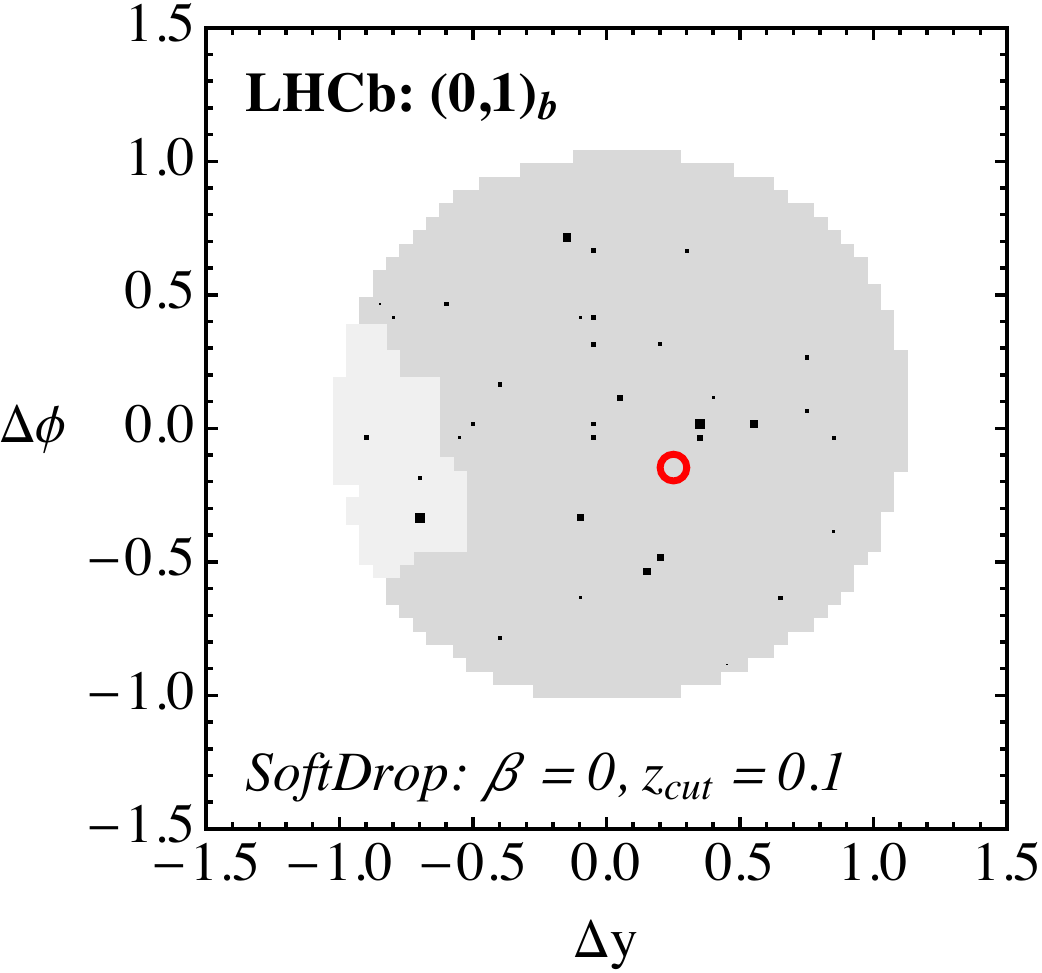}}&
\subfigure[]{\includegraphics[width=0.24\textwidth]{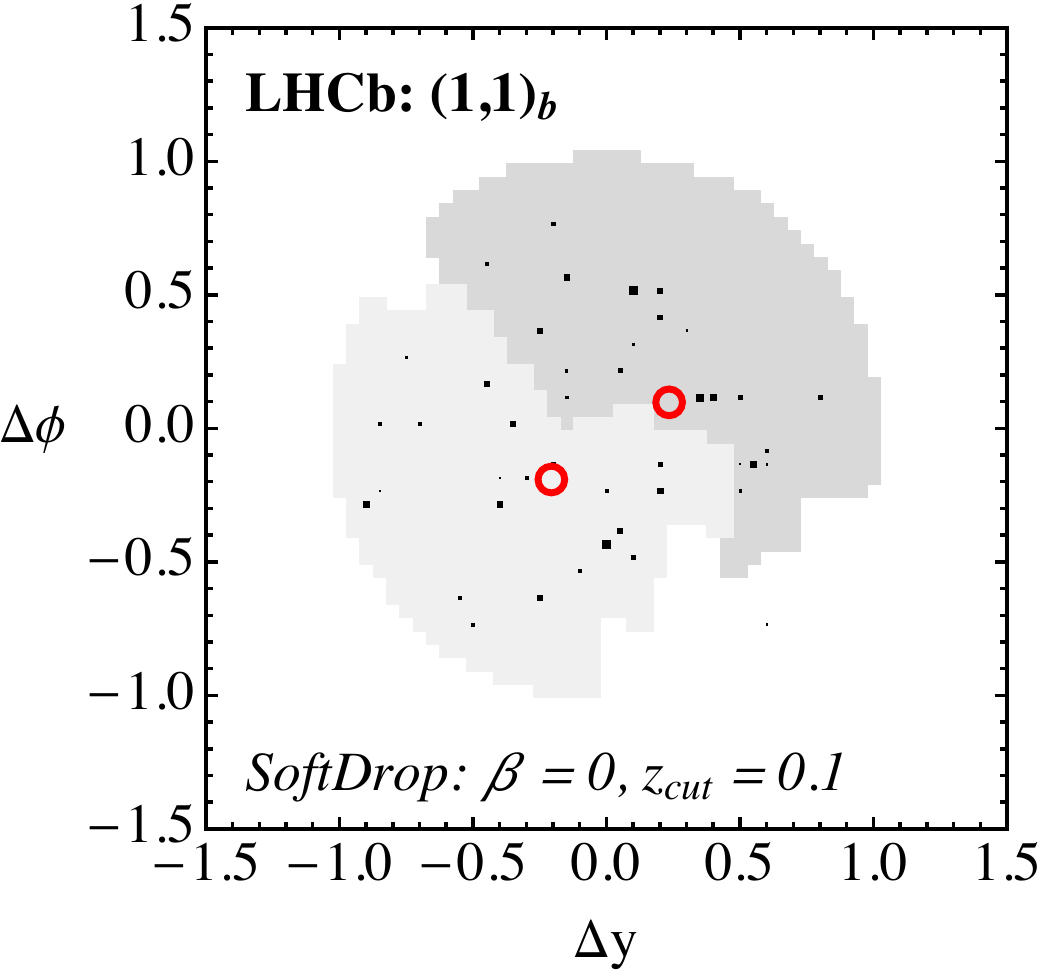}}&
\subfigure[]{\includegraphics[width=0.24\textwidth]{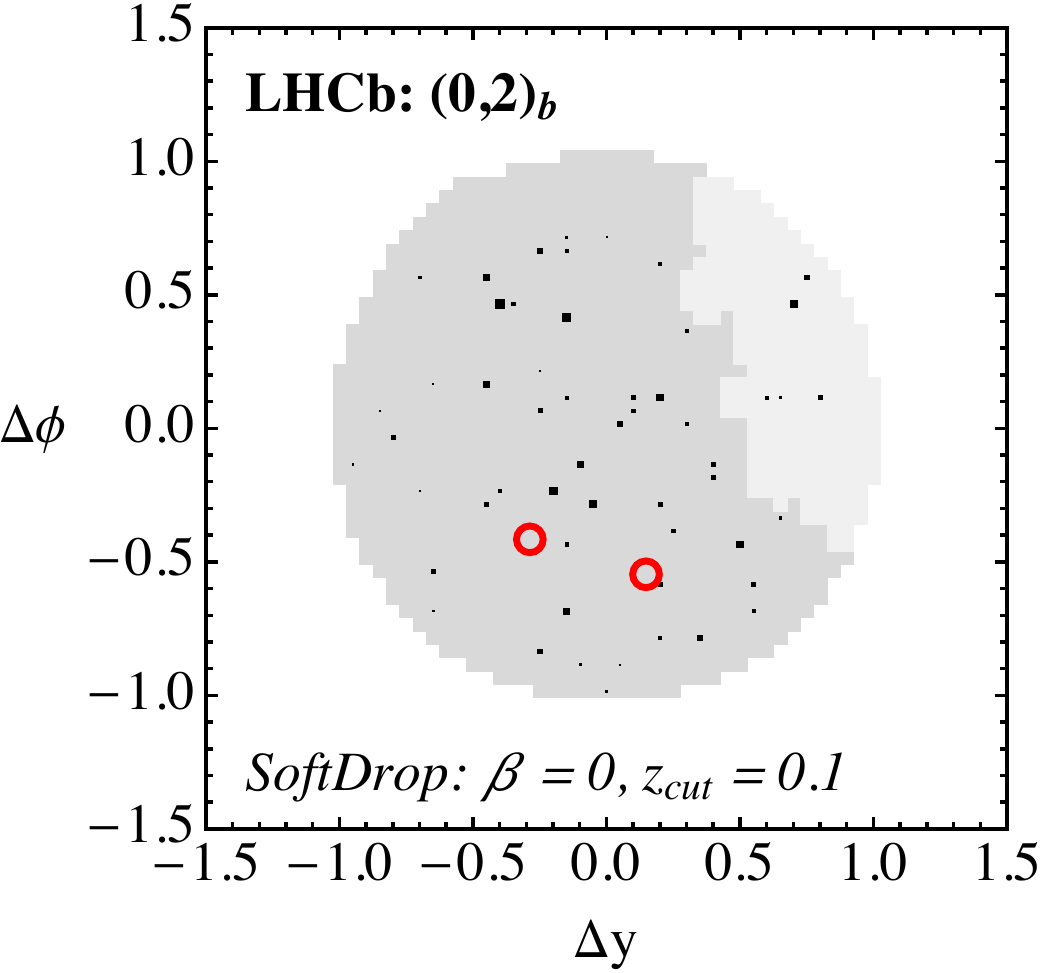}}\\
\end{tabular}
\caption{Event displays for \sdrop subjets with (a) no flavor-tagged hadrons, (b,c,d) $c$-tagged hadrons, and (e,f,g) $b$-tagged hadrons.  The fat \akt jet axis defines $\Delta y = \Delta \phi = 0$.  The filled black boxes represent particles clustered by \akt, where the area of the box is proportional to the \pt of the particle.  The flavor-tagged ghost particles are shown as open circles, with blue  for charm and red for beauty.  The (darker) leading and (lighter) subleading \sdrop subjets are displayed as shaded gray regions.
Because these are relatively low-\pt jets that are heavily contaminated by underlying event activity, the \sdrop procedure often terminates at the first stage of declustering such that no particles are removed from the \akt jet.}
\label{fig:sdrop_events}
\end{figure*}

To study the splitting kinematics, we use the flavor-tagged ghosts to label the \sdrop subjets.
Because the ghosts participate in the original \akt clustering and subsequent \sdrop declustering, they can become constituents of the subjets.
By counting the number of ghosts within each subjet that satisfy $z_{\rm tag} > 0.05$, we associate flavor labels and interpretations to the fat jets.
In cases where the ghost particles are removed by the \sdrop procedure, we simply ignore them when assigning flavor labels.\footnote{The dropped ghost tags are interesting in their own right, since they can be used to diagnose counterintuitive kinematical scenarios.  The main reason not to consider them for this study is to avoid a proliferation of curves on the following plots.}

Specifically, we label our jets as $(n_1,n_2)$, where $n_1$ is the number of heavy-flavor hadrons tagged in the first subjet and $n_2$ the number tagged in the second subjet, defined such that $n_1 \leq n_2$.
The cases we focus on are:
\begin{itemize}
\item \textit{No tagged subjets} (0,0):  Labeling light quarks generically as $q$, this category comes from $g \to gg$, $q \to q g$, and $g \to q \bar{q}$.
\item \textit{One single tag} (0,1):  This most likely arises from a heavy quark emitting a gluon, $Q \to Qg$.
\item \textit{Two single tags} (1,1):  Here, both subjets are associated with heavy flavor, which usually arises from gluon splitting, $g \to Q \bar{Q}$.
\item \textit{One double tag} (0,2):  Double-tagged subjets come from cascaded splittings such as $g \to gg$ followed by $g \to Q \bar{Q}$, making their interpretation in terms of splitting functions more complicated.
\end{itemize}
We also rarely find jets with $n_1 + n_2 > 2$, which are ignored in the analysis presented below.
For simplicity, we only treat one flavor label at a time, such that the $c$-tagged categories do not include an explicit veto on $b$-tagged objects, and vice versa.
In \Fig{fig:sdrop_events}, we show some example \sdrop event displays.

\subsection{Splitting Function Interpretation}
\label{subsec:splittingfuncinterpret}

Because the above flavor-tagged categories are based on identified hadrons, they can be directly implemented in an experimental analysis.
Of course, at the level of idealized partons, there can be category migration if one of the flavor tags is removed by the \sdrop procedure or fails the $z_{\rm tag}$ condition, and this has to be accounted for when interpreting the observed distributions.
For example, a (0,2) jet with one dropped tag becomes a (0,1) jet.
In addition, soft $g \to Q \bar{Q}$ splittings can contaminate the flavor labels, though this effect is highly suppressed by the $z_{\rm tag}$ condition.
Being mindful of migration, it is instructive to discuss the expected kinematical distributions for each of the flavor-tagged categories.

We are specifically interested in the  momentum sharing $z_g$ between the \sdrop subjets, and adopt a modified definition of $z_g$ compared to the literature  
\be
\label{eq:zg_def}
z_g \equiv \frac{p_{\rm T1}}{p_{\rm T1}+p_{\rm T2}},
\ee
where the $1$ and $2$ subjet labels are derived from the $(n_1, n_2)$ flavor-tagged label, instead of being ordered by \pt, such that $z_g \in [z_{\rm cut}, 1-z_{\rm cut}]$.  In cases where $n_1 = n_2$, we randomize the ordering of the subjets resulting in a $z_g$ distribution that is symmetric about $z_g = 1/2$.

For the (0,0) case, which has no flavor tags, this is essentially massless QCD with $N_f = 3$.  As shown in \Ref{Larkoski:2015lea} and experimentally measured by CMS\,\cite{CMS:2016jys} and STAR\,\cite{StarTalk}, the $z_g$ distribution is closely related to the massless-QCD splitting kernels.  Specifically, for $\beta = 0$ and to lowest order in $\alpha_s$, the probability distribution for $z_g$ is given by
\be
p_i(z_g) = \frac{\overline{P}_i(z_g)}{\int_{z_{\rm cut}}^{1/2} \mathrm{d} z' \overline{P}_i(z')} \Theta(z > z_{\rm cut}),
\ee
where $i$ labels the initiating parton for the jet.  Here, $\overline{P}_i(z)$ are symmetrized versions of the QCD splitting functions for parton $i$ summed over all final state partons,
\be
\overline{P}_i(z) = \sum_{jk} \bigl(P_{i \to jk}(z) + P_{i \to jk}(1-z) \bigr).
\ee
Because we are not distinguishing between quark and gluon (sub)jets in this analysis, the measured $p(z_g)$ distribution probes a combination of all massless splittings: $g \to gg$, $q \to qg$, and $g \to q \bar{q}$.  For $N_f = 3$, the symmetrized splitting functions for quarks and gluons are identical to this order,
\be
\label{eq:massless_splitting}
\overline{P}_q(z) \simeq \overline{P}_g(z) \simeq  \frac{1- z}{z} + \frac{z}{1-z} + \frac{1}{2}.
\ee
Note that $\overline{P}$ does not include the Casimir factor ($C_q = 4/3$ and $C_g = 3$), which drops out from the $p(z_g)$ distributions at lowest order in $\alpha_s$.

For the (0,1) case of one flavor tag, the dominant contribution comes from $Q \to Q g$.
In this case, the $z_g$ distribution depends on the quasi-collinear splitting function \cite{Catani:2000ef}, which is \emph{not} symmetrized over the two subjets:\footnote{Note that we are using the reversed convention of $z$ versus $1-z$ in the splitting function in order to match the definition of $z_g$.}
\be
\label{eq:massive_splitting}
P_{Q \to Qg}(z) = \frac{1-z}{z} + \frac{z}{2} - 2 \mu^2_{Qg}.
\ee
Here, the mass ratio is
\be
\mu^2_{Qg} = \frac{m_Q^2}{m_{Qg}^2 - m_Q^2},
\ee
and $m_{Qg}$ is the invariant mass of the heavy quark plus gluon system.  Taking the $m_Q \to 0$ limit and symmetrizing $z \to 1-z$, one recovers \Eq{eq:massless_splitting} as expected.  By comparing the (0,1) and (0,0) distributions, it is possible to test the splitting-function form in \Eq{eq:massive_splitting}.\footnote{\pythia implements the heavy-flavor splitting functions using a matrix-element correction \cite{Norrbin:2000uu} instead of the $-2\mu^2_{Qg}$ term in \Eq{eq:massive_splitting}.}

\begin{figure*}[t]
\centering
\begin{tabular}{c}
\subfigure[]{\includegraphics[scale=0.4]{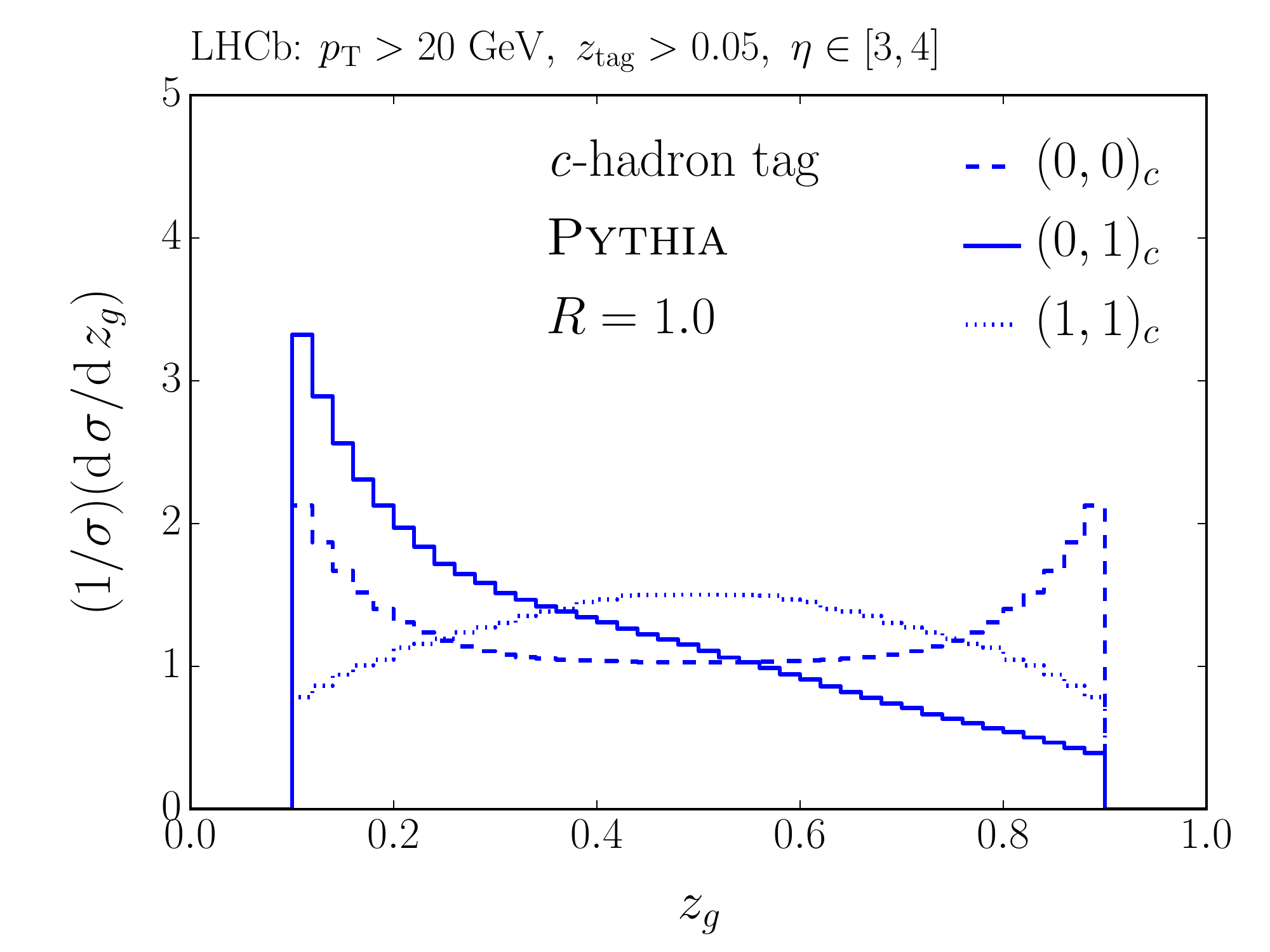}} \hspace{0.1in}
\subfigure[]{\includegraphics[scale=0.4]{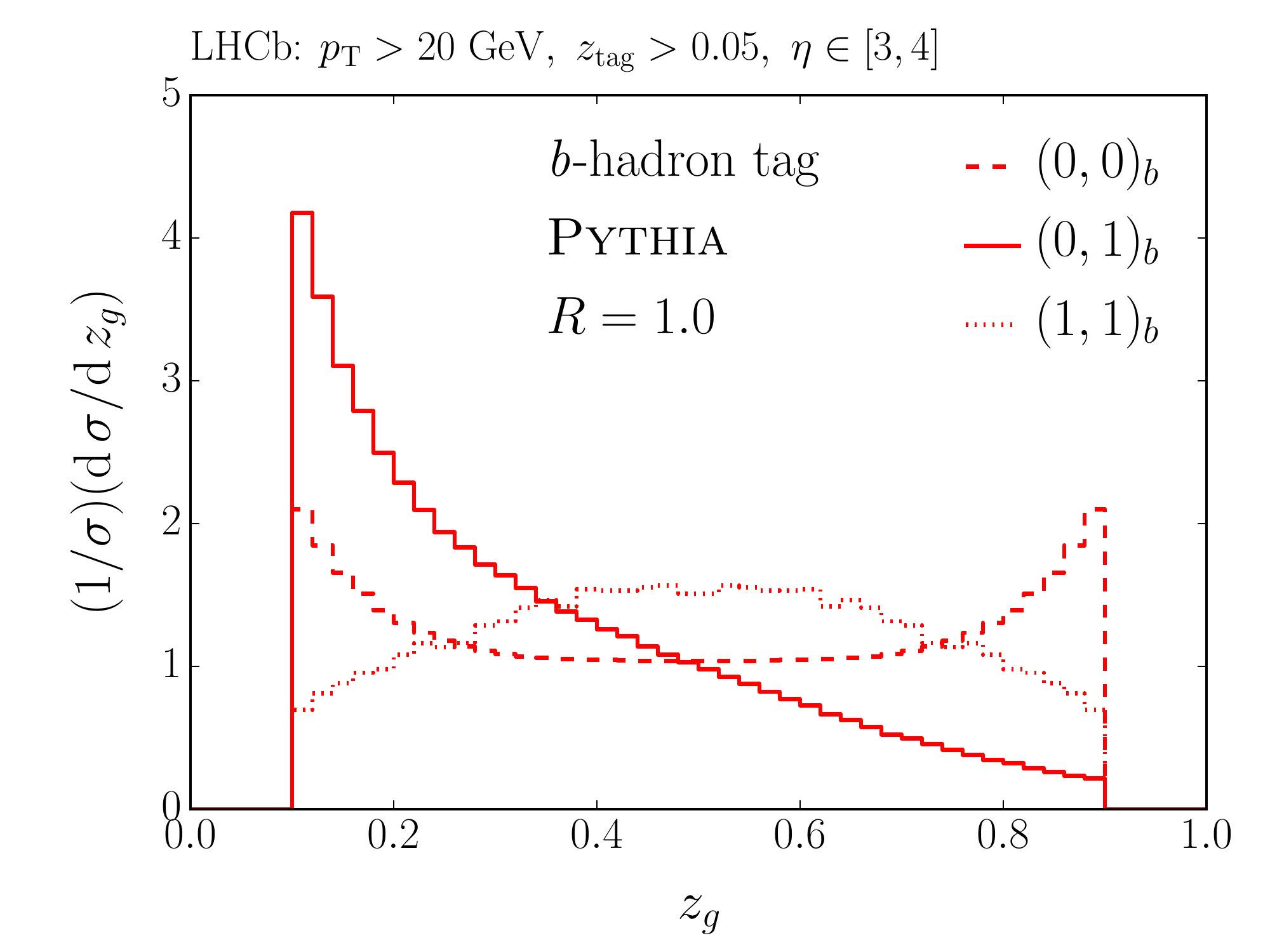}}
\end{tabular}
\caption{
The \sdrop $z_g$ distributions for the (a) $c$-tagged and (b) $b$-tagged categories.  Shown here are the results for the (0,0), (0,1), and (1,1) categories obtained from \pythia.
}
\label{fig:Kinz}
\end{figure*}

For the (1,1) category with one flavor tag in each subjet, the dominant process is $g \to Q \bar{Q}$.
The quasi-collinear splitting function for this case is \cite{Catani:2000ef}
\be
\label{eq:gQQsplitting}
P_{g \to Q \bar{Q}}(z) = z^2 + (1-z)^2 +  \mu^2_{Q\bar{Q}},
\ee
the mass ratio is
\be
\mu^2_{Q\bar{Q}} = \frac{2 m_Q^2}{m_{Q \bar{Q}}^2},
\ee
and $m_{Q \bar{Q}}$ is the invariant mass of the heavy-quark pair.\footnote{In the \pythia implementation, the $\mu^2_{Q\bar{Q}}$ term is multiplied by an additional factor of $4 z (1-z)$ \cite{Sjostrand:2014zea}.  This explains why the $(1,1)$ category in \Fig{fig:Kinz} exhibits a downturn towards $z \to 0$ and $z \to 1$.  As one goes to higher jet $\pt$, this additional factor is less important, and one recovers the expected upturn from \Eq{eq:gQQsplitting}; see \Figs{fig:GPDKinz:a}{fig:GPDKinz:b}.}  Note the absence of any singular behavior in the $z \to 0$ or $z \to 1$ limits, as expected since this process does not have a soft singularity.

Finally, the (0,2) category, where one subjet has two flavor tags, does not have a simple interpretation in terms of $1\to 2$ splitting functions.
In a parton shower, this configuration can be obtained from $g \to gg$ followed by $g \to Q \bar{Q}$.
More intuitively, one can think of the double-tagged subjet as being a color-octet configuration that radiates soft gluons via $(Q \bar{Q})_8 \to (Q \bar{Q})_8 g$.
In this color-octet interpretation, the (0,2) distribution is expected to look like the (0,1) case with the replacement $m_Q \to 2m_Q$, since the different Casimir factors do not appear in $P(z)$ at lowest order.
It is of particular interest to compare the (0,2) category to the quarkonium case studied in \Sec{subsec:quarkonium}.

In addition to $z_g$, the other natural kinematic observable for \sdrop jets is $R_g$, the opening angle between the two subjets.
For massless partons, the $R_g$ distribution was calculated to next-to-leading-logarithmic accuracy in \Ref{Larkoski:2014wba}.
The $R_g$ distribution for massive partons has not been calculated in the literature, though \Ref{Maltoni:2016ays} used a variant of $R_g$ to test the dead cone effect for boosted top quarks.
We do not show the perturbative predictions here, since for the jet \pt  range of interest for LHCb, the $R_g$ distribution is dominated by nonperturbative physics and is relatively insensitive to the flavor content of the jets.
For completeness, we show the $R_g$ distributions in \App{app:extraplotsSD}.

In the analysis below, we treat the jet fragmentation process as if it were rotationally symmetric about the jet axis.
As recently discussed in \Ref{Anderle:2017qwx}, though, it is interesting to study the angle between the jet production plane and the subjet decay plane.
For the case of $g \to Q \bar{Q}$, this angle is sensitive to gluon polarization, motivating future multi-differential studies of the full \sdrop subjet decay phase space.

\begin{table}[t]
\begin{center}
\begin{tabular}{ l @{\hskip 0.4in} l @{\hskip 0.4in}  l }
\hline \hline
& \hspace{-0.15in}$\sigma(\pythia)$ [$\mu$b] & \hspace{-0.25in}$\sigma(\herwig)$ [$\mu$b] \\ \hline
$(0,0)_c$ &$9.96\times10^ {2}$ & $5.28\times10^ {2}$ \\ 
$(0,1)_c$ &$7.56\times10^ {1}$ & $2.64\times10^ {1}$ \\
$(1,1)_c$ &$6.87\times10^ {0}$ & $2.87\times10^ {0}$ \\
$(0,2)_c$ &$1.00\times10^ {1}$ & $5.64\times10^ {0}$ \\
other$_c$ &$8.86\times10^{-1}$ & $2.47\times10^{-1}$ \\
\hline
$(0,0)_b$ & $1.07\times10^ {3}$ & $5.52\times10^ {2}$ \\
$(0,1)_b$ & $1.34\times10^ {1}$ & $9.58\times10^ {0}$ \\
$(1,1)_b$ & $8.40\times10^{-1}$ & $5.03\times10^{-1}$ \\
$(0,2)_b$ & $9.50\times10^{-1}$ & $5.94\times10^{-1}$ \\
other$_b$ & $1.13\times10^{-2}$ & $7.75\times10^{-3}$ \\
\hline
$(0,1)_{J/\psi}$  & $3.03\times10^{-1}$ & -- \\
$(0,1)_{\Upsilon}$ & $1.54\times10^{-2}$ & -- \\
\hline \hline
\end{tabular}
\end{center}
\caption{The cross sections for each of the fat-jet flavor-tagged categories determined from \pythia and \herwig, where the total cross section is normalized to the nominal inelastic cross section of $100~\mathrm{mb}$.
 Because we only consider one flavor label at a time, the sum of the $c$-categories equals the sum of the $b$-categories.
We also show cross sections for quarkonium production in \pythia.}
\label{table:tag_rate}
\end{table}

\begin{figure*}[t]
\centering
\begin{tabular}{c}
\subfigure[]{\includegraphics[scale=0.4]{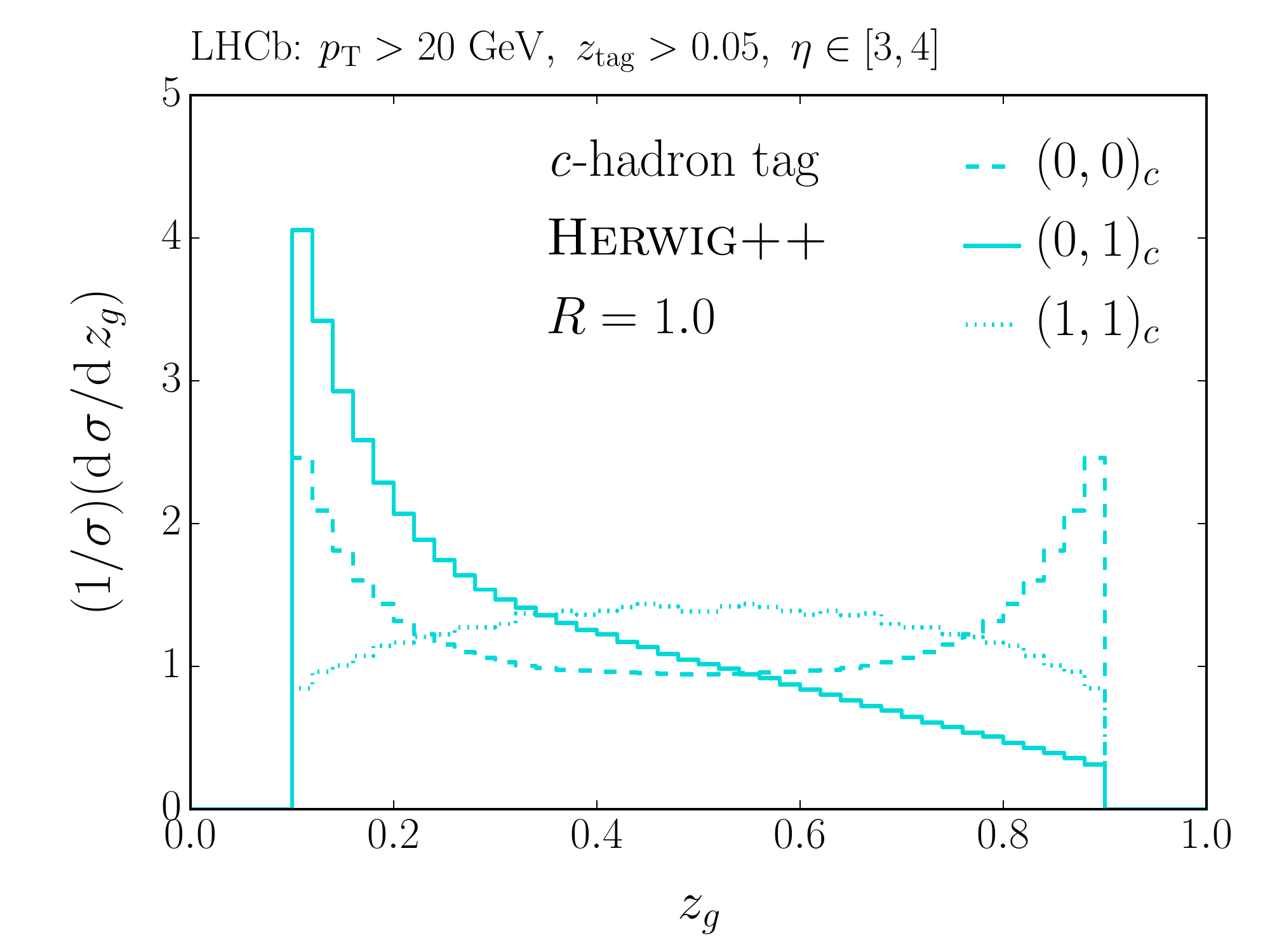}} \hspace{0.1in}
\subfigure[]{\includegraphics[scale=0.4]{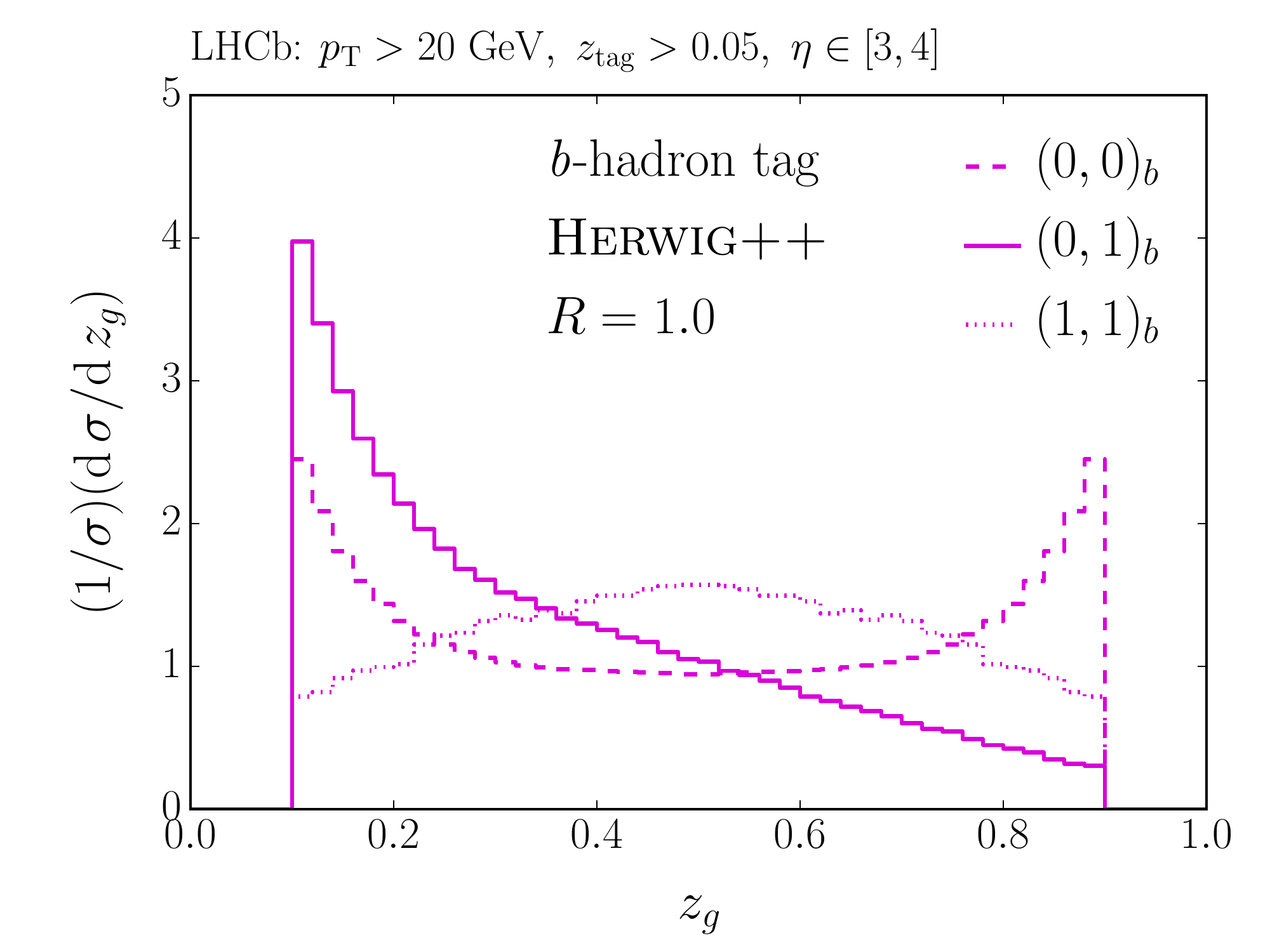}}
\end{tabular}
\caption{Same as \Fig{fig:Kinz} but for \herwig.}
\label{fig:Kinz_herwig}
\end{figure*}

\subsection{Results: Heavy-Quark Splittings}

Using this \sdrop jet-declustering strategy, we first consider the inclusive cross section for each of the flavor-tagged categories in \Tab{table:tag_rate}.
Quantitatively, the predictions obtained from \pythia and \herwig do not agree: both the absolute and relative cross sections show sizable discrepancies.
There is qualitative agreement, however, as both generators predict that the $(0,0)$ category with no flavor tags dominates the total rate, followed by the $(0,1)$ category which is largely due to $Q\to Qg$.
The $(1,1)$ and $(0,2)$ categories, which arise from $g\to Q\bar{Q}$ and cascaded splittings, respectively, are predicted to have similar rates, while events with $n_1 + n_2 > 2$ are rare as expected.

\begin{figure*}[t]
\centering
\begin{tabular}{c}
\subfigure[]{\includegraphics[scale=0.4]{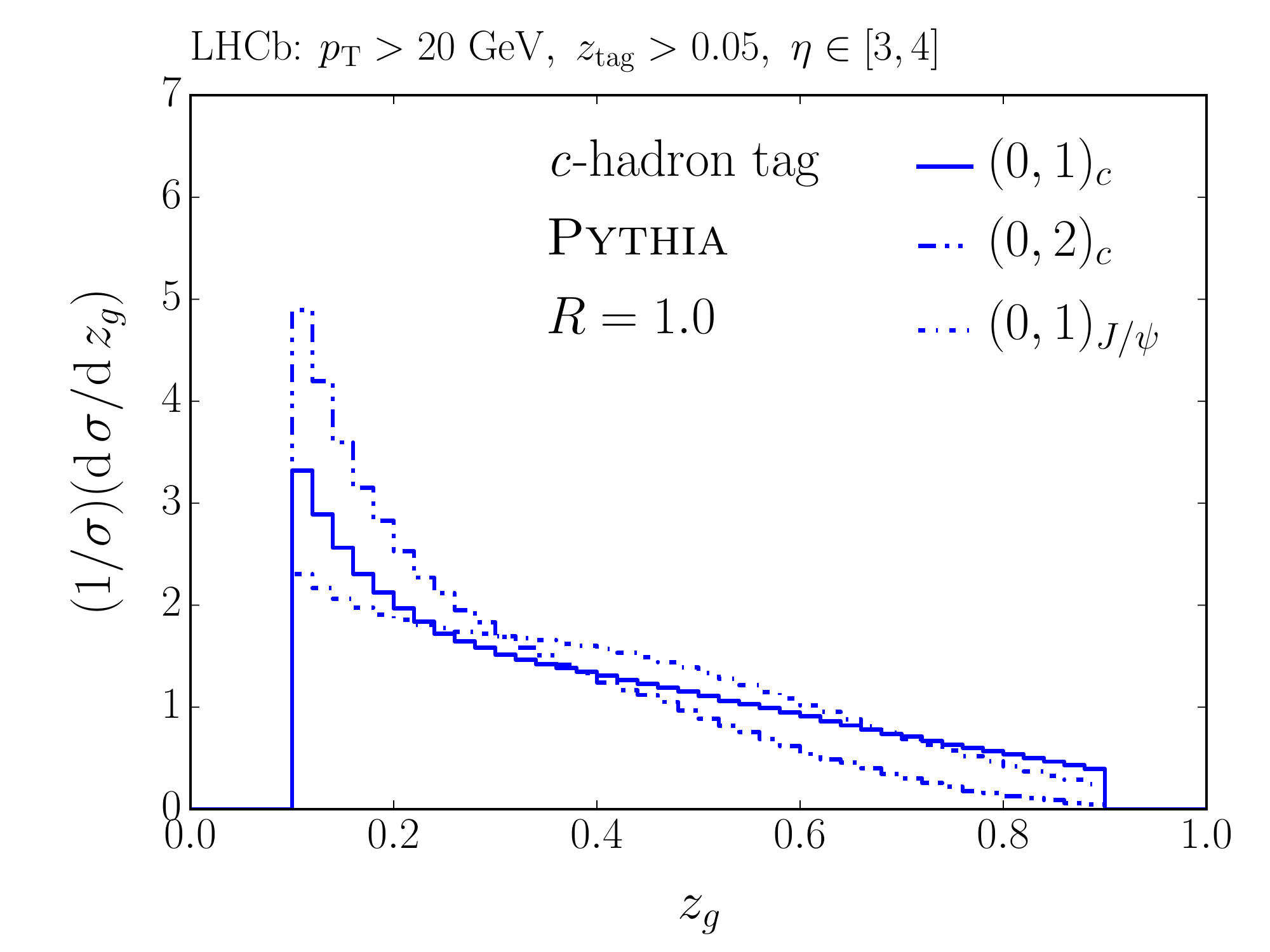}}  \hspace{0.1in}
\subfigure[]{\includegraphics[scale=0.4]{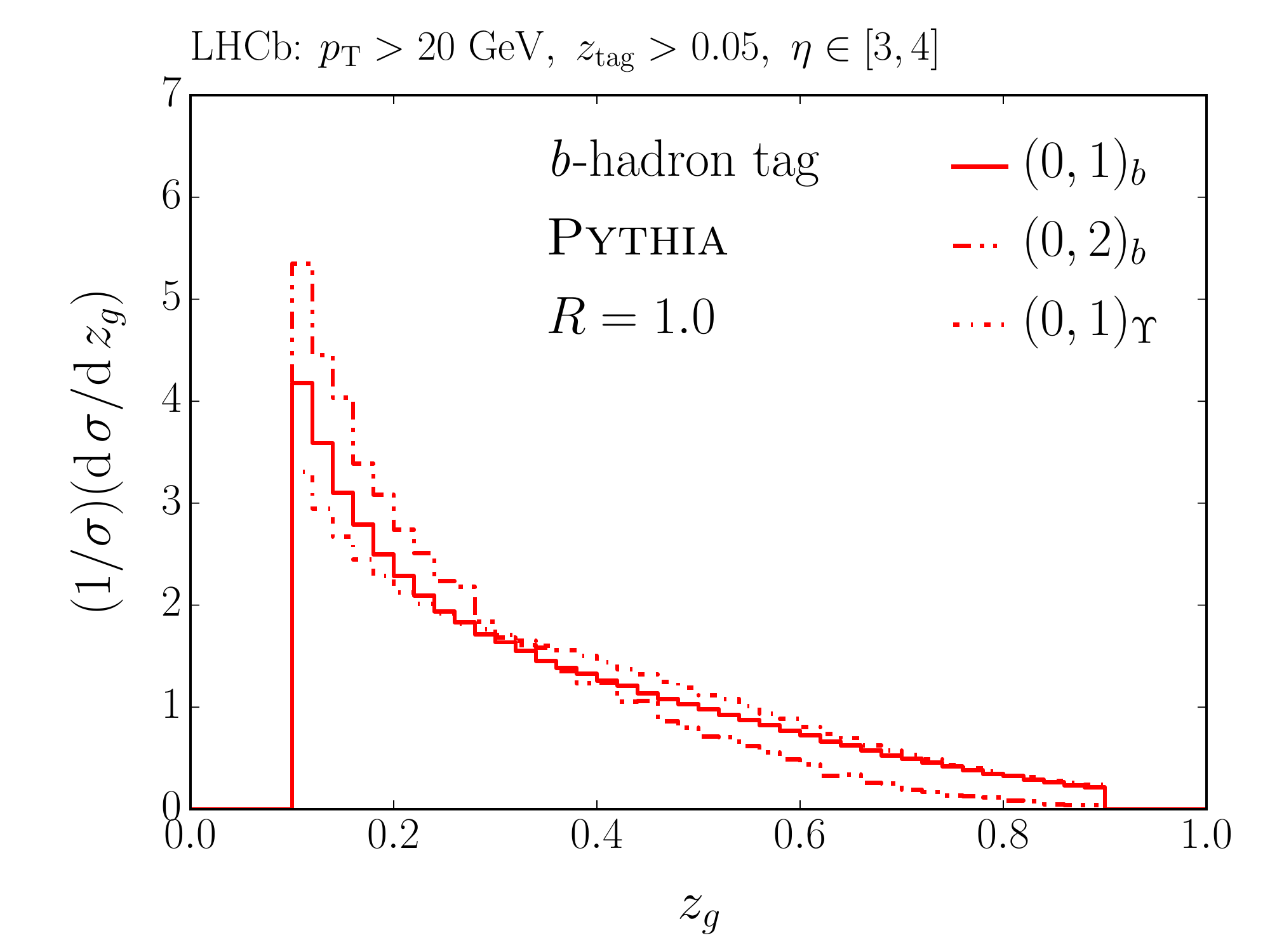}}
\end{tabular}
\caption{Same as \Fig{fig:Kinz} for \pythia, but comparing quarkonium-tagged jets to flavor-tagged jets in the (0,1) and (0,2) categories.   Note that the solid line for the (0,1) category matches \Fig{fig:Kinz}.}
\label{fig:Kinjpsiupsilon}
\end{figure*}

\begin{figure*}[t]
\centering
\begin{tabular}{c}
\subfigure[]{\includegraphics[scale=0.4]{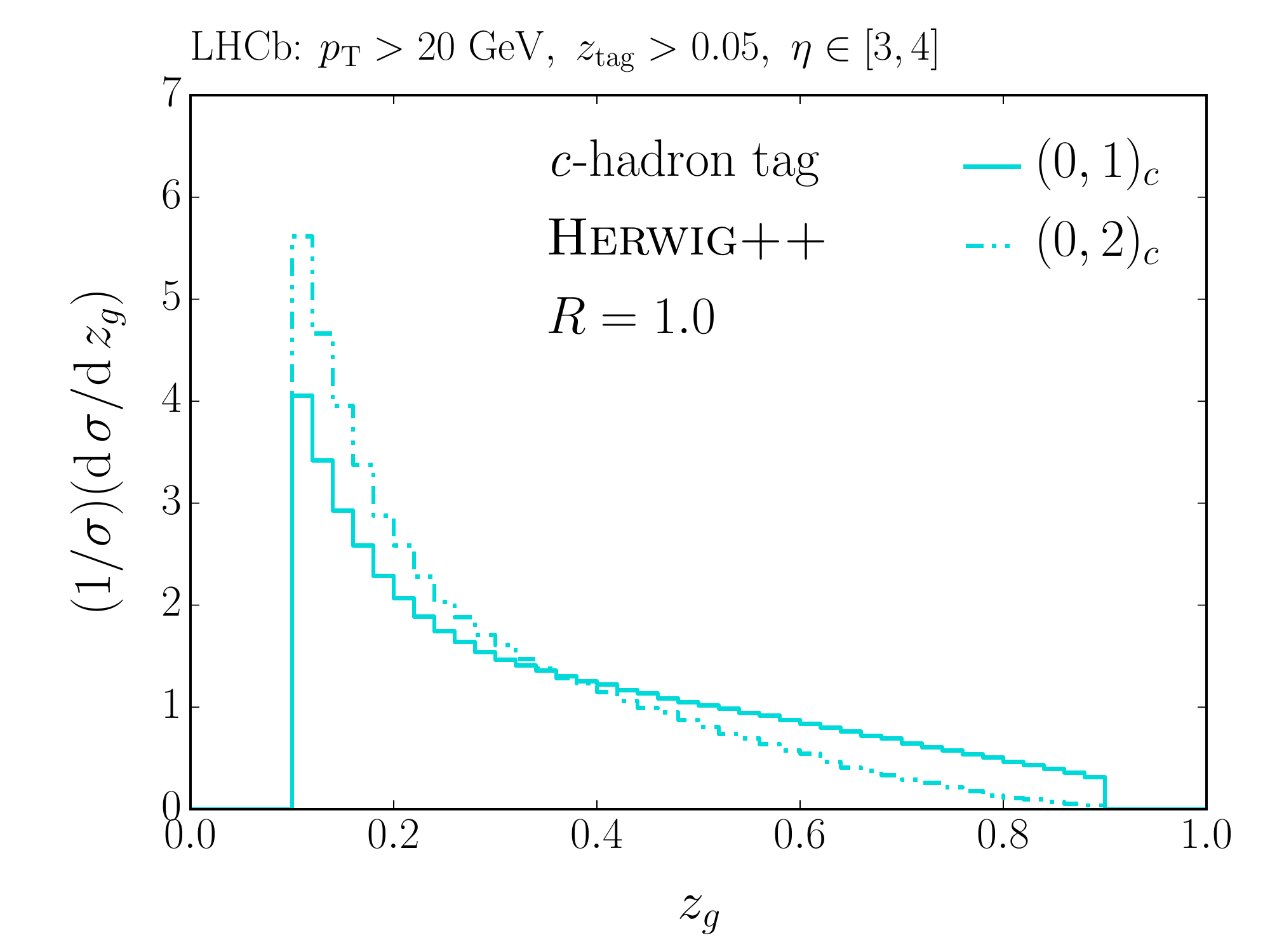}}  \hspace{0.1in}
\subfigure[]{\includegraphics[scale=0.4]{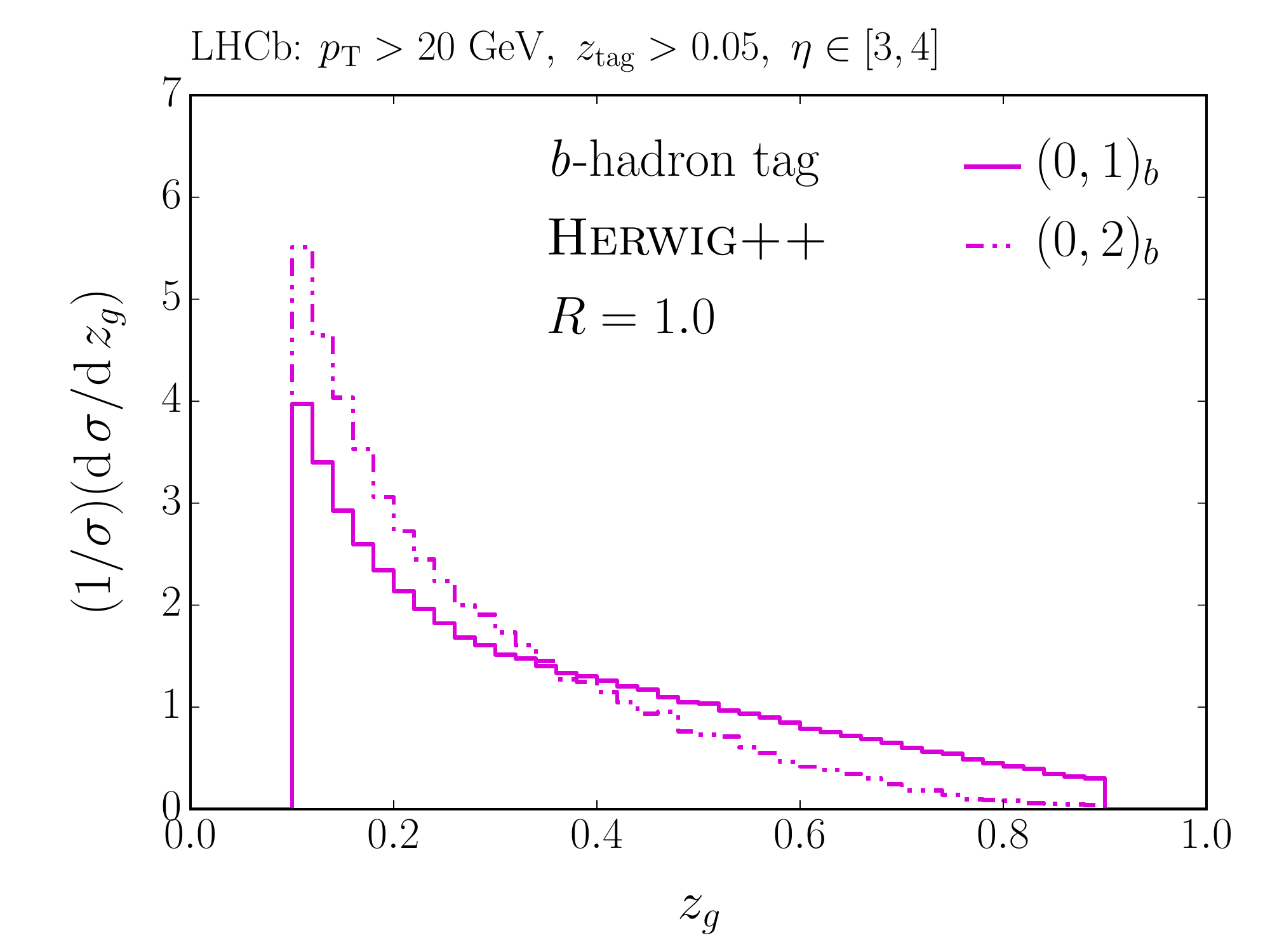}}
\end{tabular}
\caption{Same as \Fig{fig:Kinjpsiupsilon} but for \herwig. Quarkonium production is not available in \herwig.}
\label{fig:Kinjpsiupsilon_herwig}
\end{figure*}

We next show \sdrop distributions for $z_g$, as defined in \Eq{eq:zg_def}, for both $c$-tagged and $b$-tagged fat jets.
We focus on the (0,0), (0,1), and (1,1) categories here, leaving  the (0,2) category to \Sec{subsec:quarkonium}.
In \Fig{fig:Kinz}, we show results from \pythia.
The (0,0) distribution, which has no flavor tags, agrees with those already found in \Ref{Larkoski:2015lea}, with the important caveat that $z_g$ is defined here to be symmetric about $z_g = 1/2$ (instead of stopping at $z_g = 1/2$).
For the (0,1) category, the $z_g$ distribution agrees qualitatively with the $Q \to Q g$ splitting function, peaking towards $z \to 0$ as expected from \Eq{eq:massive_splitting}.
The (1,1) category, which is largely due to $g\to Q\bar{Q}$, has no singular structures near $z \to 0$ or $z \to 1$.
Figure~\ref{fig:Kinz_herwig} shows that the analogous distributions obtained using \herwig are similar to those obtained from \pythia, with some small differences observed near the  endpoints.

All three categories exhibit distinct behavior that qualitatively agrees with predictions from perturbative QCD.
While the $1 \to 2$ splitting functions of massive QCD are well known theoretically, they have never been probed in such a direct manner experimentally.
By exploiting the ability to flavor-tag \sdrop subjets, our approach provides the opportunity to directly probe the splitting kinematics of $Q \to Q g$ and $g \to Q \bar{Q}$.
 Having confirmed that our parton-shower results agree qualitatively with the expected theoretical predictions in \Sec{subsec:splittingfuncinterpret}, we look forward to tests of these $z_g$ distributions in data.

\subsection{Results: Quarkonium Production}
\label{subsec:quarkonium}

The \sdrop jet-declustering strategy is also applicable to fat jets containing identified quarkonium states.
In terms of flavor content, jets with a quarkonium-tagged subjet are similar to the (0,2) category defined above, so it is interesting to compare their $z_g$ distributions to see whether the underlying physics is similar or not.
Traditionally, quarkonium production within a jet is studied using fragmentation functions, which describe the momentum fraction carried by a \jpsi or $\Upsilon$  hadron within a resolved jet.
Here, we pursue a complementary approach using $z_g$, which gives the momentum fraction carried by a \jpsi-tagged or $\Upsilon$-tagged subjet within a \sdrop fat jet.

As a preamble, it is worth emphasizing that there are considerable theoretical uncertainties in both the production and splittings associated with the \jpsi and $\Upsilon$.
A standard approach to study these quarkonium states is to use nonrelativistic QCD (NRQCD)\,\cite{Bodwin:1994jh,Cho:1995vh,Cho:1995ce}, though there is a long-standing quarkonium-polarization puzzle associated with this approach; see, for example, \Ref{Brambilla:2010cs} and references therein.
More recently, \Refs{Baumgart:2014upa,Bain:2016clc} used the method of fragmenting jet functions (FJF)\,\cite{Procura:2009vm} as an alternate method to calculate \jpsi production.
Specifically, \Ref{Bain:2016clc} considered the kinematics of \jpsi production within a resolved jet, finding that their theoretical predictions for the showering, and hence splitting functions, associated with the \jpsi disagreed with those implemented in \pythia.
In \pythia, a \jpsi produced in the color-octet channel is treated as a loosely bound $c \bar{c}$ state, and its total showering is the sum of the showers originating from either quark.
 By contrast, in the FJF approach, a produced \jpsi is showered through DGLAP evolution of the splitting kernels from $2 m_c$ up to the jet energy scale.
There have been other treatments of quarkonia showers discussed in the literature; see, for example, \Ref{Ernstrom:1996am}.
Given these theoretical uncertainties, we expect measurements of $z_g$ to help clarify the mechanism of quarkonium production within jets.
Furthermore, LHCb recently published a study of prompt \jpsi production in jets~\cite{LHCb-PAPER-2016-064}. Their results are consistent with the predictions of the FJF-based approach, and in stark disagreement with \pythia, providing additional motivation to measure $z_g$ for quarkonium production.

In \Tab{table:tag_rate} we give the predicted rates for \jpsi-tagged and $\Upsilon$-tagged jet production.
These rates are more than an order of magnitude smaller than those of the double-flavor-tagged $(0,2)$ category with the same quark flavor.
In \Fig{fig:Kinjpsiupsilon}, we compare the $z_g$ distributions for quarkonium-tagged jets to the $(0,1)$ and $(0,2)$ categories.
In the context of \pythia, soft-gluon radiation from heavy-flavor quarks primarily differs from soft-gluon radiation from quarkonium only in the overall color factor.
The quarkonium distributions are not included in the equivalent \herwig plot shown in \Fig{fig:Kinjpsiupsilon_herwig}, since quarkonium production is not available in the version of \herwig used in these studies.
Since the $z_g$ observable is insensitive to Casimir factors at lowest order, the distributions in \Fig{fig:Kinjpsiupsilon} are all similar.
Given the calculation in \Ref{Bain:2016clc}, and the LHCb results in \Ref{LHCb-PAPER-2016-064}, it will be fascinating if this prediction is borne out in data.

\section{\fcone Jets to Disentangle Heavy-Flavor Production}
\label{sec:Rate}

We now transition from studying heavy-flavor within a single jet to heavy-flavor production in the event as a whole.
There are multiple production channels for heavy flavor in QCD, which leads to various complications in predicting heavy-flavor rates.
Typically, one considers the three main production mechanisms shown in \Fig{fig:LeadingOrderProduction}, with the caveats that the  $\alpha_s$ order at which these diagrams appear depends on the PDF scheme employed, and that at higher orders there is no gauge-invariant definition of these categories.
Still, making a gluon-splitting versus non-gluon-splitting distinction provides helpful intuition about the relevant phase-space regions populated by these diagrams.
The main challenge of using traditional jet algorithms in the gluon-splitting regime is jet merging.
In this section, we first review the jet-merging issue, and then introduce our \fcone jet algorithm designed to resolve overlapping heavy-flavor jets.

\subsection{Traditional Approach to Heavy-Flavor Rates}

At the LHC, it is well-known that the gluon-splitting and flavor-excitation processes can dominate the total rate over flavor creation; see, for example, \Refs{Zanderighi:2007dy,Aad:2012ma,Voutilainen:2015lqa}.
These two dominant processes are challenging to calculate since, as emphasized by their $2 \to 3$ representations in \Figs{fig:LeadingOrderProduction:a}{fig:LeadingOrderProduction:b}, they involve multiple emission scales.
The cross sections for these channels have been calculated at NLO in perturbative QCD; see, for example, \Refs{Nason:1987xz,Nason:1989zy,Beenakker:1990maa,Olness:1997yc,Frixione:1997ma,Cacciari:1998it,Nason:1999ta,Tung:2001mv,Czakon:2007wk} and references therein. 
Fixed-order results have been interfaced with parton-shower programs to provide predicted rates for the LHC \cite{Norrbin:2000zc,Banfi:2007gu,Cacciari:2012ny}.
Perturbative QCD predictions for heavy-flavor production have been tested experimentally for $b$ quarks---at the Tevatron \cite{Abe:1993sj,Abe:1993hr,Abbott:1999se,Abbott:2000iv}, and at the LHC by ATLAS \cite{ATLAS:2011ac,Aaboud:2016jed,Aad:2011kp,Aad:2013vka,Aad:2011jn,Aad:2014dvb}, CMS \cite{Khachatryan:2011wq,Chatrchyan:2012dk,Chatrchyan:2013uza,Chatrchyan:2012vr,Chatrchyan:2013zja,Chatrchyan:2014dha}, and LHCb\,\cite{LHCb-PAPER-2014-023}---and for a combination of flavors at ATLAS \cite{Aad:2012ma}.
Many of these studies have noted a tension between the theoretical predictions and the experimental data, especially in the gluon-splitting regime.

\begin{figure*}[t]
\centering
\begin{tabular}{cccc}
\subfigure[]{\includegraphics[width=0.24\textwidth]{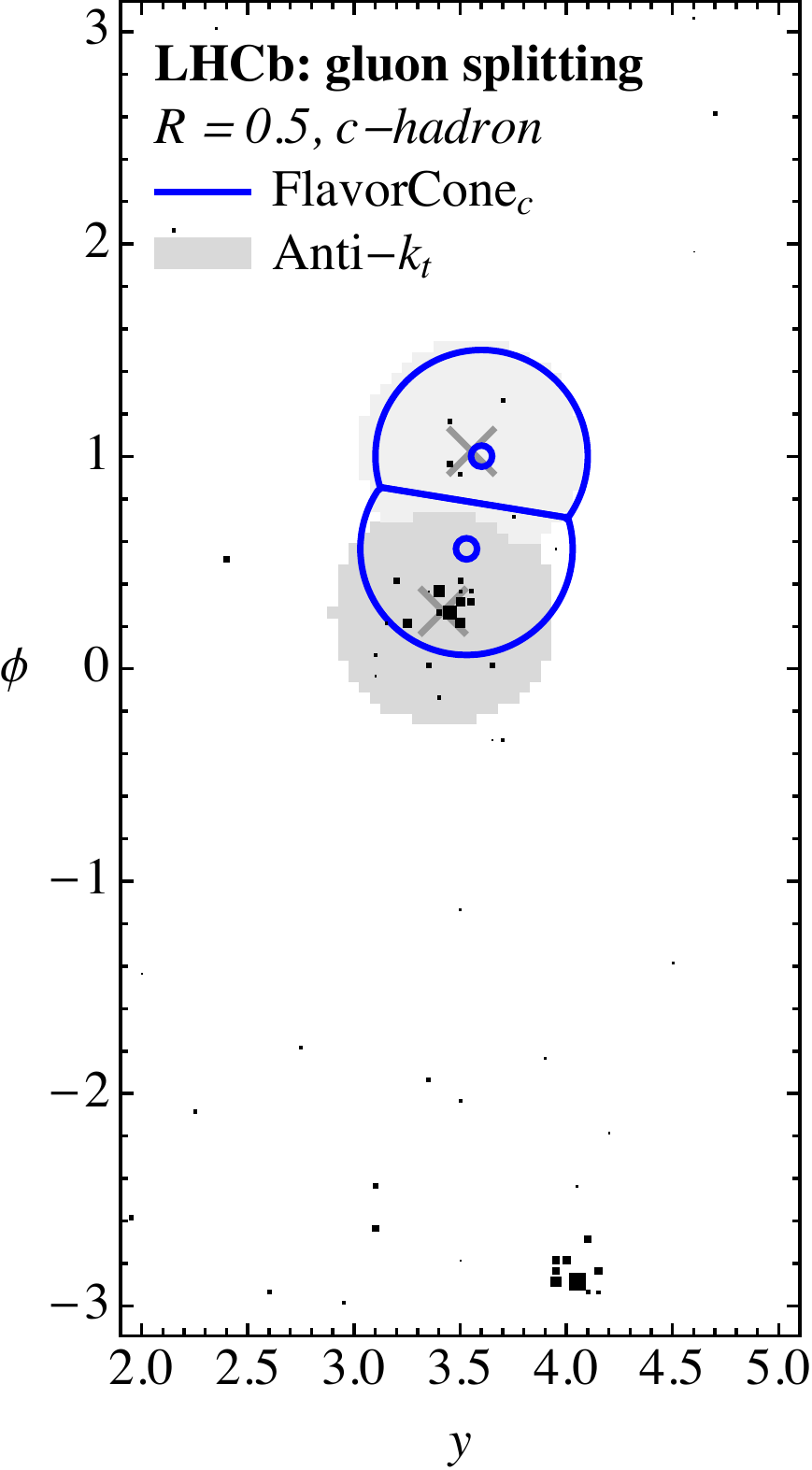}}&
\subfigure[]{\includegraphics[width=0.24\textwidth]{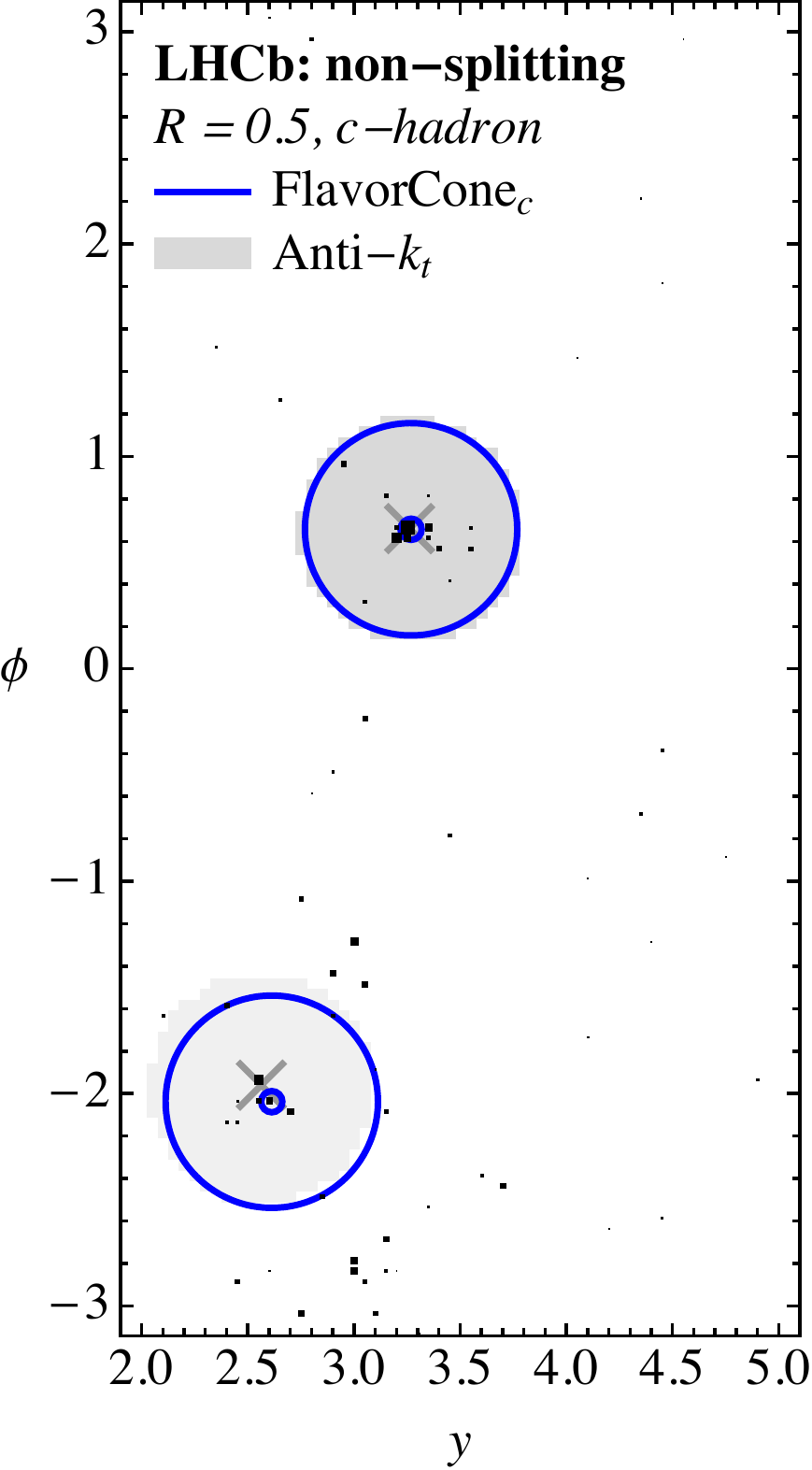}}&
\subfigure[]{\includegraphics[width=0.24\textwidth]{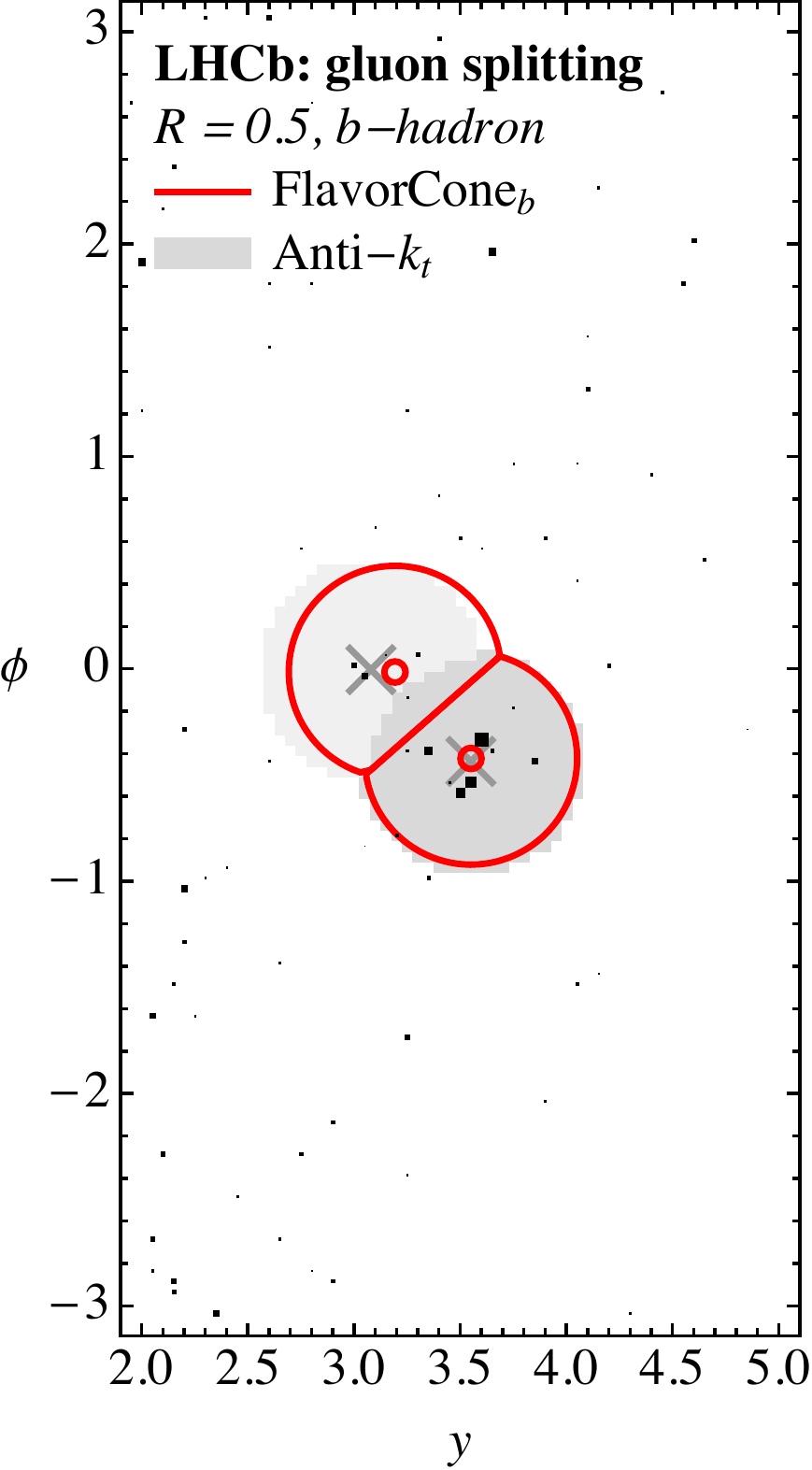}}&
\subfigure[]{\includegraphics[width=0.24\textwidth]{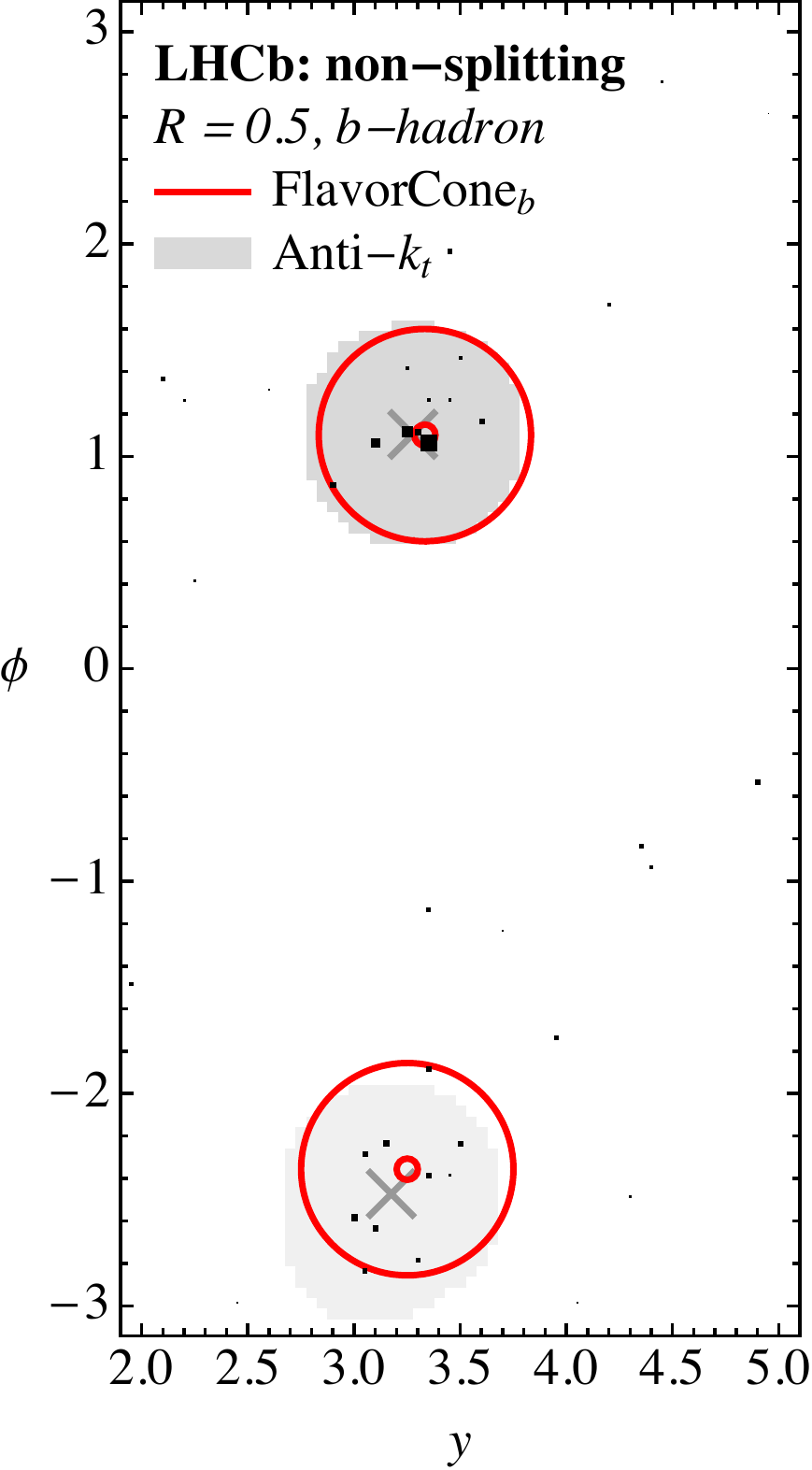}}
\end{tabular}
\caption{Event displays using the \fcone algorithm, for (a) a gluon-splitting $c\bar{c}$ event, (b) a non-splitting $c\bar{c}$ event, (c) a gluon-splitting $b\bar{b}$ event, and (d) a non-splitting $b\bar{b}$ event.  The flavor-tagged ghost particles are shown as small open circles, with blue for charm and red for beauty, and the \fcone jet boundaries are shown in the corresponding color.  The shaded gray regions show the two hardest flavor-tagged \akt jets for comparison.  Here we have selected gluon-splitting events where the two \akt jets remain resolved, although we emphasize that in many such events the \akt jets in fact merge.}
\label{fig:fcone_events}
\end{figure*}

The phase-space region where the disagreement is largest is also where analysis strategies based on traditional jet algorithms break down.
In gluon-splitting events, the heavy quarks tend to have small angular separation due to the gluon propagator.
Standard jet algorithms, such as \akt, have difficulty in this situation, since events where two heavy quarks fall into a single jet are typically cut from the analysis.\footnote{\label{footnote:doubleflavortag}Alternatively, one could separately perform a subjet analysis on \akt jets with multiple flavor tags \cite{ATLAS-CONF-2012-100,CMS-PAS-BTV-13-001,CMS-PAS-BTV-15-002,ATLAS-CONF-2016-039}. This strategy, however, does not result in a smooth transition between the gluon-splitting and non-gluon-splitting regimes.}
This limitation is unavoidable for \akt, since the jet axes cannot get any closer than the jet radius $R$, usually taken to be $R=0.4$ or $0.5$.\footnote{Other cone-like algorithms can even lead to jet merging below $\sqrt{2} R$ or $2R$; see discussion in \Ref{Thaler:2015uja}.}
Despite this, almost all of the LHC analyses referenced above use \akt, and suffer the associated loss of performance.

It is worth mentioning an alternative strategy for studying heavy-flavor production based on flavored jet algorithms \cite{Banfi:2006hf,Banfi:2007gu}.
These algorithms, which work equally well on heavy-flavor quarks or hadrons, attempt to {\em neutralize} gluon-splitting topologies through a recursive clustering strategy.
In this way, events with $g \to Q \bar{Q}$ are not even categorized as heavy-flavor production, giving complementary information to \fcone jets.

\subsection{A New Approach: \fcone}

We now introduce a simple jet algorithm aimed at reconstructing the gluon-splitting phase space.
In an event with $n$ heavy-flavor-tagged hadrons, we define $n$ \fcone jets of radius $R$ as follows.
\begin{enumerate}
\item The flight direction of each flavor-tagged hadron defines a separate jet axis.
\item Any particle that is further away than $R$ from a jet axis is left unclustered.
\item Each remaining particle is clustered to the nearest jet axis.
\item The momentum of each jet is then determined by the summed momenta of its constituents.
\end{enumerate}
As is appropriate for $pp$ collisions, we use the rapidity-azimuth distance $\Delta R = \sqrt{(\Delta y)^2 + (\Delta \phi)^2}$ to determine the unclustered region and the nearest jet axis.\footnote{In an experimental analysis, it may be preferable to use pseudorapidity $\eta$ instead of rapidity $y$ to avoid complications of assigning masses to reconstructed particles.  See \Sec{sec:ImplementationAtLHCb} for further discussion.}
Example events clustered with the \fcone algorithm are shown in \Fig{fig:fcone_events}.

By construction, the \fcone algorithm does not include a merging step, so there are always exactly $n$ jets in the event, regardless of whether the $n$ flavor-tagged hadrons are widely separated or almost collinear.
In the well-separated regime, the resulting jets are perfect cones centered on the flight directions of the flavor-tagged hadrons, yielding results that are similar to \akt.
Crucially, there is no limitation on the jet axes getting closer than $R$, and abutting jet regions are determined by nearest-neighbor partitioning.
Of course, this feature relies fundamentally on the ability of experiments like LHCb to accurately reconstruct the flight direction of heavy-flavor hadrons, even when the hadrons are almost collinear, as discussed further in \Sec{sec:ImplementationAtLHCb}.

The partitioning scheme used for \fcone is motivated by the XCone jet algorithm \cite{Stewart:2015waa,Thaler:2015xaa}, which is also designed to yield a fixed, predetermined number of jets.
Beyond just the simplicity of the \fcone algorithm, its main advantage over XCone is that flavor-tagged hadrons always end up in separate jets, whereas XCone can still allow merging.
The XCone algorithm is infrared and collinear (IRC) safe, since it starts from a set of IRC-safe seed-jet axes and then applies an iterative procedure to find the jet regions that (locally) minimize an $N$-jettiness measure \cite{Stewart:2010tn}.
The \fcone algorithm is also IRC safe, since additional soft particles or collinear splittings cannot impact the flight direction of a flavor-tagged hadron, to the extent that $m_{b,c} \gg \Lambda_{\rm QCD}$.\footnote{Similar to XCone, \fcone can be sensitive to the IRC regime, since the algorithm will identify $n$ jets even if the $n$ flavor-tagged hadrons of interest have very low $p_T$.  For this reason, one may wish to impose additional requirements on the \fcone jets to ensure that one is in the perturbative regime.  Here, we set a minimum $\pt$ on the flavor tag and then place an additional cut on the overall jet $\pt$, which introduces some mild dependence on the $b$-quark fragmentation function.  Alternatively, one could place a cut on the relative $\pt$ of the tag and the jet, in the same spirit as the $z_{\rm tag}$ condition in \Eq{eq:ztagdef}.}

In general, the \fcone jet axis and the \fcone jet momentum are not aligned.
In this respect, the \fcone axes are similar in spirit to the winner-take-all axes\,\cite{Larkoski:2014uqa,Bertolini:2013iqa,Salam:unpub}, especially since flavor-tagged hadrons often carry a large fraction of the jet momentum in the non-gluon-splitting regime.
If desired, one could apply an iterative procedure to find $n$ mutually stable cones using the $n$ flavor-tagged-hadron directions as seed axes.
As mentioned in \Sec{subsec:alternative}, we find better performance by simply using flavor-tagged-hadron axes directly, since this avoids the issue of abutting jets repelling (or merging into) each other after iteration, which tends to wash out the gluon-splitting phase space.

\subsection{Event Selection and Classification}

Using this \fcone algorithm, we now outline an analysis strategy to disentangle the mechanisms for heavy-flavor production.  Our focus here is on the LHCb experiment, though we show related distributions for ATLAS and CMS in \App{app:CMSATLAS}.
Our analysis workflow is as follows.
\begin{itemize}
\item We select events that have at least two heavy-flavor hadrons with $\eta \in [2.5,4.5]$, so that the full \fcone (with $R = 0.5$) is within LHCb acceptance.  For concreteness in the plots below, we require these hadrons to have $\pt > 2\gev$, though the specific selection criteria would depend on the implementation details; see \Sec{sec:ImplementationAtLHCb}.
\item These heavy-flavor hadrons are ordered in \pt, and the two hardest ones are used to form \fcone jets with $R = 0.5$.\footnote{Although reconstructing the full four-momenta of the hadrons is challenging at the LHC, ordering them by \pt is more feasible. That said, events with more than two reconstructed flavor-tagged hadrons are expected to be rare; see \Sec{sec:ImplementationAtLHCb}.}
\item The leading \fcone jet is required to have $\pt > 20 \gev$, though larger \pt thresholds would likely be needed at ATLAS or CMS.
\item The subleading \fcone jet is required to have $z_{\rm sub} > 0.1$, where we define
\be
z_{\rm sub} = \frac{p_{\rm T}^{\rm sub}}{p_{\rm T}^{\rm lead} + p_{\rm T}^{\rm sub}} \,. \label{eq:subfrac}
\ee
This cut avoids highly asymmetric events that are more difficult to predict from perturbation theory.
\end{itemize}
We perform this analysis separately for $b \bar{b}$ and $c \bar{c}$ final states.  In principle, one could also look at mixed $bc$ events, but we do not pursue that here.

For comparison, we also consider events clustered using \akt with $R=0.5$, where flavor tagging is performed by treating the flight directions of the hardest two heavy-flavor hadrons as ghost particles in the clustering.
The two resulting \akt flavor-tagged jets are then treated in the same way as the \fcone jets, with the same cut on the leading jet \pt and subleading $z_{\rm sub}$.
In keeping with the traditional strategy, \akt jets that contain two flavor-tagged hadrons are rejected from the analysis.

When using the \pythia parton shower, it is possible  to classify generated events as being either \textit{splitting} or \textit{non-splitting} using the event record.
If the two flavor-tagged hadrons can be traced back to a common gluon parent from the shower, then the event is labelled as splitting.
Every other event is classified as non-splitting, which includes prompt production via flavor creation and excitation, as well as gluon-splitting-like events where the two heavy-flavor hadrons do not come from a common gluon parent from the shower.\footnote{We also tested alternative classification schemes based on the hard-production vertex.  We found cases, however, where the hard production is labeled as flavor creation or excitation, but the actual heavy-flavor hadrons within the LHCb acceptance come from a subsequent gluon splitting.}
Of course, this splitting/non-splitting distinction is not physical, as it cannot be implemented experimentally and is ambiguous beyond the strongly-ordered parton-shower limit.
Nonetheless, it allows us to isolate the way that \pythia treats the gluon-splitting regime and test whether the \fcone algorithm is sufficiently sensitive to the underlying physics.

\subsection{Results:  Clear Separation of Gluon Splitting}

The strategy above selects events with two flavor-tagged jets.
To probe the physics of heavy-flavor production, we consider the cross section differential in four dijet observables:  $\Delta y$, $\Delta \phi$, $\Delta R$, each determined using the jet axes; and the invariant mass of the dijet system $m_{jj}$.
For \akt, the jet axis and momentum are aligned, whereas for \fcone, the jet axis is aligned with the flight direction of the flavor-tagged hadron.

In order to determine the potential separation power between splitting and non-splitting events, we first calculate the \textit{classifier separation} as implemented in TMVA\,\cite{Hocker:2007ht}.\footnote{An alternative classifier metric is mutual information \cite{Larkoski:2014pca}, which is closely related to classifier separation\,\cite{Badger:2016bpw}.}
Given two event categories $A$ and $B$, and probability distributions $p_A(\lambda)$ and $p_B(\lambda)$ for the observable $\lambda$, the classifier separation $\delta_{\lambda}$ is defined as
\be
\delta_{\lambda} = \frac{1}{2} \int \mathrm{d} \lambda \, \frac{\left( p_A(\lambda) - p_B(\lambda) \right)^2}{p_A(\lambda) + p_B(\lambda)}\,. \label{eq:ClassSep}
\ee
The value of $\delta_{\lambda}$ always lies within $[0,1]$, where $\delta_{\lambda} = 0$ means that $\lambda$ has no discriminating power and  $\delta_{\lambda} = 1$ indicates ideal separation.

\begin{figure*}[t]
\centering
\begin{tabular}{c}
\subfigure[]{\includegraphics[scale=0.4]{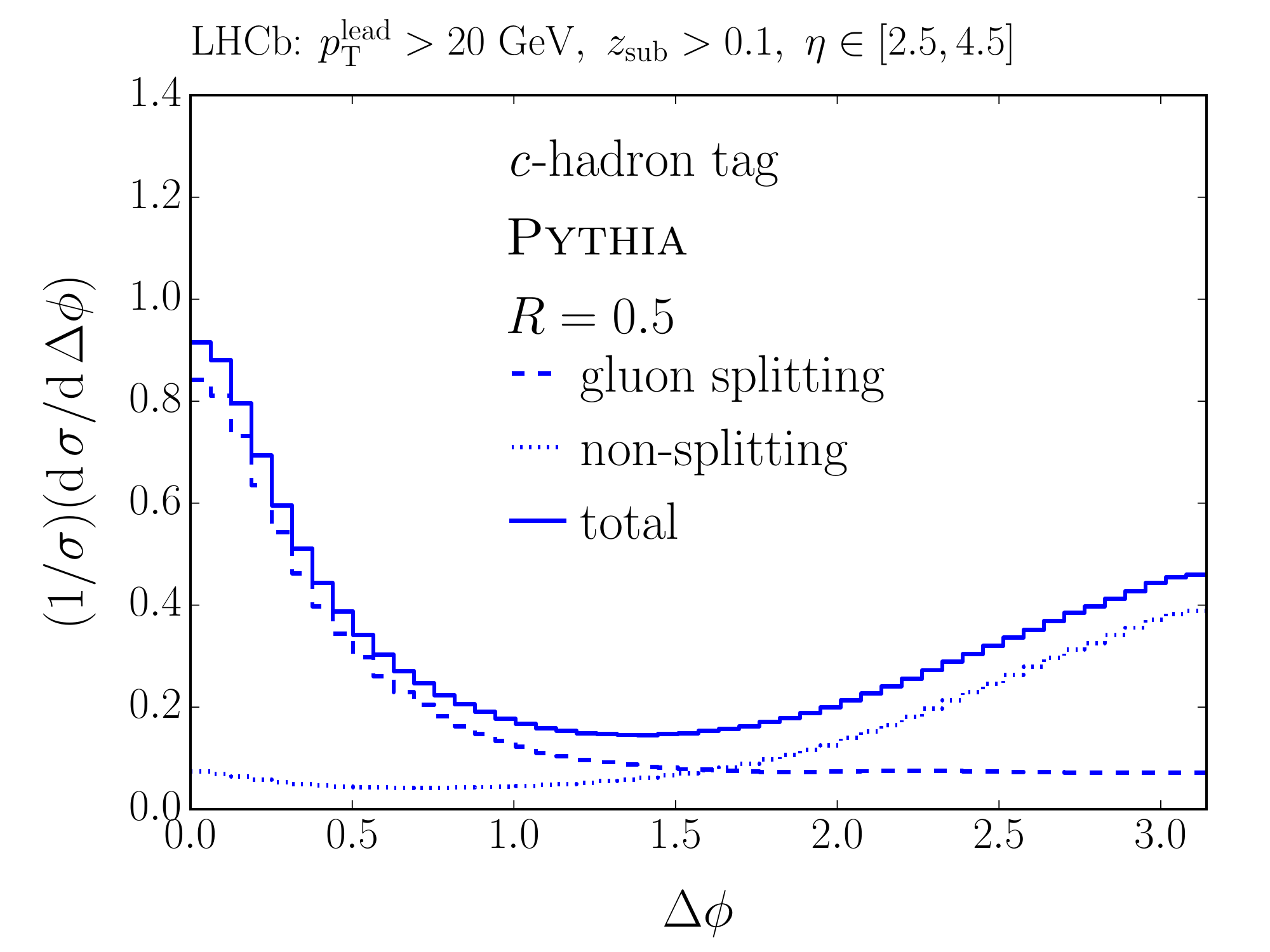}} \hspace{0.1in}
\subfigure[]{\includegraphics[scale=0.4]{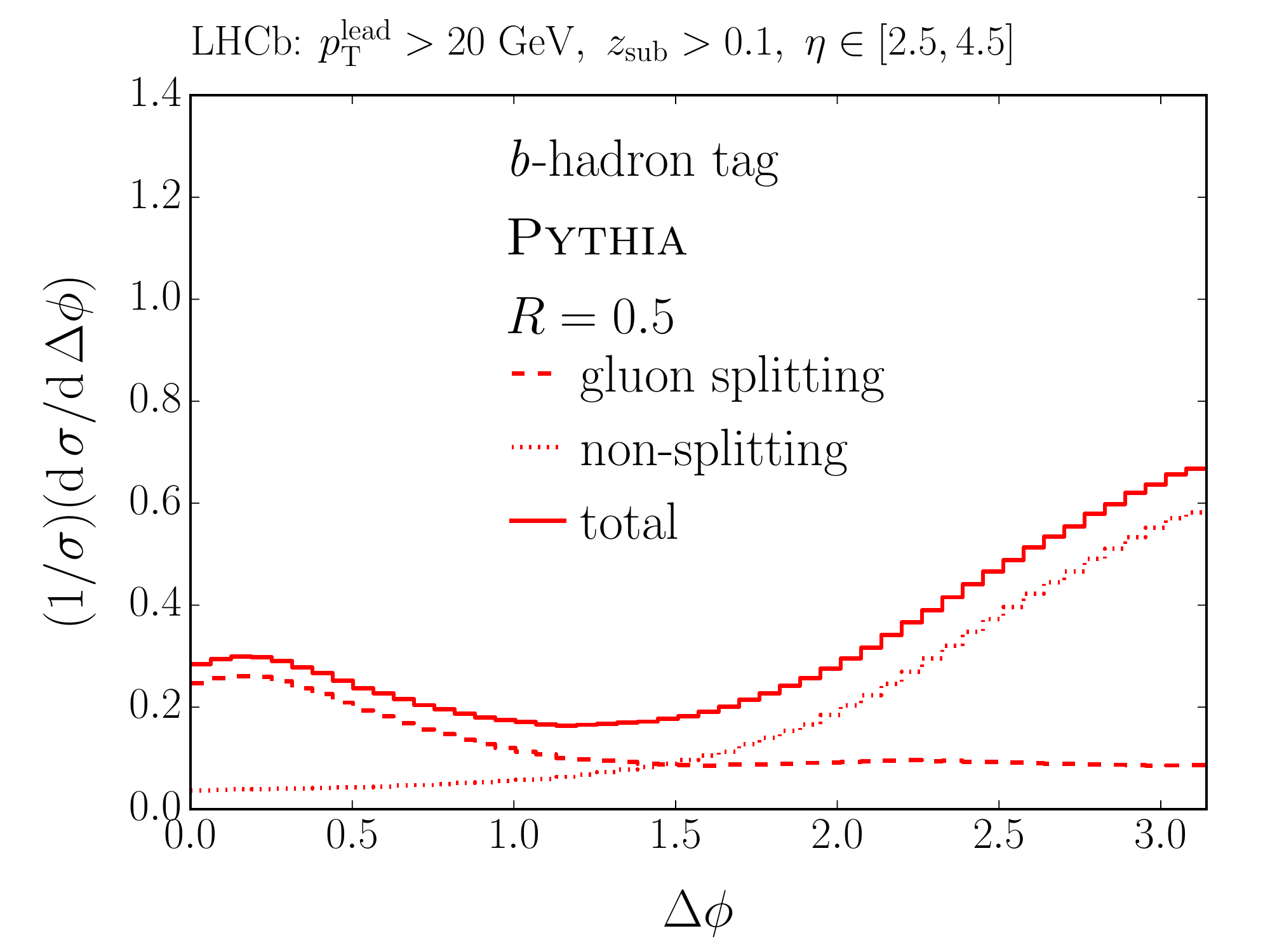}}
\end{tabular}
\caption{The azimuthal separation between the two \fcone axes seeded by (a) $c$-tagged and (b) $b$-tagged hadrons.   The normalized distributions are shown for \pythia, which allows a useful (but ambiguous) categorization into splitting and non-splitting events.  Low values of $\Delta \phi$ probe the phase-space region dominated by gluon splitting. The \fcone algorithm is ideally suited to study this region, with no anomalous features at the jet radius $R = 0.5$.
}
\label{fig:RateSplitNonSplit}
\end{figure*}

\begin{figure*}[t]
\centering
\begin{tabular}{c}
\subfigure[]{\includegraphics[scale=0.4]{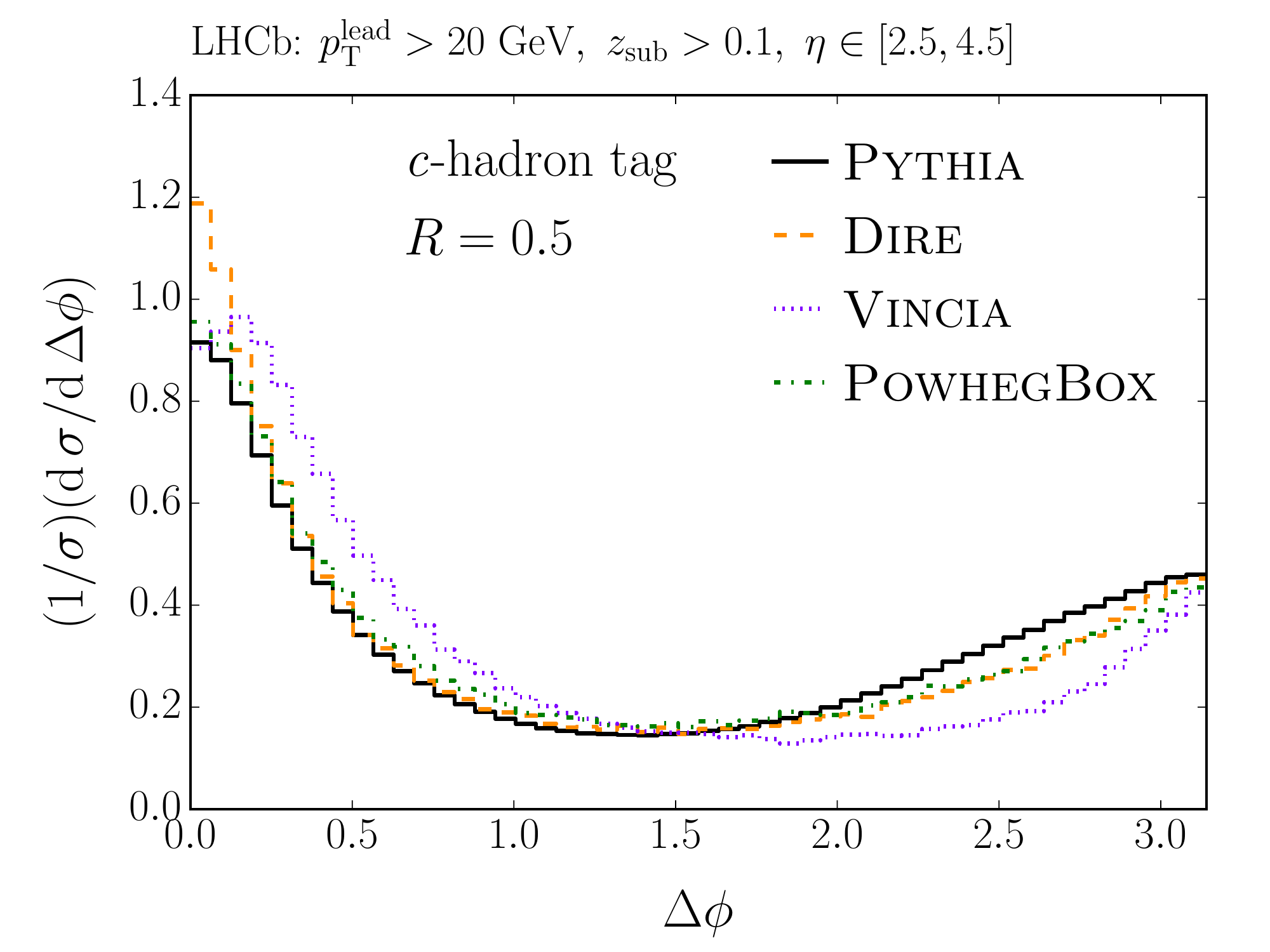}} \hspace{0.1in}
\subfigure[]{\includegraphics[scale=0.4]{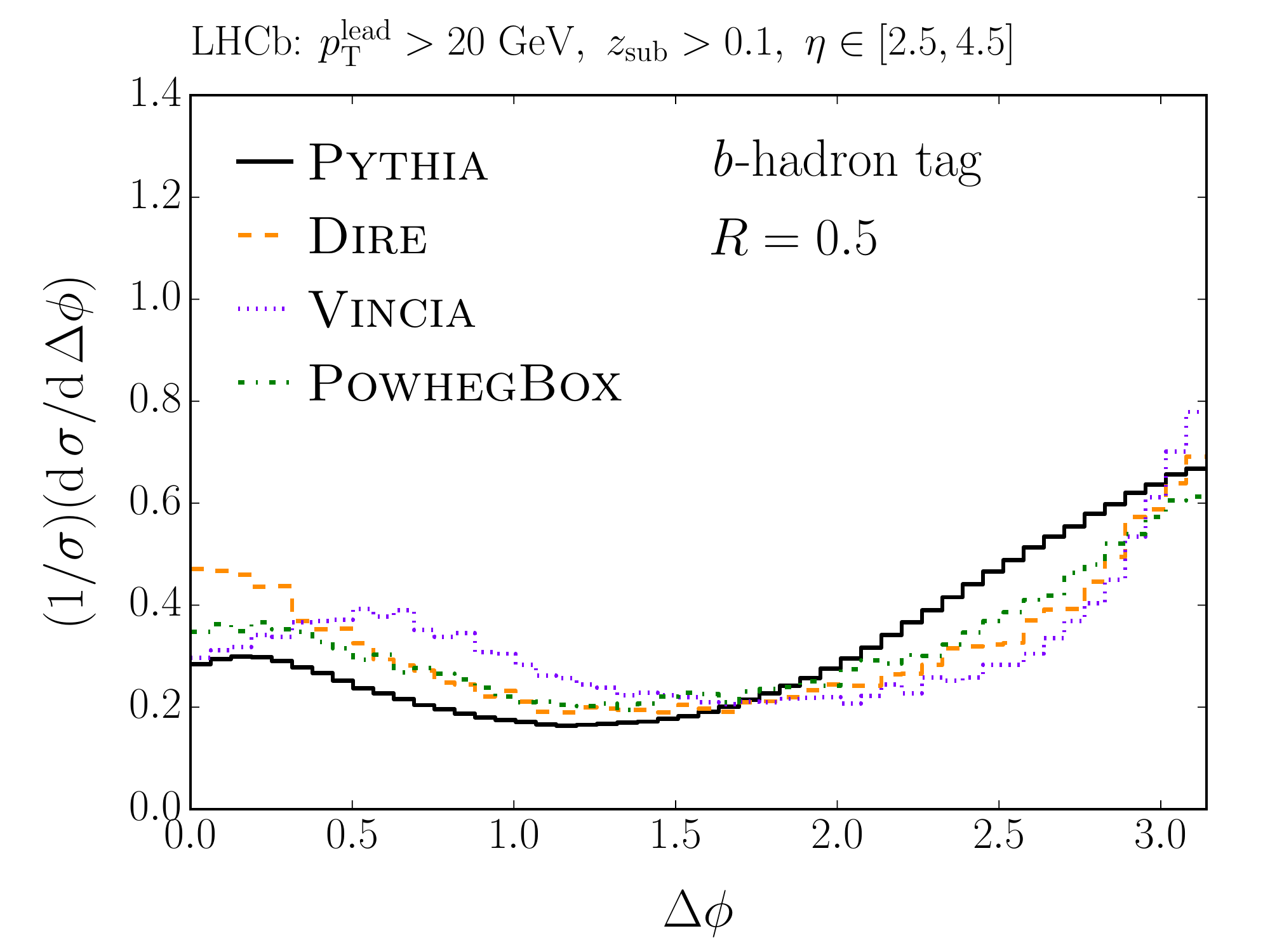}}
\end{tabular}
\caption{Same observable and event selection as \Fig{fig:RateSplitNonSplit}, but now comparing four different predictions.  The qualitative behavior is similar between the generators, but quantitatively they differ at a level that should be testable at the LHC.  These are normalized distributions; see \Tab{table:ClasSep} for the differences in the  absolute cross section.  
}
\label{fig:RateGenerator}
\end{figure*}

\begin{table}[t]
\begin{center}
\begin{tabular}{ r @{\hskip 0.3in} c  c @{\hskip 0.3in}   c  c }
\hline \hline
& \multicolumn{2}{ c@{\hskip 0.3in}}{$c \bar{c}$} & \multicolumn{2}{  c }{$b \bar{b}$} \\
& \fcone & \Akt & \fcone & \Akt \\ \hline
$\Delta y$    & 0.09 & 0.03 & 0.05 & 0.02 \\
$\Delta \phi$ & 0.41 & 0.30 & 0.31 & 0.24 \\
$\Delta R$    & 0.43 & 0.31 & 0.32 & 0.25 \\
$m_{jj}$      & 0.42 & 0.30 & 0.29 & 0.22 \\ \hline \hline
\end{tabular}
\end{center}
\caption{The discrimination power between splitting and non-splitting events in \pythia for $c\bar{c}$ and $b \bar{b}$.  For each dijet observable, the values shown correspond to classifier separation $\delta_\lambda$ from \Eq{eq:ClassSep}, where larger values indicate better performance.  Regardless of the choice of discriminant, the \fcone  approach outperforms the traditional \akt approach.}
\label{table:ClasSep}
\end{table}

In \Tab{table:ClasSep}, we show the $\delta$ values for each of the dijet observables as computed from \pythia.
As expected, $\Delta y$ is not a good discriminant, since back-to-back jets from flavor creation can also have a small rapidity separation.
The remaining three observables show good performance in separating the splitting and non-splitting categories.
These observables are, of course, strongly correlated, and we checked that combining them in a multivariate analysis provides little improvement.
\Tab{table:ClasSep} clearly shows that the \fcone approach outperforms \akt, yielding a 30--40\% gain in performance as measured by $\delta_\lambda$.
This is largely due to heavy-flavor merging by the anti-$k_t$ algorithm.\footnote{In the spirit of footnote \ref{footnote:doubleflavortag}, one could consider double-tagged anti-$k_t$ jets as a separate event category to be included in the calculation of $\delta_\lambda$.  Depending on the precise double-tagged selection criteria one uses, the resulting performance can be comparable to \fcone. Alternatively, one could use a smaller jet radius to reduce the impact of jet merging.}

To highlight the separation power achievable using the \fcone algorithm, we show the full distribution for $\Delta \phi$ in \Fig{fig:RateSplitNonSplit}, with the other three observables given in \App{app:extraplotsFC}.
We present both $b \bar{b}$ and $c \bar{c}$ events, broken down into the splitting and non-splitting categories.
In order to isolate the phase-space region dominated by gluon splitting, one can select events with $\Delta \phi \lsim 1$.
Note that the $\Delta \phi$ distribution smoothly approaches zero, with no change of behavior as the angle approaches the jet radius $R = 0.5$.
It is in this region that the \fcone algorithm performs particularly well, while traditional jet algorithms result in jet merging (see \Sec{subsec:alternative}).

\begin{table}[t]
\begin{center}
\begin{tabular}{ r @{\hskip 0.3in} c @{\hskip 0.3in}  c }
\hline \hline
& $\sigma(c \bar{c})$ [$\mu$b] & $\sigma(b \bar{b})$ [$\mu$b] \\ \hline
\pythia    & 2.02 & 0.97 \\
\vincia    & 1.40 & 0.59 \\
\dire      & 2.55 & 0.93 \\
\powhegbox & 1.27 & 0.55 \\ \hline \hline
\end{tabular}
\end{center}
\caption{The cross section in the LHCb fiducial region, defined with the nominal \fcone algorithm, for $c\bar{c}$ and $b\bar{b}$ production from four different predictions.  Note that there is a greater disagreement in these rates than there is in the differential shapes in \Fig{fig:RateGenerator}.}
\label{table:Prodxsec}
\end{table}

In addition to \pythia, we apply this analysis to three alternative predictions---\vincia, \dire, and \powhegbox---all interfaced to \pythia for hadronization.
Already from the total cross sections in \Tab{table:Prodxsec}, one can see substantial differences between the generators, but the origin of the disagreement is not clear from the rates alone.
In the normalized distributions in \Fig{fig:RateGenerator}, we can see in more detail the different predictions for the splitting-enriched and non-splitting-enriched regions of phase space.
While there is qualitative agreement about the peaks at $\Delta \phi = 0$ and $\pi$, quantitatively the predictions are sufficiently different that direct comparison to LHC data should result in improved modeling of heavy-flavor production in parton-shower generators.

\subsection{Alternative Approaches}
\label{subsec:alternative}

\begin{figure*}[t]
\centering
\begin{tabular}{c}
\subfigure[]{\includegraphics[scale=0.4]{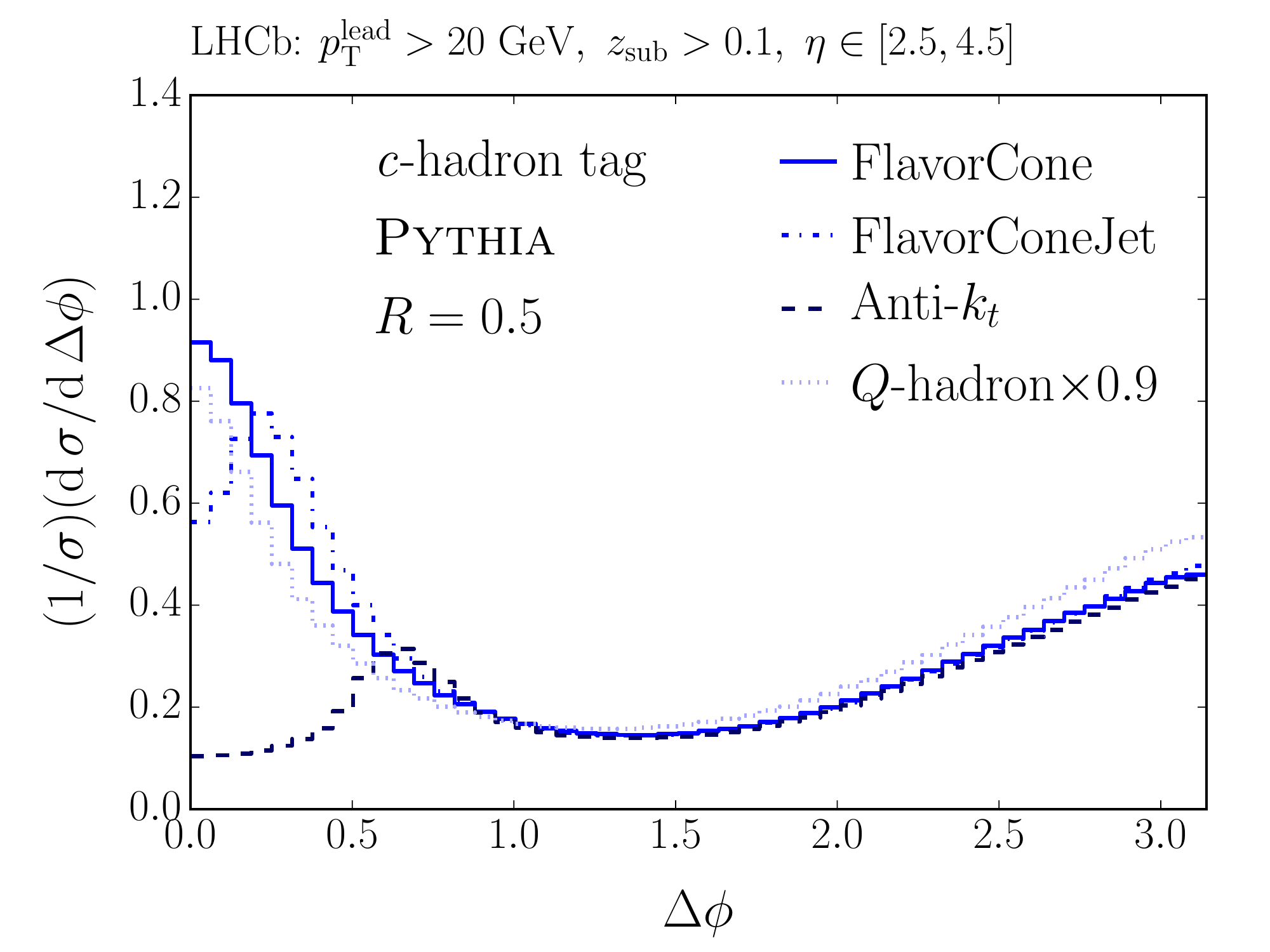}} \hspace{0.1in}
\subfigure[]{\includegraphics[scale=0.4]{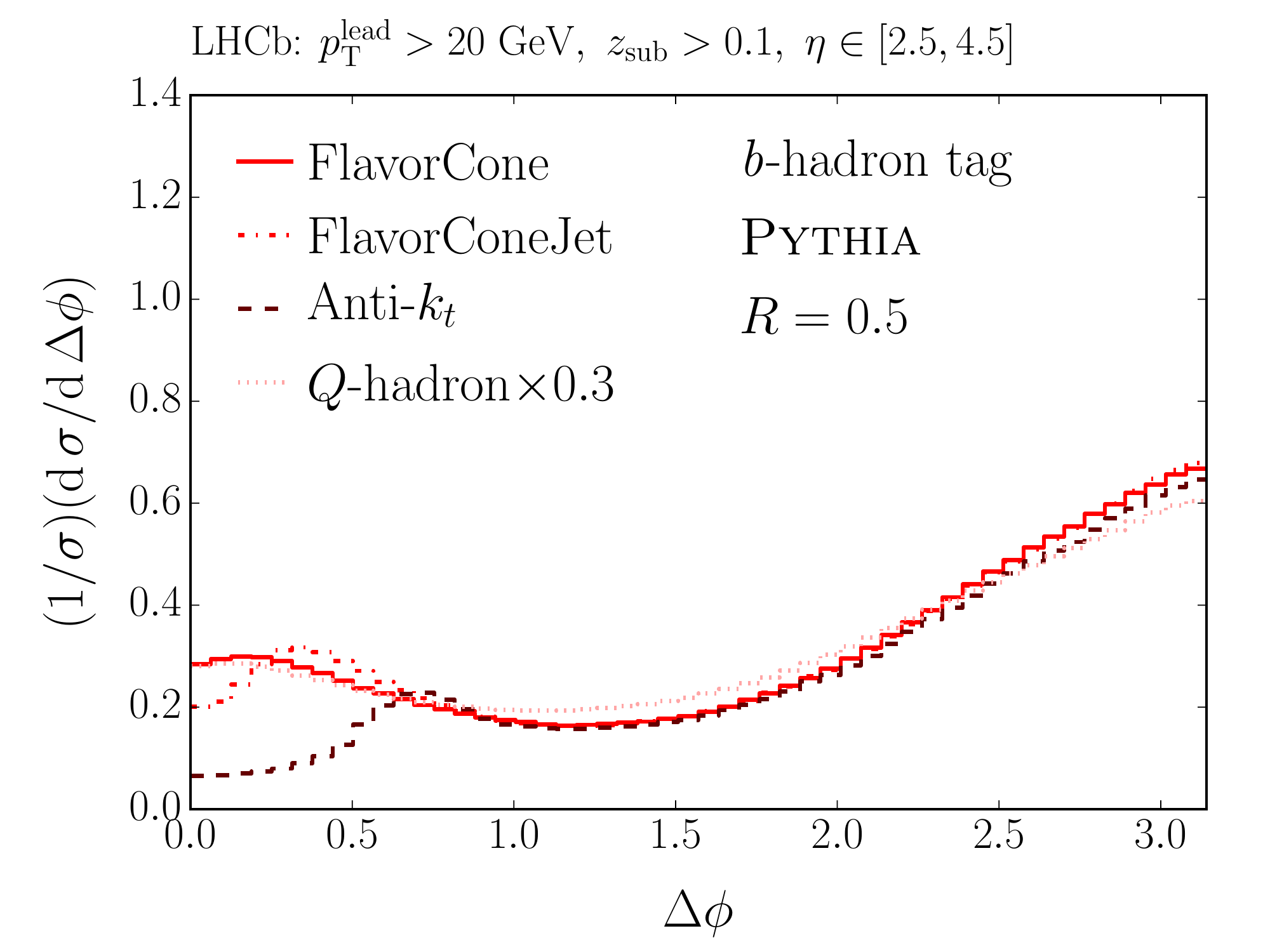}}
\end{tabular}
\caption{Same event selection as \Fig{fig:RateSplitNonSplit}, but now comparing four different definitions for the $\Delta \phi$ observable.  Here, the distributions are normalized to the \fcone approach in order to highlight the loss in efficiency for the \akt method below $\Delta \phi \simeq R$.  For $Q$-hadron, we put an additional offset factor to partially account for the difference in $p_{\rm T}$ acceptance for hadrons versus jets.  For FlavorConeJet, the feature at $\Delta \phi \simeq 0.3$ is due to the misalignment of the jet axis and jet momentum for abutting jets.
}
\label{fig:RateJetAlg}
\end{figure*}

To better understand the behavior of \fcone jets, it is instructive to make a comparison to alternative reconstruction strategies.   In \Fig{fig:RateJetAlg}, we compare the $\Delta \phi$ distributions obtained using the following four methods.
\begin{enumerate}
\item \emph{\fcone}:  The default \fcone approach, where $\Delta \phi$ is determined from the jet axes (which are aligned with the flavor-tagged hadrons).
\item \emph{FlavorConeJet}:  An alternative \fcone approach, where $\Delta \phi$ is determined by the jet momenta ({\em i.e.}\ the vector sum of the constituent momenta of each jet).
\item \emph{\Akt}:  The default \akt approach using $E$-scheme recombination, where the jet axes and jet momenta align.%
\item \emph{$Q$-hadron}:  Omitting any jet clustering and determining $\Delta \phi$ from the flavor-tagged hadrons alone. Here, we require that the heavy-flavor hadrons have $\pt > 16\gev$ to roughly select the same phase-space regime as the jet-based approaches.
\end{enumerate}
We also tested an alternative \akt approach where $\Delta \phi$ is determined from the flavor-tagged ghosts, but that has nearly identical behavior as the \akt option tested above.

All four approaches result in similar behavior at large $\Delta\phi$, but significant differences are clearly visible in the gluon-splitting regime.
The power of the FlavorConeJet method degrades at small $\Delta\phi$, because the jet momenta recoil against each other as the cones begin to overlap.
The \akt algorithm suffers a precipitous drop in efficiency in the gluon-splitting regime due to jet merging.
Of course, one could reduce the impact of jet merging by using a smaller jet radius, at the expense of introducing more out-of-jet radiation.
Finally, because the $Q$-hadron and \fcone distributions are based on the same flavor-tagged hadrons, it is not surprising that they exhibit the same qualitative features. 
Without any jet reconstruction, however, the $Q$-hadron method requires reconstructing the full \pt of the heavy-flavor hadrons, and not just their flight directions.
Experimentally, this requirement inherently leads to a much lower flavor-tagging efficiency.

Beyond \Fig{fig:RateJetAlg}, we also considered two additional methods, but neither is as powerful as \fcone.
In the spirit of stable cone algorithms, we studied an iterated \fcone, where the two flavor-tagged hadrons provide seed axes that are iteratively adjusted until they align with the jet momenta.\footnote{Specifically, we start from the original \fcone jets.  In each iteration step, we define new jet axes based on the jet momenta, and repartition the hadrons to the closest jet axis within the jet radius $R$.  This process is guaranteed to terminate in a finite number of steps \cite{Stewart:2015waa}.}
This approach gave similar results to the nominal \fcone, but occasionally the iteration procedure caused the two heavy-flavor hadrons to merge into the same jet, leading to a loss of performance in the gluon-splitting regime.
We also tried various exclusive-generalized-$k_{t}$ strategies\,\cite{Catani:1993hr,Ellis:1993tq,Catani:1991hj}, but none worked as well as the \fcone algorithm.\footnote{Like XCone, recursive exclusive jet algorithms also ensure a fixed number of jets in the final state.  In the generalized $k_{t}$ algorithm, the merging and dropping distance measures scale as $p_{\rm T}^{2q}$.
We considered $q=0.5$, $1.0$, and $2.0$ in exclusive mode with $R=0.5$, where the algorithm terminates when there are exactly two undropped jets in the final state.
We tried using the resulting exclusive $k_{t}$ jets directly in the analysis, as well as using them as jet axes for cone finding and as seeds for iterative stable cone finding.}
While exclusive $k_{t}$ does allow the jet axes to become arbitrarily close in principle, in practice there is still considerable heavy-flavor merging in the gluon-splitting regime.
Considering all of these results, we advocate the \fcone approach as being best suited to studying heavy-flavor production in the gluon-splitting regime.

\section{Implementation at LHCb}
\label{sec:ImplementationAtLHCb}

The parton-shower studies presented above are very encouraging, but they ignore the realities of detector-response effects.
Efficient reconstruction of flavor-tagged hadrons is the most important ingredient needed to carry out these analyses, and tagging multiple heavy-flavor hadrons in close proximity is a challenge.
Furthermore, both the \sdrop and \fcone analyses require the flight directions of the flavor-tagged hadrons to be well measured.
Here, we briefly sketch a possible implementation of these methods at the LHCb detector, which has excellent heavy-flavor reconstruction capabilities.

We start with the case of $c$-hadron tagging.
Charm quarks primarily hadronize into a $D^0$, $D^{+}$, or $D_s^+$ meson, a $\Lambda_c^+$ baryon, or any of their antiparticles.
Each of these charm hadrons has at least one all-charged-particle decay with a sizable branching fraction.\footnote{For example, the folowing decays can each be cleanly and efficiently reconstructed at LHCb: $\mathcal{B}(D^0 \!\to\! K^-\pi^+) \approx 4\%$, $\mathcal{B}(D^+ \!\to\! K^-\pi^+\pi^+) \approx 9\%$, $\mathcal{B}(D_s^+ \!\to\! K^-K^+\pi^+) \approx 5\%$, and $\mathcal{B}(\Lambda_c^+ \!\to\! p K^- \pi^+) \approx 6\%$.}
Therefore, a potential strategy at LHCb is to fully reconstruct one $c$-hadron, which would provide excellent momentum resolution of $\sigma(\pt)/\pt \approx 1\%$ and $\sigma(\phi) \approx 2$\,mrad (see App.~A of \Ref{Ilten:2015hya}).
Combining this exclusive tag with the excellent vertex resolution at LHCb permits precise determination of the $c$-hadron impact parameter, making it possible to distinguish prompt-charm production from charm produced in $b\to c$ decays.
 With full reconstruction, one could choose to use the reconstructed charm hadron directly in the jet finding, instead of  using it only as a ghost particle for tagging; this would mitigate track sharing between (sub)jets.

In principle, LHCb could also fully reconstruct the second $c$-tagged hadron, but the efficiency of performing exclusive reconstruction of both $c$-hadrons will be low.
For charm tagging (as opposed to exclusive charm reconstruction), LHCb developed a $c$-jet tagging algorithm in Run~1 that is largely based on properties of the $c$-hadron.
This algorithm achieves a $c$-tagging efficiency of 20--25\%, while also providing excellent $c$--$b$ discrimination\,\cite{LHCb-PAPER-2015-016}.
Removing the two (out of ten) features that depend on properties of the jet (as opposed to the $c$-hadron) used by the LHCb machine-learning-based $c$-jet-tagging algorithm should make this tagger suitable for use in both the \sdrop and \fcone analyses.
That said, most of the discriminating power for $c$-jet tagging comes from a single feature, the {\em corrected mass},
so a simple analysis based on secondary vertices is likely sufficient.\footnote{The corrected mass takes the reconstructed secondary-vertex mass $m_{\rm sv}$, and adds a correction based on the momentum transverse to the direction of flight  $\sqrt{m^2_{\rm sv} + (p_{\rm sv}\sin{\delta\theta})^2} + |p_{\rm sv}\sin{\delta\theta}|$, where $\delta\theta$ is the angle between $\vec{p}_{\rm sv}$ and the flight vector.}
Either way, the expected resolution on $\sigma(\Delta\phi_{c\bar{c}})$ is $\mathcal{O}(10\,{\rm mrad})$\,\cite{Ilten:2015hya}, which is more than sufficient for both \sdrop and \fcone, given that none of the plots shown above resolve features finer than $\Delta\phi_{\rm min} \simeq 50\,{\rm mrad}$.

The only remaining issue is that of \pt ordering of the charm hadrons in \fcone, since only the two hardest tags are used in our analysis.
Only 7\% of events in our \fcone analysis of \pythia data contain more than two $c$-tagged hadrons with $\pt > 2\gev$, which means that $c\bar{c}$ events with three (or more) reconstructed $c$-hadron tags will be rare.
Therefore, \pt ordering in \fcone will be less important than the small correction required to account for the case where one of the two hardest $c$-hadrons is not reconstructed, but the third-hardest is.
This correction can be derived from data using events with (at least) three $c$-hadron tags.

We now turn to the case of $b$-hadron tagging.
Unfortunately, there are no all-charged-particle $b$-hadron decays with percent-level branching fractions, making exclusive reconstruction of $b$-hadrons inefficient.
Inclusive secondary-vertex-based $b$-hadron tagging, however, is highly efficient.\footnote{Another potential strategy is to reconstruct one $b$-tag using a displaced $\jpsi\to\mu^+\mu^-$ decay, and the other one using an inclusive secondary-vertex $b$-tag.}
Backgrounds from $c$-hadron decays can be highly suppressed by requiring the reconstructed secondary-vertex mass to be greater than the $c$-hadron masses; a similar strategy was employed by CMS in \Ref{Khachatryan:2011wq}.
The flight direction is measured sufficiently well at LHCb, $\sigma(\Delta\phi_{b\bar{b}}) \approx 50$\,mrad, to employ this inclusive strategy in both the \sdrop and \fcone analyses.
Events in the \pythia data sample with more than two $b$-tagged hadrons are rare enough that improper \pt-ordering of $b$-tagged hadrons will have negligible impact on the \fcone analysis.

For the relatively low-\pt jets studied at LHCb, the distinction between rapidity $y$ and pseudorapidity $\eta$ is non-negligible.
In the parton-shower studies above, we assumed access to full four-vector information for all particles, including the ghost tags.
At LHCb, it is much easier to determine $\eta$, though one can reliably estimate $y$ since LHCb has good particle identification to infer the appropriate hadron mass value.
In practice, we expect the distinction between $y$ and $\eta$ could be implemented as a simple unfolding correction.
At worst, one could use a FlavorCone variant based only on $\eta$, and we checked that this does not have a large impact on performance.

Finally, we note that the \sdrop jet-declustering analysis can be performed on fat jets clustered using only charged particles.
Given that charged-particle reconstruction at LHCb is far superior to that of neutral particles, a charged-only strategy may be desirable.
From a theoretical perspective, charged-only distributions can be treated using generalized fragmentation functions called track functions \cite{Chang:2013rca,Chang:2013iba}, appropriately adapted to treat heavy-flavor fragmentation.
From an experimental tagging perspective, inclusive flavor-tagging of hadrons is already largely based on charged particle information, so there is relatively little loss of information in only using charged particles for fat jet reconstruction.
Note that while $z_g$ itself is a dimensionless observable that is rather insensitive to the charged/neutral ratio, \sdrop depends on an angular-ordered clustering tree, which benefits from the improved angular resolution provided by charged particles.
Either way, comparing the $z_g$ distributions obtained using all particles versus only charged ones will provide a valuable cross check.\footnote{As shown in \Ref{Anderle:2017qwx}, there can be significant event-by-event fluctuations in the momentum fraction $z$ when going from all particles to charged particles.  But these fluctuations often have a mild impact on the final distribution, see \Ref{Chang:2013iba}.}

\section{Conclusion}
\label{sec:conclusion}

In this article, we outlined two analysis strategies designed to gain a better understanding of the origin and kinematics of heavy flavor at colliders.
First, we showed how \sdrop jet declustering of a fat jet can expose the well-known but as-of-yet-unmeasured splitting kernels for heavy-flavor quarks and quarkonia in QCD.
Second, we showed how the \fcone jet algorithm can help separate the gluon-splitting and non-gluon-splitting heavy-flavor production mechanisms.
Our parton-shower studies were focused on the LHCb experiment because of its superior ability to identify and reconstruct heavy-flavor hadrons, offering the best short-term prospects for carrying out these measurements.
We also expect that similar techniques can be pursued by ATLAS and CMS, especially in the higher-\pt jet range.

A key theme from this study is the value of tagging heavy-flavor hadrons, as opposed to the more traditional strategy of tagging heavy-flavor jets.
This theme has already been emphasized in jet substructure studies based on subjet flavor-tagging \cite{CMS-PAS-BTV-15-002,ATLAS-CONF-2016-039}, which use ghost association in a similar way as our $(n_1, n_2)$ flavor-categorization scheme.
The \fcone algorithm goes one step further, using flavor-tagged hadrons to define jets, in contrast to the typical approach of first finding jets and then flavor-tagging them.
Experimentally, tagging hadrons is far more challenging than tagging jets, but we think the physics opportunities justify investing in the development of hadron-based tagging strategies.
Theoretically, it will be interesting to study the behavior of fixed-order QCD calculations when using these heavy-flavor categories.

Beyond just the excellent tagging performance of LHCb, the heavy-flavor strategies presented here are motivated by the need for a more robust theoretical definition of heavy flavor.
In many collider-based studies, a jet with $g \to Q \bar{Q}$ will be tagged as a heavy-flavor jet, yet the physics of $g \to Q \bar{Q}$ is very different from that of $Q \to Q g$, with different underlying production mechanisms and different final-state kinematics.
A similar point was emphasized in the context of flavored jet algorithms \cite{Banfi:2006hf,Banfi:2007gu}.
By individually identifying heavy-flavor hadrons, one can more easily separate double-tag versus single-tag (sub)jets, mitigating the confusions that arise from gluon splitting to heavy flavor.
We expect that the \sdrop $z_g$ distributions will be helpful in validating new flavor-tagging methods, and we hope that this study inspires more sophisticated heavy-flavor categorization.

Finally, heavy-flavor production is important for stress-testing event generators.
Even though all of the generators tested here are based (in principle) on the same underlying QCD splitting kernels, the differences in their distributions are substantial, especially in the gluon-splitting regime.
It would be particularly interesting to see how heavy-flavor production evolves as a function of $\pt$, since the relative importance of each contribution varies as function of jet kinematics.
The measurements proposed here should inspire further precision QCD calculations of heavy-flavor production, and lead to improved parton-shower modeling of heavy-flavor physics.

\section{Acknowledgements}

We thank Vava Gligorov, Ben Nachman, Frank Petriello, Ira Rothstein, Gavin Salam, and Peter Skands for helpful discussions.
Feynman diagrams were drawn using \Ref{Ellis:2016jkw}. 
The work of NLR and JT is supported by the U.S. Department of Energy (DOE) under grant contract numbers DE-SC-00012567 and DE-SC-00015476. 
NLR is also supported in part by the American Australian Association's ConocoPhillips Fellowship.
PI and MW are supported by the U.S.\ National Science Foundation grant PHY-1607225. 

\appendix

\section{Additional \sdrop Distributions}
\label{app:extraplotsSD}

\begin{figure*}[t]
\centering
\begin{tabular}{c}
\subfigure[]{\includegraphics[scale=0.4]{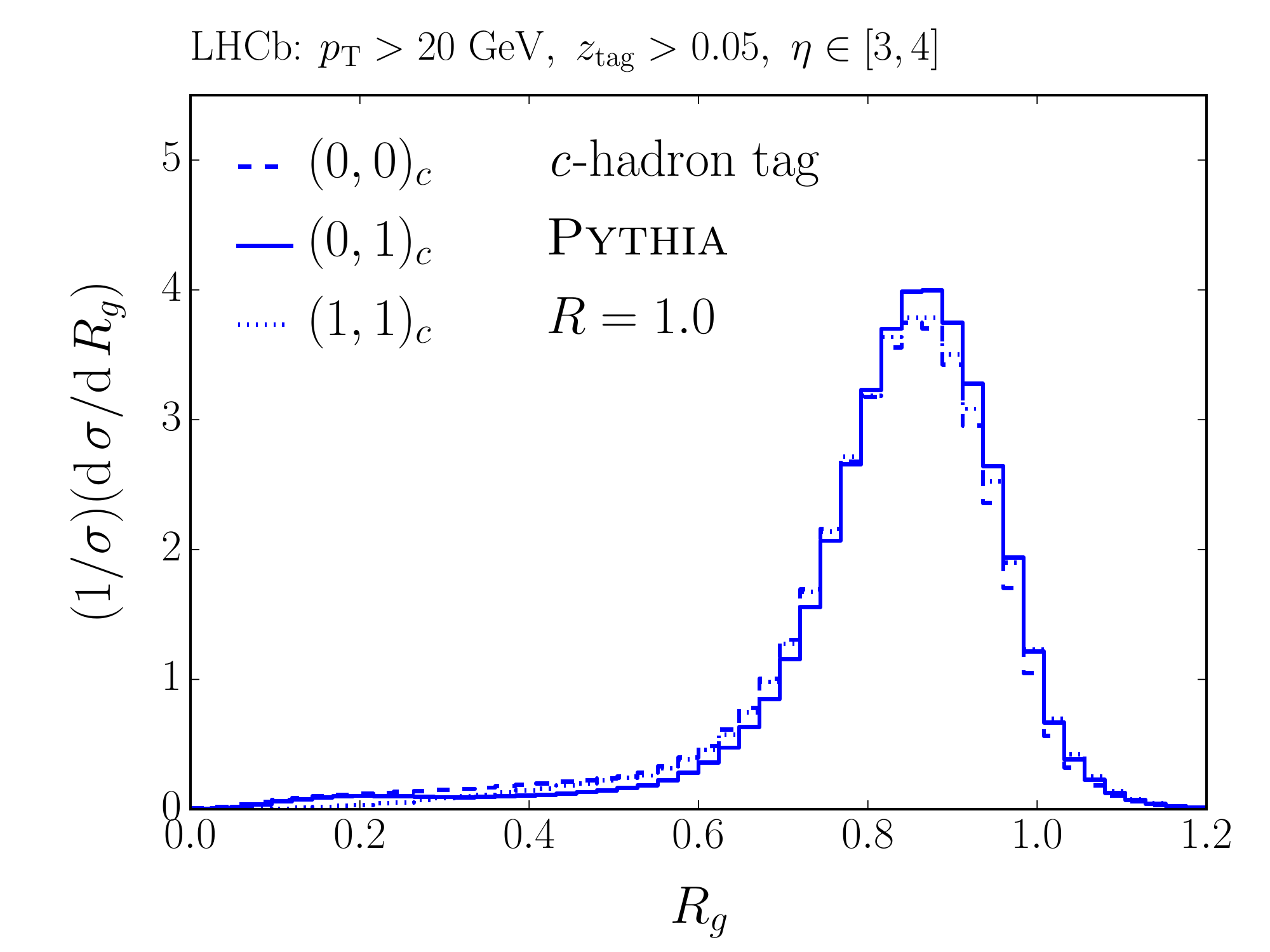}} \hspace{0.1in}
\subfigure[]{\includegraphics[scale=0.4]{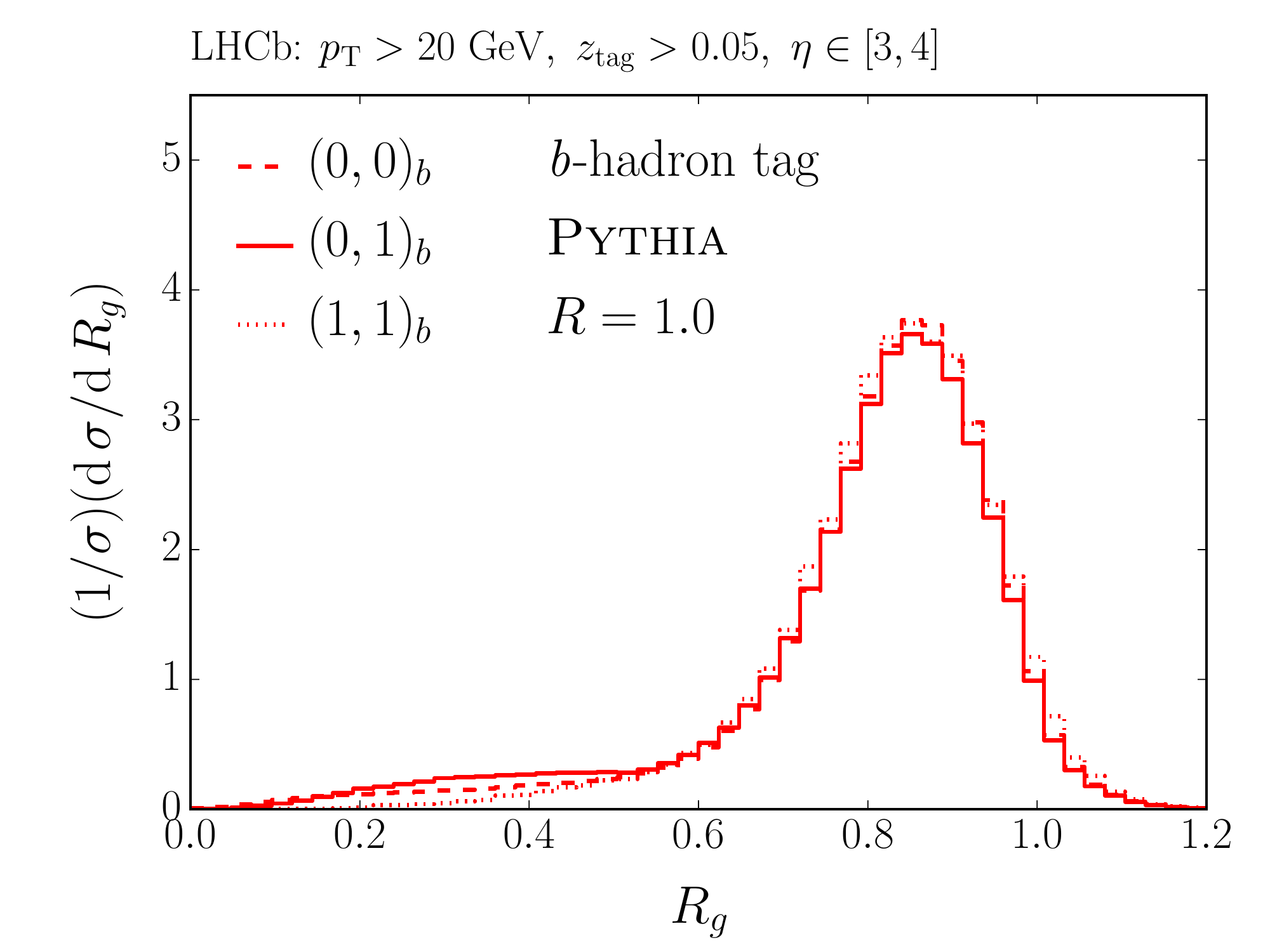}}\\
\subfigure[]{\includegraphics[scale=0.4]{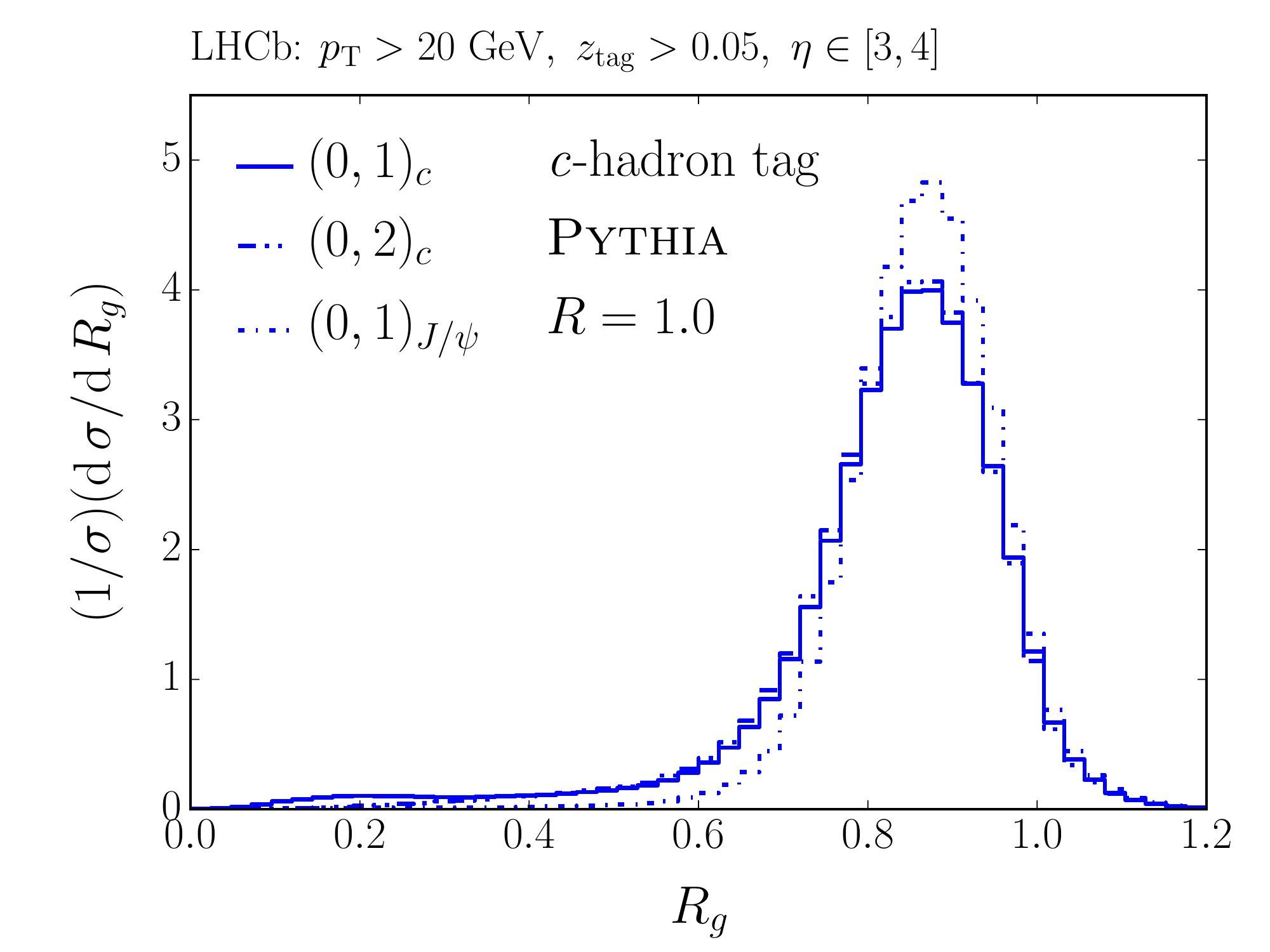}}  \hspace{0.1in}
\subfigure[]{\includegraphics[scale=0.4]{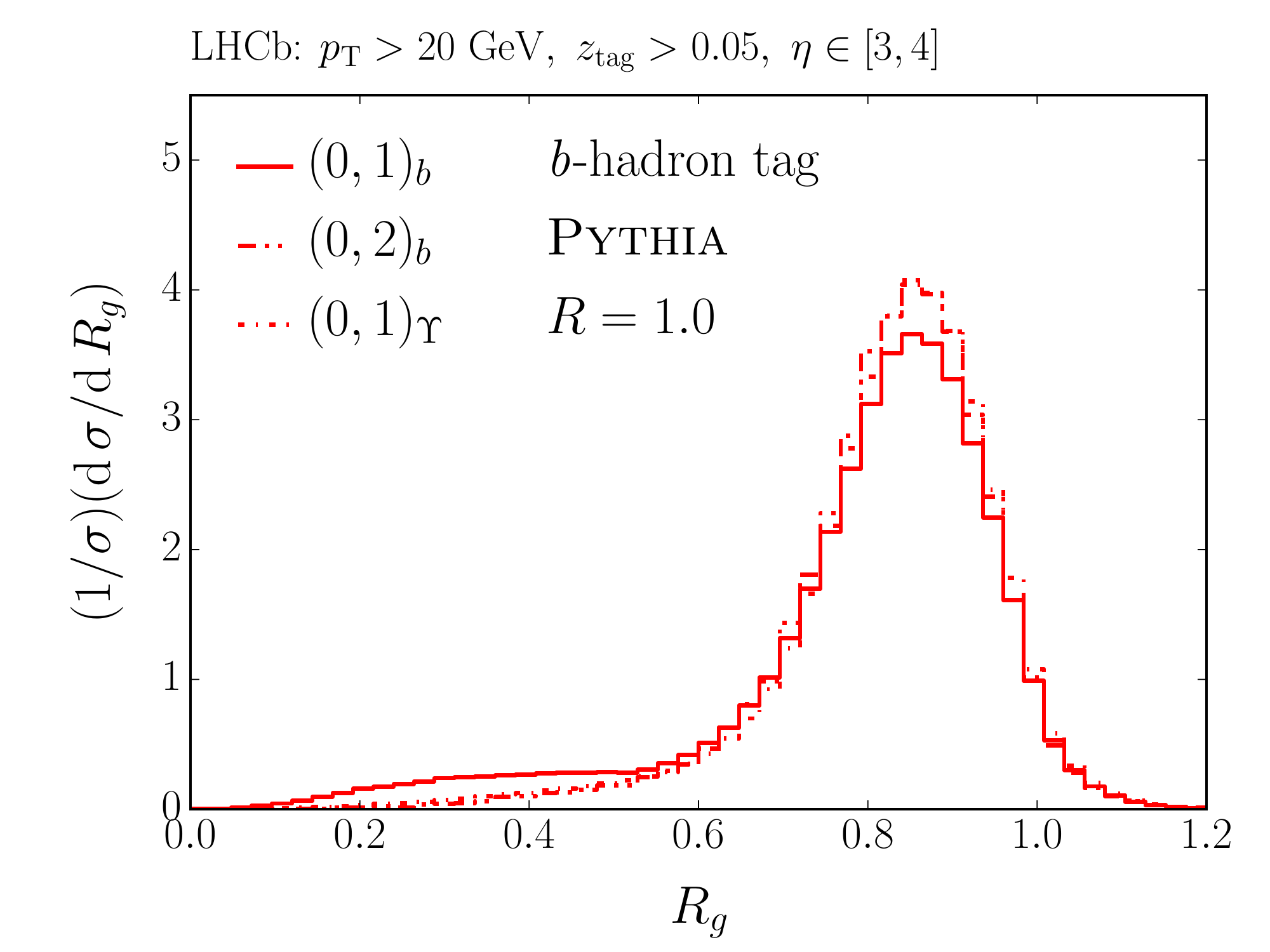}}
\end{tabular}
\caption{The distribution for \sdrop $R_g$ in (left) $c$-tagged and (right) b-tagged events.  The top row (a,b) shows the same flavor categories as \Fig{fig:Kinz}, and the bottom row (c,d) shows the same flavor categories as \Fig{fig:Kinjpsiupsilon}.}
\label{fig:KinR}
\end{figure*}

The two natural observables to describe \sdrop subjets are the momentum sharing $z_g$, defined in \Eq{eq:zg_def}, and the opening angle $R_g$.
In \Sec{sec:Kinematics}, we showed the $z_g$ distributions for our various tagging categories in \Figs{fig:Kinz}{fig:Kinjpsiupsilon}. 
In this appendix, we show the analogous distributions for $R_g$ in \Fig{fig:KinR}.

For the case of $\beta = 0$, the differential $R_g$ cross section was first calculated in \Ref{Dasgupta:2013via}, where it was shown that the $R_g$ distribution exhibits single-logarithmic behavior for high-$\pt$ jets.
For our LHCb study, however, we are working with rather low $\pt$ jets with $\pt > 20 \gev$, where the distributions are largely controlled by nonperturbative effects.
Indeed, the \sdrop algorithm often terminates at the first stage of declustering, such that $R_g \simeq R$, as reflected by peaks shown in \Fig{fig:KinR}.
Given these $R_g$ distributions, it is in some sense surprising that the $z_g$ distributions in \Figs{fig:Kinz}{fig:Kinjpsiupsilon} show no indication for large nonperturbative corrections.

\section{Additional \fcone Distributions}
\label{app:extraplotsFC}

\begin{figure*}[t]
\centering
\begin{tabular}{c}
\subfigure[]{\includegraphics[scale=0.4]{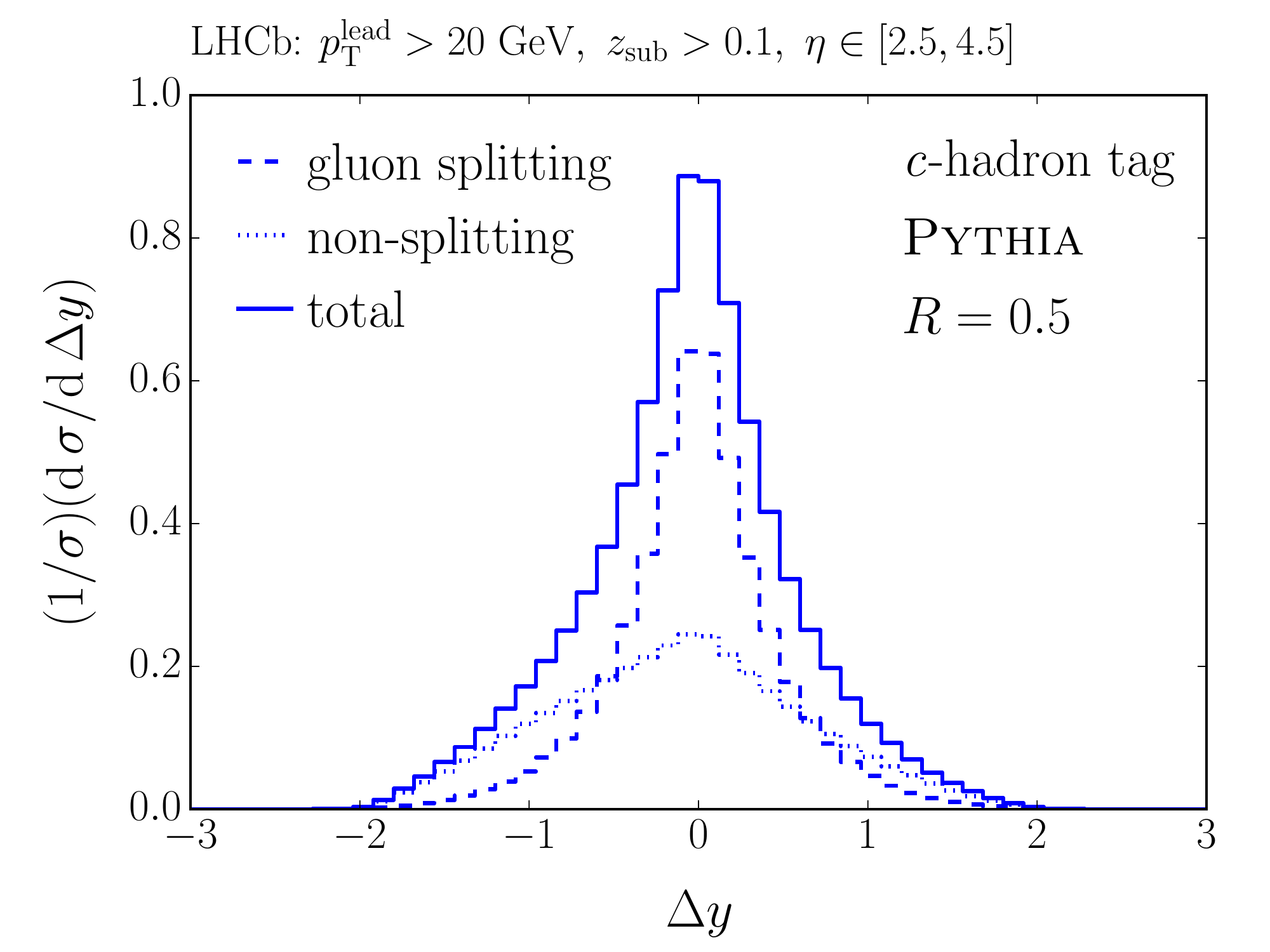}} \hspace{0.1in}
\subfigure[]{\includegraphics[scale=0.4]{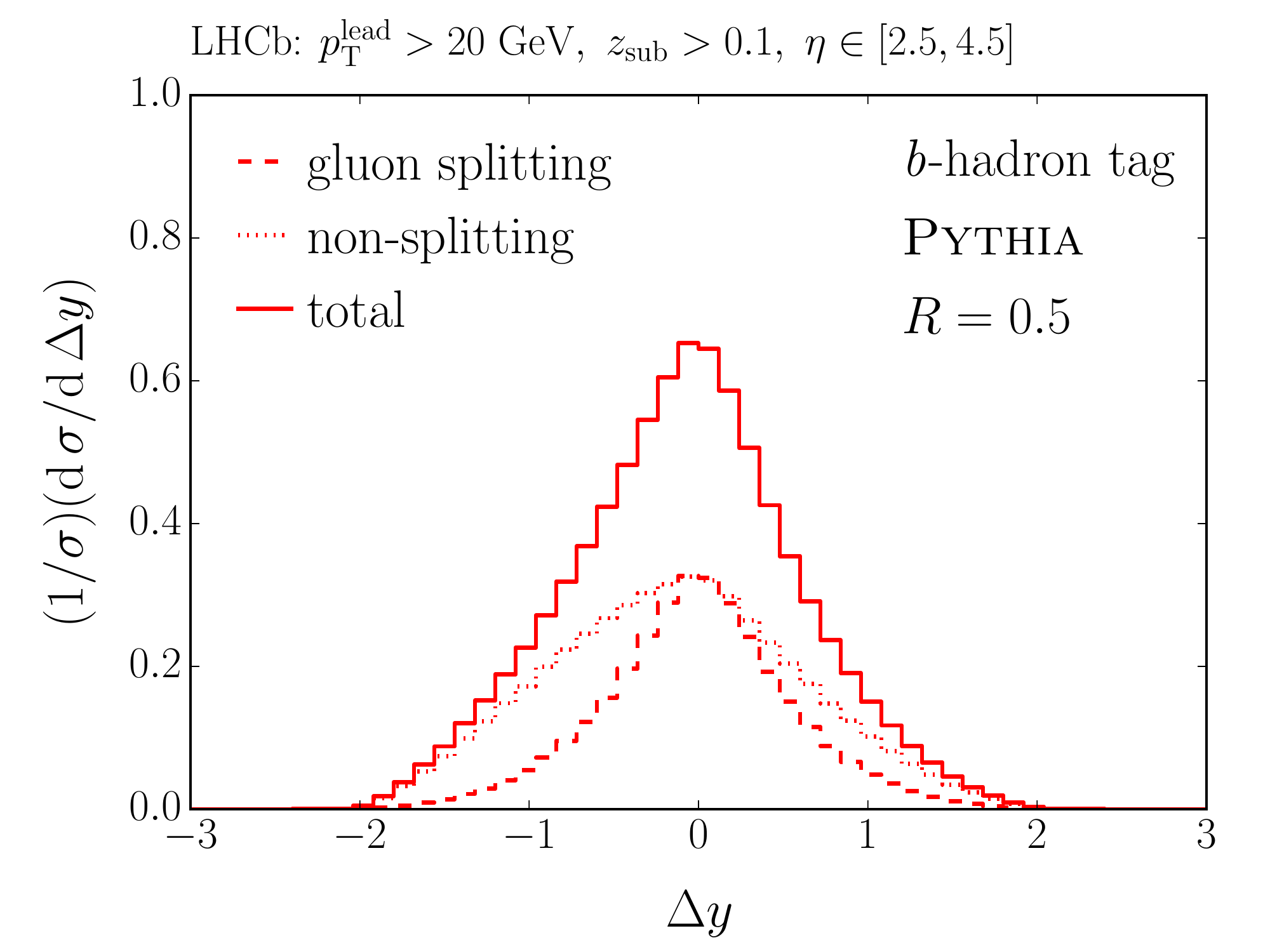}}\\
\subfigure[]{\includegraphics[scale=0.4]{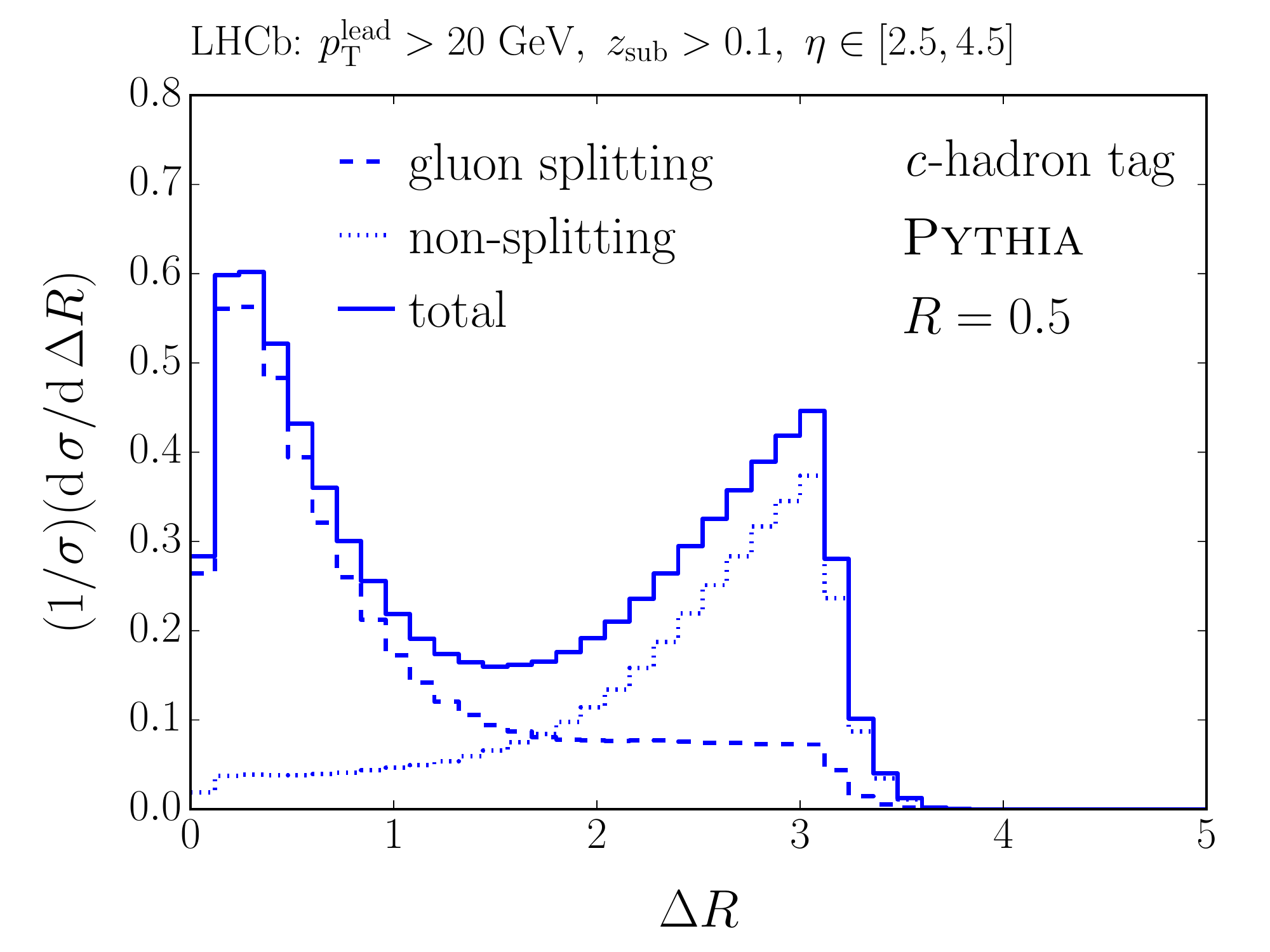}} \hspace{0.1in}
\subfigure[]{\includegraphics[scale=0.4]{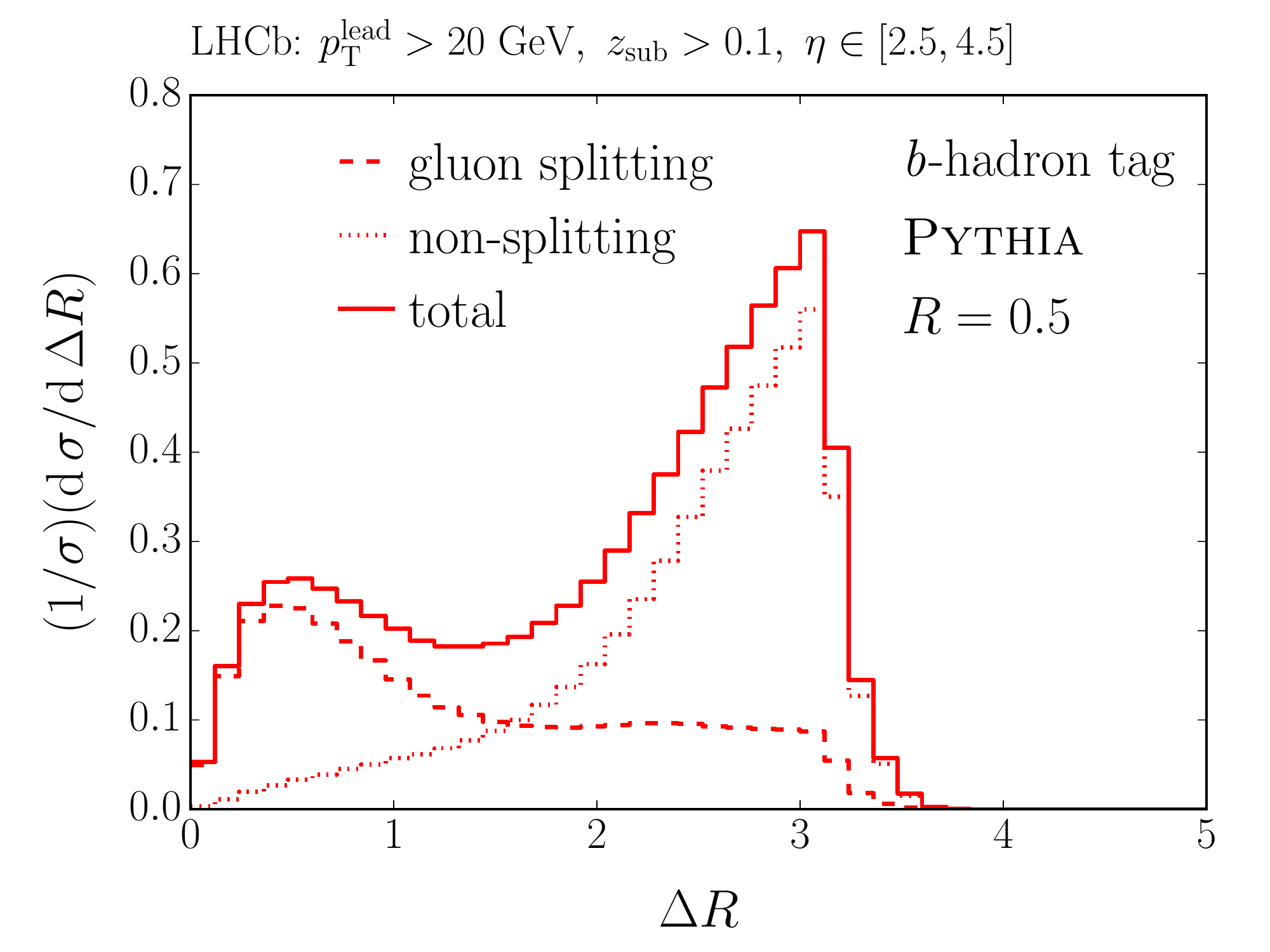}}\\
\subfigure[]{\includegraphics[scale=0.4]{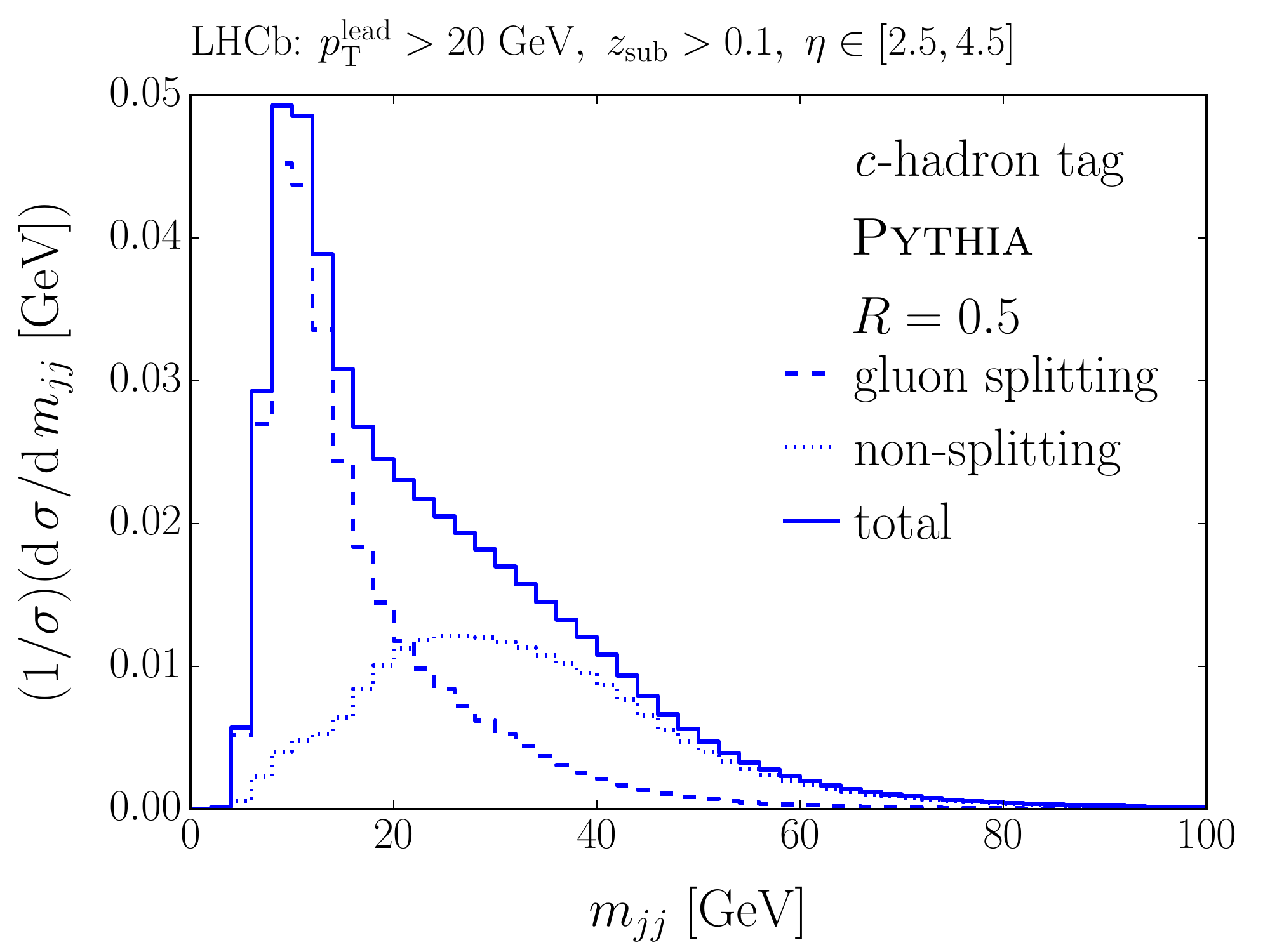}} \hspace{0.1in}
\subfigure[]{\includegraphics[scale=0.4]{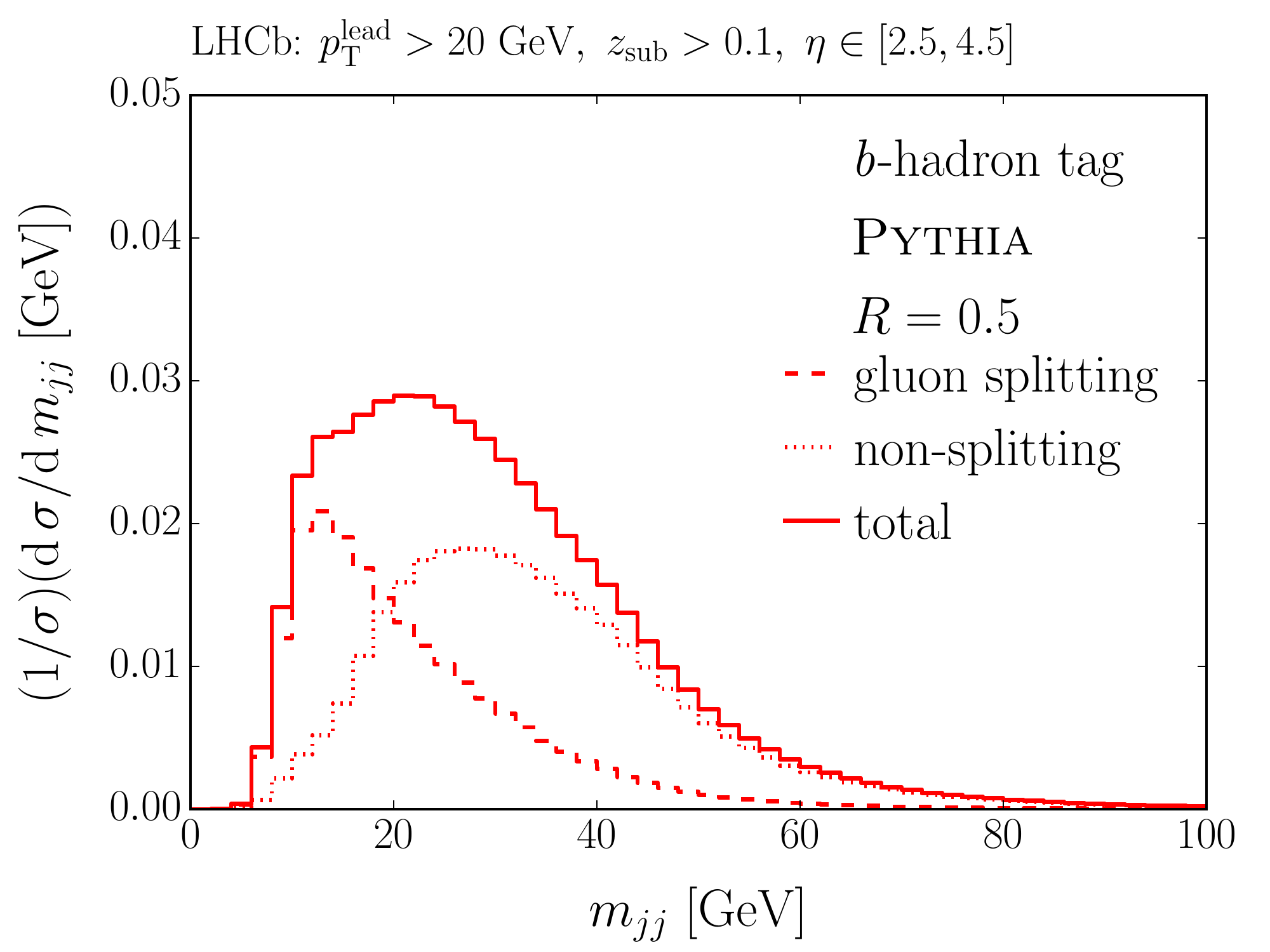}}
\end{tabular}
\caption{Separation observable from the \fcone algorithm for (left) $c$-tagged and (right) $b$-tagged hadrons.  The event selection is the same as \Fig{fig:RateSplitNonSplit}, but now showing (top row) $\Delta y$, (middle row) $\Delta R$, and (bottom row) $m_{jj}$.}
\label{fig:Ratedydrm}
\end{figure*}

In \Sec{sec:Rate}, we demonstrated the separation achievable between splitting and non-splitting events in \pythia using the \fcone algorithm.
The $\Delta \phi$ distribution was shown in \Fig{fig:RateSplitNonSplit}, and for completeness we show the $\Delta y$, $\Delta R$, and $m_{jj}$ observables in \Fig{fig:Ratedydrm}.
As demonstrated in \Tab{table:Prodxsec}, $\Delta R$ and $m_{jj}$ are effective discriminants between gluon splitting and non-splitting events, similar to $\Delta \phi$, though all three variables are highly correlated.
By contrast, $\Delta y$ is not an effective discriminant because of the prevalence of back-to-back dijets from flavor creation.

\section{Implementation at ATLAS and CMS}
\label{app:CMSATLAS}

\begin{figure*}[t]
\centering
\begin{tabular}{c}
\subfigure[]{\label{fig:GPDKinz:a} \includegraphics[scale=0.4]{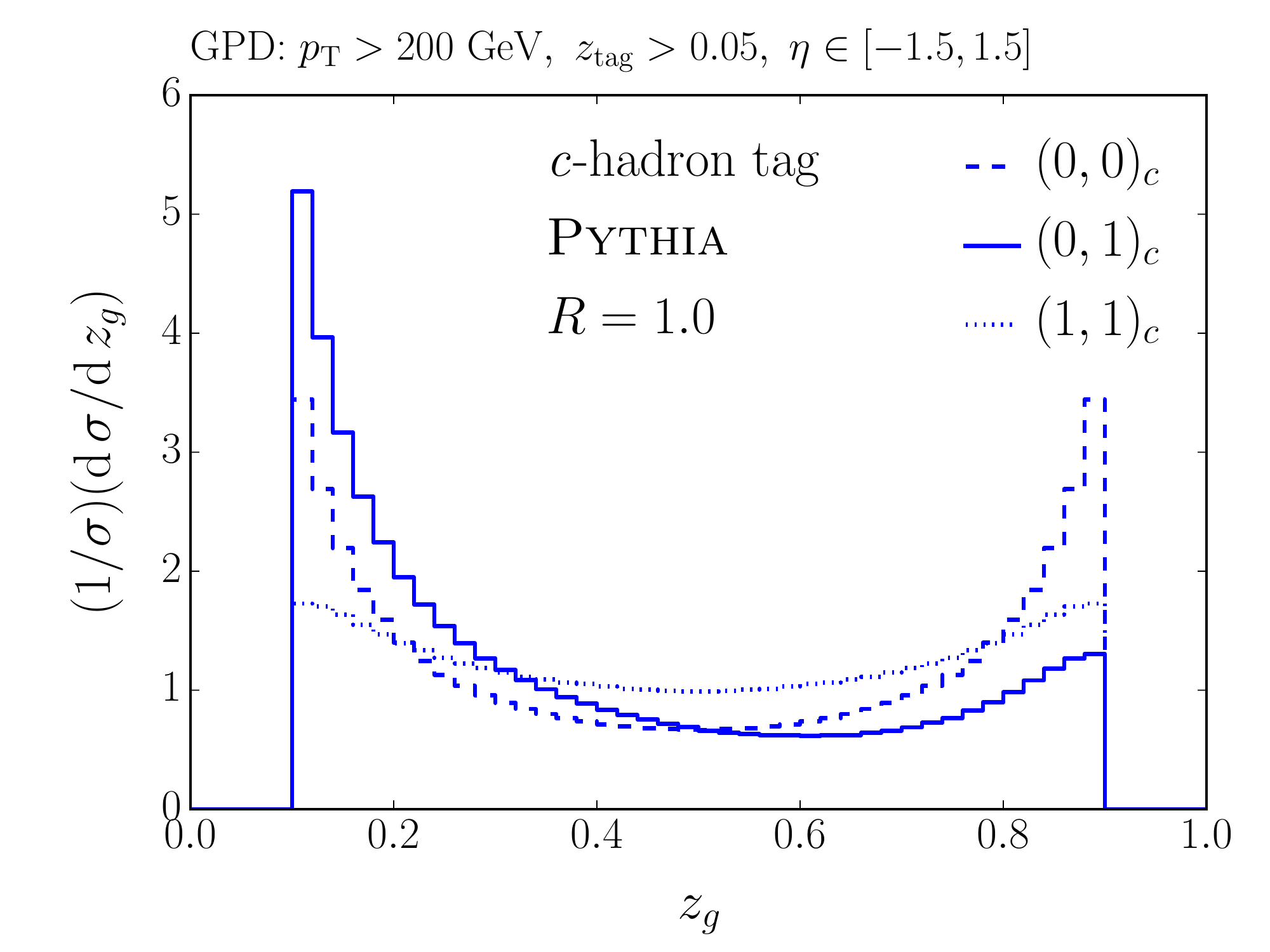}} \hspace{0.1in}
\subfigure[]{\label{fig:GPDKinz:b} \includegraphics[scale=0.4]{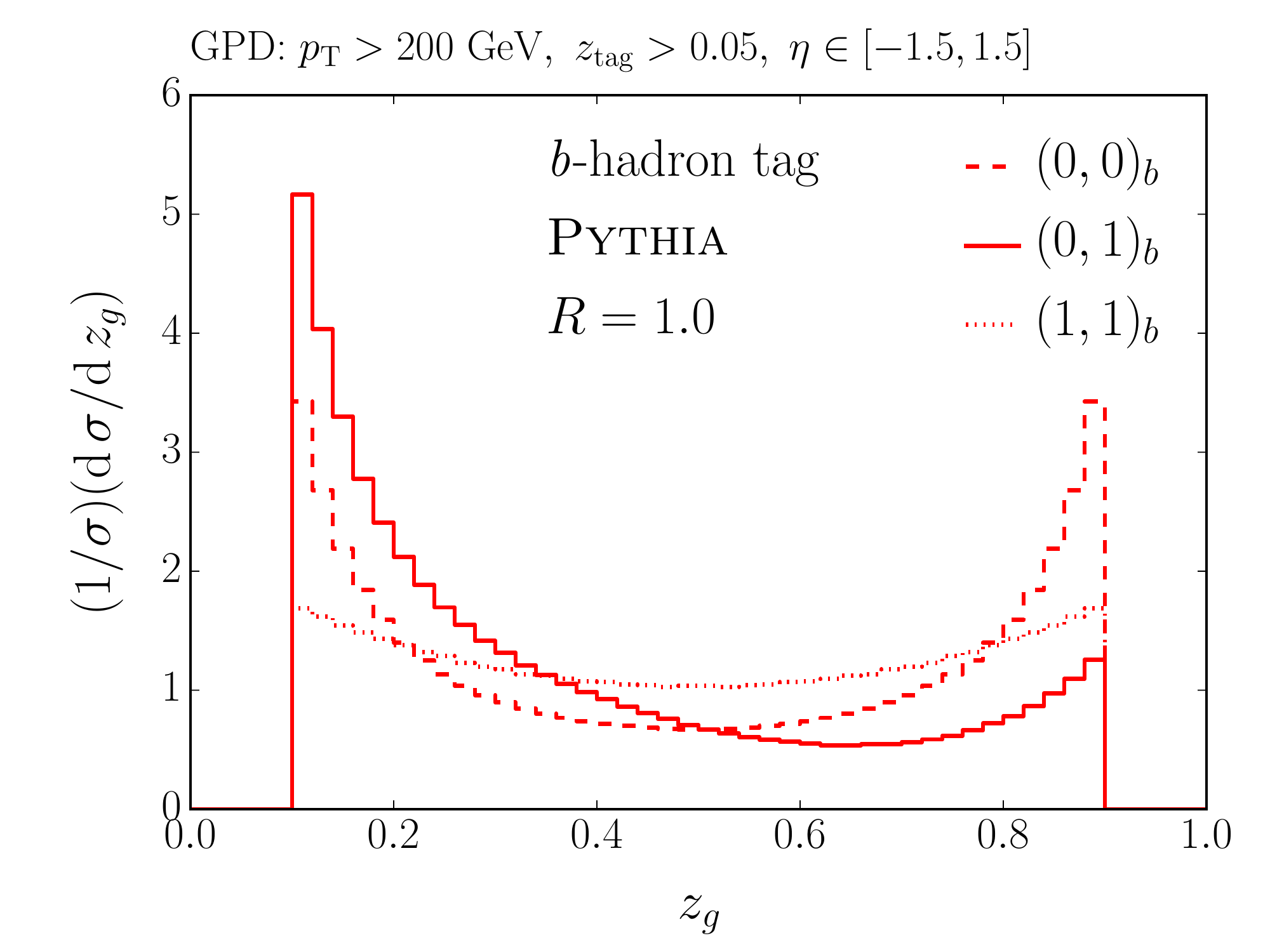}} \\
\subfigure[]{\includegraphics[scale=0.4]{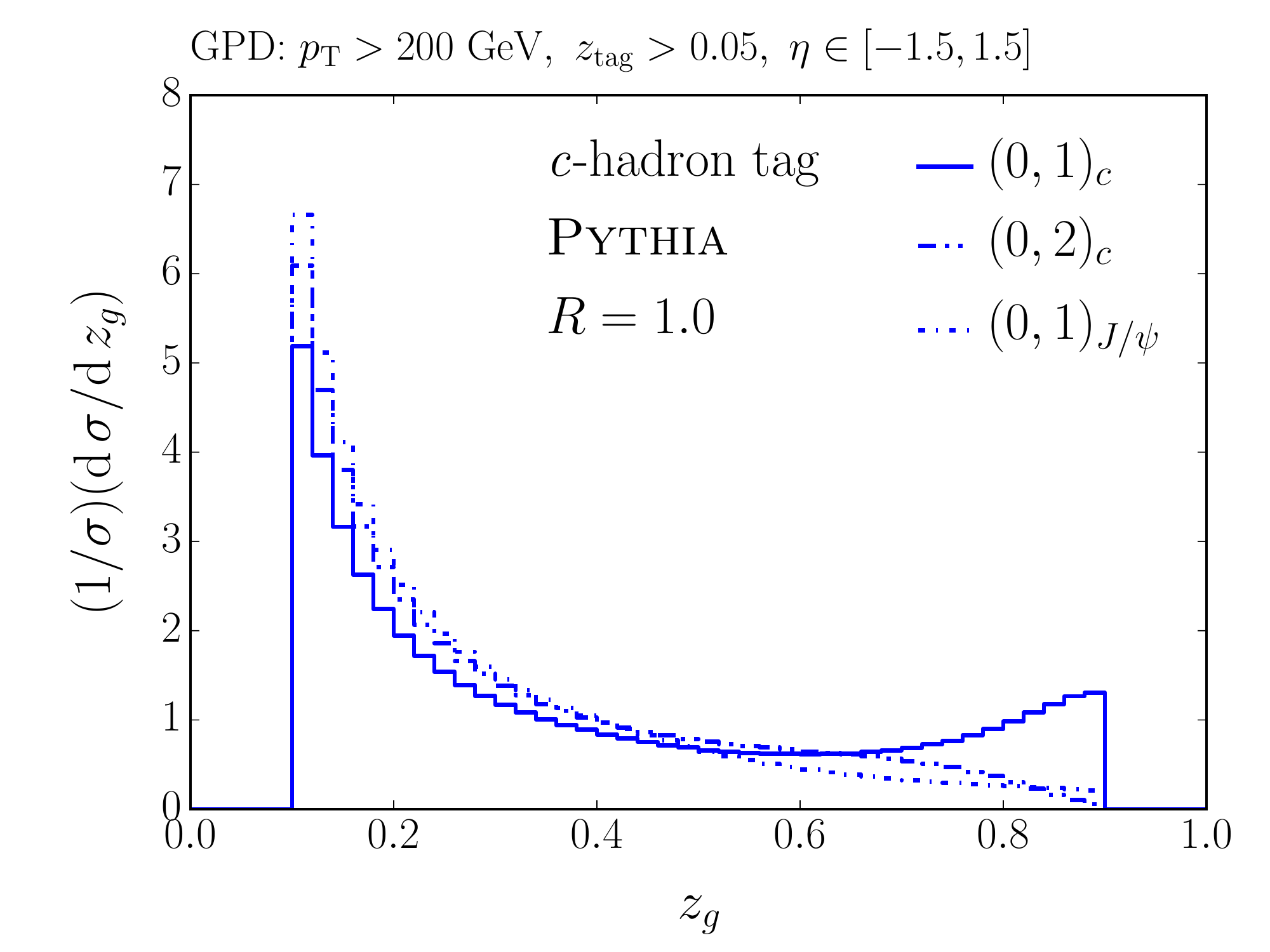}}  \hspace{0.1in}
\subfigure[]{\includegraphics[scale=0.4]{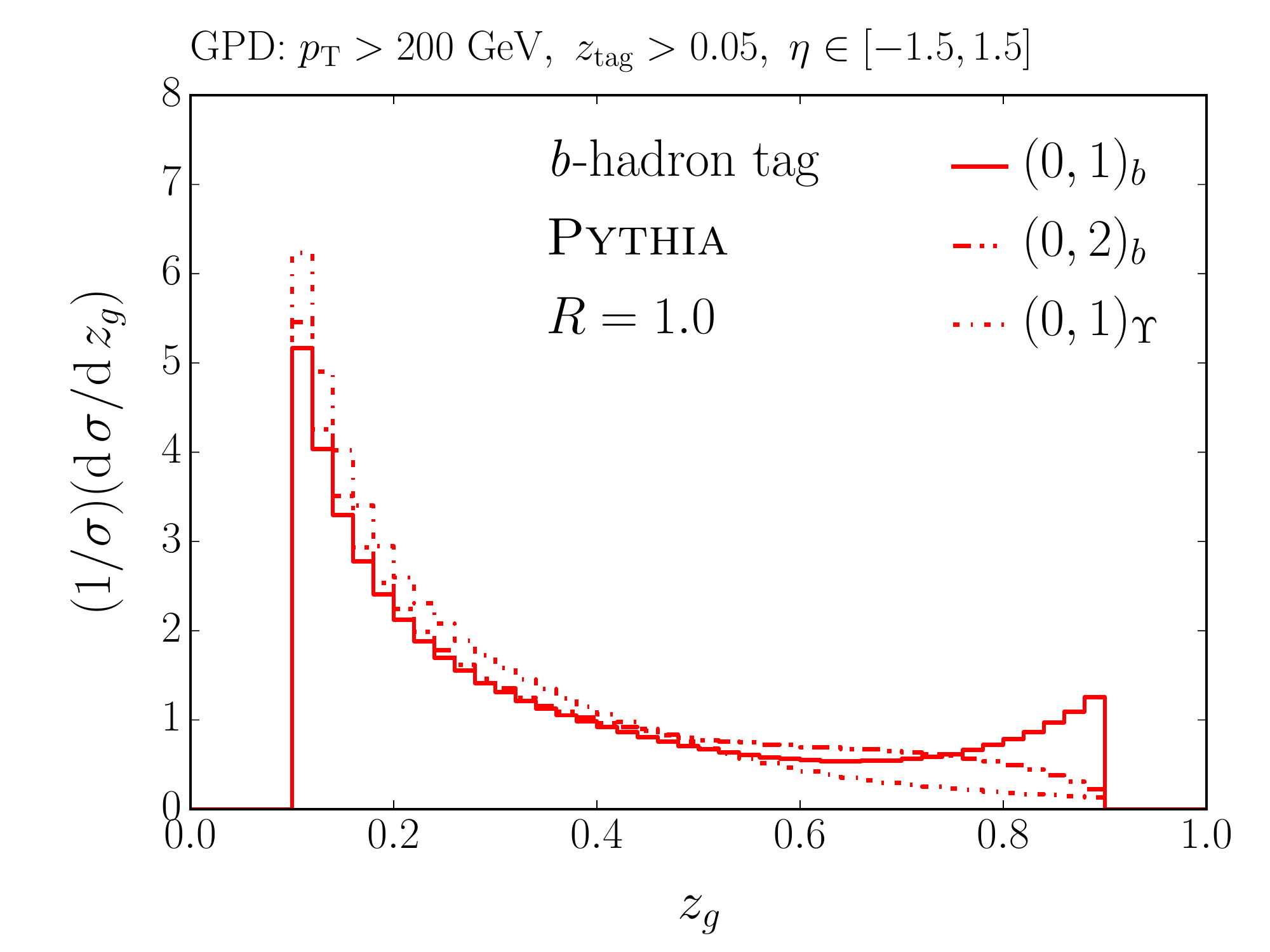}}
\end{tabular}
\caption{The distribution at a GPD for \sdrop $z_g$ in (left) $c$-tagged and (right) $b$-tagged events.  The top row (a,b) is analogous to the LHCb results in \Fig{fig:Kinz} and the bottom row (c,d) corresponds to \Fig{fig:Kinjpsiupsilon}.}
\label{fig:GPDKinz}
\end{figure*}

\begin{figure*}[t]
\centering
\begin{tabular}{c}
\subfigure[]{\includegraphics[scale=0.4]{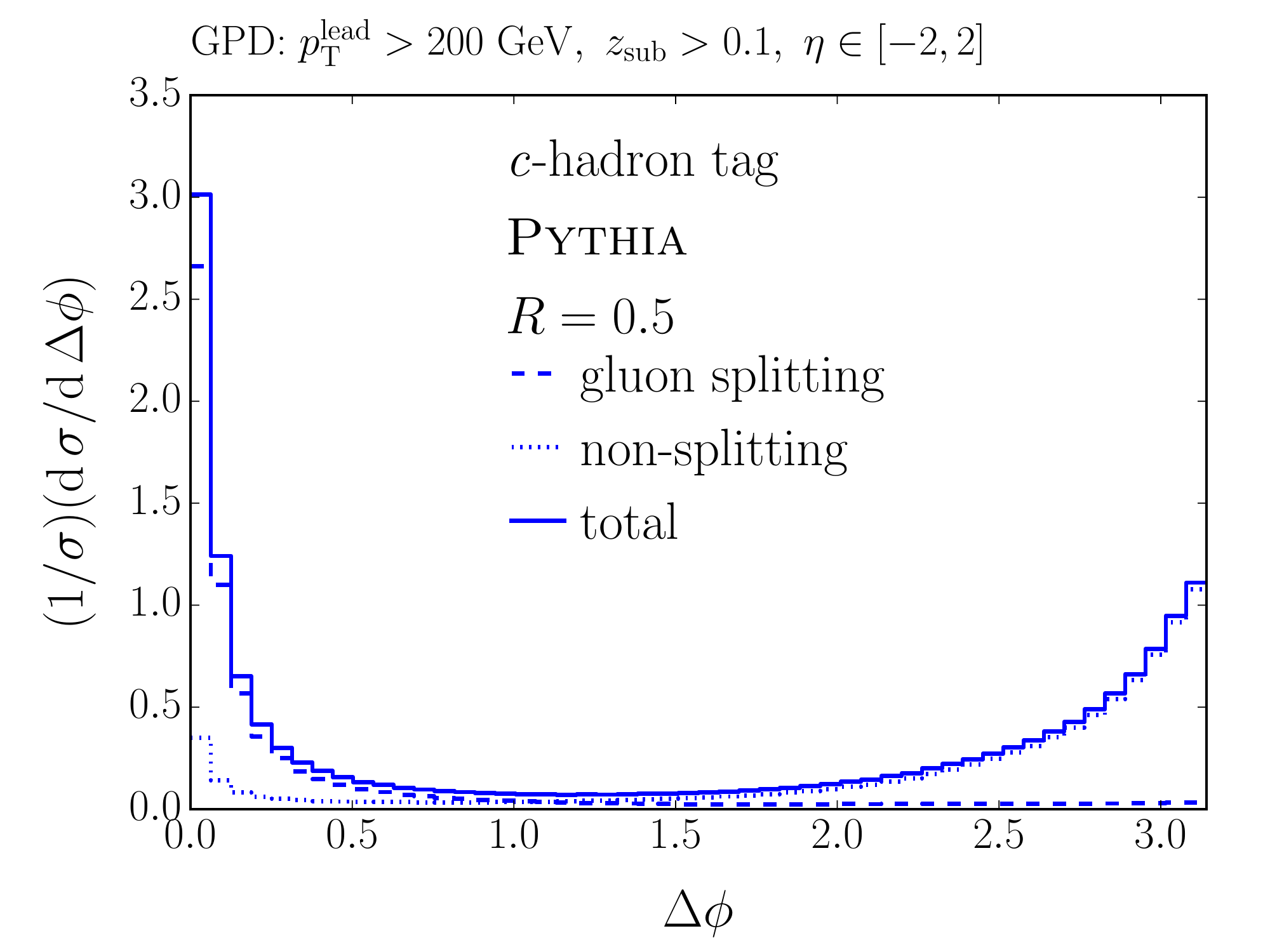}} \hspace{0.1in}
\subfigure[]{\includegraphics[scale=0.4]{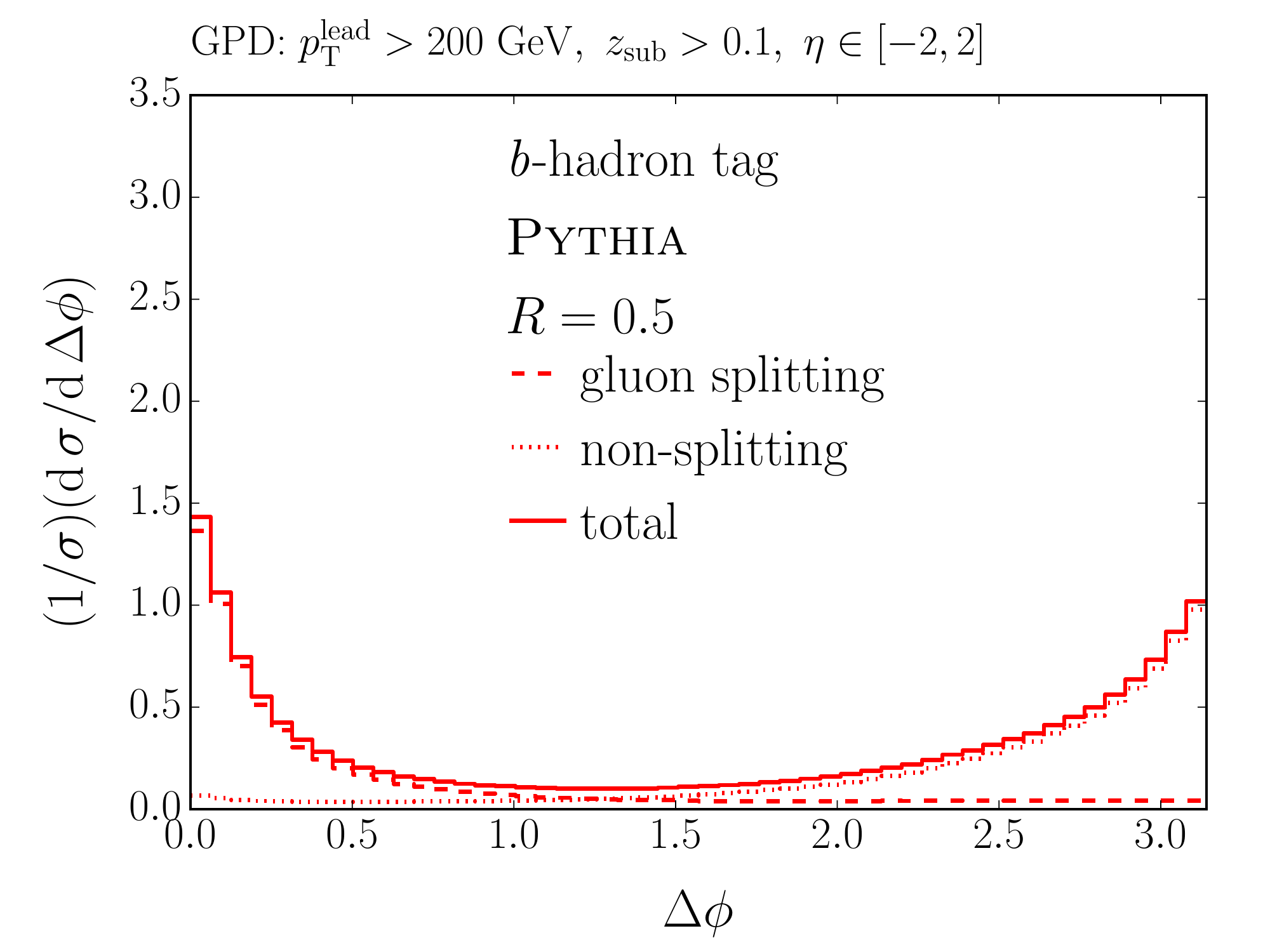}}
\end{tabular}
\caption{The distribution at a GPD for the \fcone $\Delta \phi$ observable in (a) $c$-tagged and (b) $b$-tagged events.  These plots are analogous to the LHCb results in \Fig{fig:RateSplitNonSplit}.}
\label{fig:RateSplitNonSplitGPD}
\end{figure*}

\noindent

While the focus of our study has been on the LHCb experiment, the \sdrop and \fcone strategies can also be applied at ATLAS and CMS, which we refer to as general purpose detectors (GPDs).
The primary experimental challenge at GPDs is achieving the required heavy-flavor-hadron tagging efficiencies, as well as associated $\pt$ and flight-direction measurements.
As emphasized in \Sec{sec:ImplementationAtLHCb}, it is known how these challenges can be overcome at LHCb in the near term. 
Nevertheless, CMS has already probed small angular separations between $b$-hadrons in \Ref{Khachatryan:2011wq}, so in the future, we expect the stringent heavy-flavor-tagging requirements for \sdrop and \fcone can be met at a GPD.

To show the expected performance of our methods at a GPD, we repeat the main results from our parton-shower studies, albeit with three changes to the analysis workflow.
First, we require the entire jet to be contained within the GPD acceptance of $\eta \in [-2.5,2.5]$ (instead of $\eta \in [2,5]$).
Second, we increase the \pt threshold to $200 \gev$ (instead of  $20 \gev$) to account for the typical jet scales used for triggering at a GPD.
Third, we change the $\pt$ threshold for identifying heavy-flavor hadrons to be $20 \gev$ (instead of $2 \gev$) to account for the difficulty of GPDs to resolve low $\pt$ tracks.

The \sdrop $z_g$ results for the GPD workflow are shown in \Fig{fig:GPDKinz}, to be compared to the LHCb results in \Figs{fig:Kinz}{fig:Kinjpsiupsilon}.
As one goes to the higher-$\pt$ regime of the GPDs, there are a number of important differences.
At the perturbative level, one expects relatively little change in the $z_g$ distributions as a function of $\pt$\,\cite{Larkoski:2015lea}.
That said, the nonperturbative effect of underlying event is very relevant at low jet $\pt$ values, where it tends to make $z_g$ more flatly distributed.
Thus, going to higher $\pt$ in \Fig{fig:GPDKinz}, the $z_g$ distributions exhibit stronger singularities towards $z \to 0$ and $z \to 1$ as expected.
Another important difference is related to category migration.
Because we are taking the $z_{\rm tag}$ requirement to be half of the $z_{\rm cut}$ requirement, there is phase space for subsequent $g \to Q \bar{Q}$ emissions to cause noticeable migration from the $(0,0)$ to the $(0,1)$ category.
This was not as much of an issue at low jet $\pt$, since the phase space for additional gluon radiation was more restricted.
As one goes to higher jet $\pt$, one could mitigate this category migration by choosing a higher $z_{\rm tag}$ requirement, at the expense of introducing shoulders in the $z_g$ distribution.
In practice, one would probably want to make measurements of $z_g$ with multiple $z_{\rm tag}$ values, to test the stability of the distributions to the tagging conditions.

The \fcone $\Delta \phi$ distributions for the GPD workflow are shown in \Fig{fig:RateSplitNonSplitGPD}, to be compared to the LHCb results in \Fig{fig:RateSplitNonSplit}.
Here, the relative size of the splitting and non-splitting categories are different, with \fcone providing even greater separation than achieved at lower \pt.
We conclude that the GPDs should be able to successfully use these analysis strategies to distill the kinematics and rates associated with heavy flavor, probing a complementary phase space compared to LHCb.

\bibliography{heavy_flavor_lhcb}

\end{document}